\documentclass[review,12pt]{elsarticle}

\usepackage{amssymb}
\usepackage{amsthm}
\usepackage{amsmath}
\usepackage{mathrsfs}
\usepackage{graphicx}
\usepackage{graphics}
\usepackage{epstopdf}
\usepackage{float}
\usepackage{caption}
\usepackage{subcaption}
\usepackage{bm}
\usepackage{url}
\usepackage{bbm}
\usepackage{mathrsfs}
\usepackage{cleveref}
\usepackage{soul}
\usepackage{tikz} 
\usepackage{upgreek}
\usepackage{array}
\usepackage[margin=2cm]{geometry}
\usetikzlibrary{shapes.geometric, arrows} 
\usepackage{adjustbox}
\usepackage{algorithm}
\usepackage{algpseudocode}
\usepackage{booktabs}
\biboptions{sort&compress}

\tikzstyle{startstop} = [rectangle, rounded corners, minimum width=1cm, minimum height=1cm,text centered, draw=black, fill=red!5]
\tikzstyle{io} = [trapezium, trapezium left angle=70, trapezium right angle=110, minimum width=1cm, minimum height=1cm, text centered, draw=black, fill=blue!5]
\tikzstyle{process} = [rectangle, minimum width=1cm, minimum height=1cm, text centered, draw=black, fill=orange!5]
\tikzstyle{decision} = [diamond, minimum width=1cm, minimum height=0.5cm, text centered, draw=black, fill=green!5]
\tikzstyle{arrow} = [thick,->,>=stealth]

\journal{Computer Methods in Applied Mechanics and Engineering}

\makeatletter
\def\@author#1{\g@addto@macro\elsauthors{\normalsize%
    \def\baselinestretch{1}%
    \upshape\authorsep#1\unskip\textsuperscript{%
      \ifx\@fnmark\@empty\else\unskip\sep\@fnmark\let\sep=,\fi
      \ifx\@corref\@empty\else\unskip\sep\@corref\let\sep=,\fi
      }%
    \def\authorsep{\unskip,\space}%
    \global\let\@fnmark\@empty
    \global\let\@corref\@empty  
    \global\let\sep\@empty}%
    \@eadauthor={#1}
}
\makeatother

\makeatletter
\def\thickhline{%
  \noalign{\ifnum0=`}\fi\hrule \@height \thickarrayrulewidth \futurelet
   \reserved@a\@xthickhline}
\def\@xthickhline{\ifx\reserved@a\thickhline
               \vskip\doublerulesep
               \vskip-\thickarrayrulewidth

             \fi
      \ifnum0=`{\fi}}
\makeatother

\newlength{\thickarrayrulewidth}
\begin{document}

\begin{frontmatter}



\title{A phase field model for hydraulic fracture: Drucker-Prager driving force and a hybrid coupling strategy}


\author{Yousef Navidtehrani\fnref{Uniovi}}
\author{Covadonga Beteg\'{o}n \fnref{Uniovi}}
\author{Javier Vallejos\fnref{UniChile}}
\author{Emilio Mart\'{\i}nez-Pa\~neda\corref{cor1}\fnref{OXFORD}}
\ead{emilio.martinez-paneda@eng.ox.ac.uk}

\address[Uniovi]{Department of Construction and Manufacturing Engineering, University of Oviedo, Gij\'{o}n 33203, Spain}

\address[UniChile]{Advanced Mining Technology Center, Mining Engineering Department, University of Chile, Chile}

\address[OXFORD]{Department of Engineering Science, University of Oxford, Oxford OX1 3PJ, UK}

\cortext[cor1]{Corresponding author.}

\begin{abstract} 
Recent years have seen a significant interest in using phase field approaches to model hydraulic fracture, so as to optimise a process that is key to industries such as petroleum engineering, mining and geothermal energy extraction. Here, we present a novel theoretical and computational phase field framework to simulate hydraulic fracture. The framework is general and versatile, in that it allows for improved treatments of the coupling between fluid flow and the phase field, and encompasses a universal description of the fracture driving force. Among others, this allows us to bring two innovations to the phase field hydraulic fracture community: (i) a new hybrid coupling approach to handle the fracture-fluid flow interplay, offering enhanced accuracy and flexibility; and (ii) a Drucker-Prager-based strain energy decomposition, extending the simulation of hydraulic fracture to materials exhibiting asymmetric tension–compression fracture behaviour (such as shale rocks) and enabling the prediction of geomechanical phenomena such as fault reactivation and stick-slip behaviour. Four case studies are addressed to illustrate these additional modelling capabilities and bring insight into permeability coupling, cracking behaviour, and multiaxial conditions in hydraulic fracturing simulations. The codes developed are made freely available to the community and can be downloaded from \url{https://mechmat.web.ox.ac.uk/}.\\

\end{abstract}


\end{frontmatter}



\section{Introduction}
\label{sec:Introduction}

Hydraulic fracturing plays a pivotal role in industries such as petroleum engineering \cite{barati2014review}, mining \cite{Rojas2017}, geothermal energy extraction \cite{LEI2023102692}, and various subsurface operations. Due to the necessity of accurately predicting fracture behaviour, extensive research has been dedicated to hydraulic fracturing using theoretical \cite{SAMPATH2018251}, numerical \cite{LECAMPION201866}, and experimental approaches \cite{2205_85_Vallejos,2205_81_Amorer}. Among the various numerical methods used to simulate hydraulic fracture, the phase field approach has become particularly popular due to its ability to simulate complex fracture phenomena in a mesh-objective and robust fashion, for arbitrary geometries and dimensions, and without requiring explicit crack tracking \cite{YI2020113396}. \\

Phase field models for hydraulic fracture were pioneered by Bourdin \textit{et al.} \cite{Bourdin2012a}, laying a robust foundation that has been further developed in numerous subsequent works \cite{Wheeler2014,mikelic2015phase,Mikelic2015,mikelic2015quasi}. While discrete methods allow for explicit computation of the displacement jump (crack opening), continuum-based approaches such as the phase field fracture method require specialised treatment \cite{hageman2023phase}. Bourdin \textit{et al.} \cite{Bourdin2012a} introduced an integration method that approximates crack opening by integrating the displacement field and weighting it by the phase field gradient. Building on this framework, Miehe \textit{et al.} \cite{miehe2015minimization,Miehe2016} incorporated a modified Darcy’s law to model fluid flow between crack surfaces as a Poiseuille-type flow. Their approach estimates fluid flux by solving laminar flow equations between parallel surfaces, deriving an anisotropic permeability tensor from crack opening computations, which account for element size and phase field gradient direction. Wilson and Landis \cite{Wilson2016} addressed element size effects on crack opening by introducing a viscosity-scaling factor. Ehlers and Luo \cite{EHLERS2018429} proposed a crack-opening indicator to address the challenges associated with the phase field modelling of closed fractures or fractures that reclose after formation, where fluid flow transitions between Darcy-type and Navier–Stokes-type regimes. Heider and Markert \cite{Heider2017} integrated phase field fracture models with the Theory of Porous Media (TPM) to simulate the multiphase behaviour of saturated porous media. Alternatively, Lee \textit{et al.} \cite{Lee2016} employed auxiliary fields to segment the domain into reservoir, transient, and fracture regions, estimating material properties via linear interpolation across these regions, eliminating the need for explicit crack opening computations. This methodology has been widely adopted in subsequent research \cite{Zhou2018c,LI2021107887,LEE2025126487,WANG2025117750}. Additionally, Lee \textit{et al.} \cite{lee2017iterative} introduced a level-set method for calculating crack opening. Later, Yoshioka \textit{et al.} \cite{yoshioka2020crack} compared the line integral method and the level-set method for computing crack opening within the phase field framework. While the line integral method is theoretically robust, its implementation poses challenges. In contrast, the level-set method is more practical, albeit requiring parameter adjustments to achieve acceptable results. Santillán et al. \cite{https://doi.org/10.1002/2016JB013572} developed a phase field approach to simulate fluid-driven fractures in elastic materials, employing an immersed-fracture formulation to accurately capture fracture propagation. Formulations able to capture the role of inertia were developed by Zhou \textit{et al.} \cite{Zhou2019c} and Shahoveisi \textit{et al.} \cite{SHAHOVEISI2024117113}. Efforts have also been directed towards reducing the computational cost. For example, Lusheng \textit{et al.} \cite{yang2024phase} used the length-scale insensitive degradation function developed by Lo \textit{et al.} \cite{Lo2023} to tackle large-scale hydraulic fracture problems, while Aldakheel and co-workers \cite{Aldakheel2021} proposed a global–local approach, confining fracture computations to a local domain linked to the global domain via a Robin-type interface condition. Additionally, phase field fracture has been employed to model the initiation and propagation of desiccation fractures in porous media \cite{maurini2014crack, LUO2023115962}. For a comprehensive review of phase field hydraulic fracture the reader is referred to Refs. \cite{Heider2021, chen2022review}.\\

While these recent developments have brought significant progress, establishing phase field modelling as the leading technique in simulating hydraulic fractures, there are aspects of the formulation that need further development to enable accurate and versatile predictions, as needed to capture real site behaviour \cite{2063_15_Rimmelin,Pardo201611}. In this work, we present a formulation that encompasses relevant developments of hydraulic phase field fracture in a single framework, and adds two novel and important contributions. First, we present a new hybrid coupling approach to link the phase field evolution equation with pore pressure more effectively. As demonstrated in the numerical experiments conducted, this approach enhances both flexibility and accuracy in capturing the interactions between fracture and fluid flow in complex environments. Second, building upon our recent work \cite{Navidtehrani2022}, we enrich existing models with a general decomposition of the phase field fracture driving force. This is of key importance and a popular topic in the phase field fracture community as there is a need to enrich models with arbitrary failure surfaces to capture the nucleation and growth of cracks exhibiting asymmetrical tension–compression fracture behaviour \cite{molnar2020toughness,de2022nucleation,lopez2025classical}. Rocks and other quasi-brittle materials exhibit failure criteria that are well-described by Drucker-Prager or Mohr-Coulomb type of failure surfaces and thus a general treatment of hydraulic fractures in shale rocks requires this development to capture both tensile and shear-dominated failures. Accordingly, we particularise our generalised model to a Drucker-Prager-based decomposition of the strain energy density, the fracture driving force, which allows us to simulate geomechanical phenomena like stick-slip behaviour and fault activation. Insight is also gained on the role of the fracture driving force on the crack trajectory and the peak pore pressure in problems involving interactions between multiple cracks. The manuscript is organised as follows. First, in Section \ref{Sec:Theory}, we present our phase field-based formulation for hydraulic fracture. We begin by discussing the phase field description of crack evolution, through appropriate constitutive choices and various approaches to decompose the strain energy density. Then, we discuss fluid flow theory in porous media, including Darcy’s law and Biot’s poroelasticity. Three distinct coupling methods for phase field fracture and fluid flow are introduced and evaluated. The numerical implementation, which takes advantage of the analogy between the heat transfer and the fluid flow and phase field equations, is given in Section \ref{Sec:NumImplementation}. In Section \ref{Sec:Examples}, four case studies are presented, demonstrating the practical application of the proposed framework and highlighting the importance of the novel ingredients of the model. Hence, the numerical experiments encompass permeability coupling, stick-slip behaviour, crack interaction issues, and multiaxial stress conditions. These case studies illustrate the robustness and adaptability of our generalised framework in modelling hydraulic fractures across diverse geomechanical scenarios. Finally, concluding remarks are given in Section \ref{Sec:Conclusions}. 

\section{A phase field-based model for hydraulic fracture}
\label{Sec:Theory}

The coupled equations of phase field hydraulic fracture are presented in this section. Consider an elastic body occupying an arbitrary domain $\Omega \subset {\rm I\!R}^n$ $(n \in [1,2,3])$, with an external boundary $\partial \Omega$, where the outward unit normal is denoted by $\mathbf{n}$. The primary variables considered are the displacement vector field $\mathbf{u}$, the phase field variable $\phi$, and the pore pressure of the fluid $p$. Assuming small strain and isothermal conditions, the strain tensor is defined as $\boldsymbol{\varepsilon}= \left(\nabla\mathbf{u}^T+\nabla\mathbf{u}\right)/2$.\\

The damage process is described by a smooth scalar field $\phi \in [0,1]$, referred to as the phase field. In this model, $\phi=0$ represents the undamaged material, while $\phi=1$ corresponds to a fully cracked state. The phase field value transitions smoothly between these two extremes, representing intermediate states of damage. The length scale parameter $\ell$ controls the extent of crack regularization, allowing for a diffuse approximation of cracks. The phase field formulation introduces a crack density function $\gamma(\phi,\nabla\phi)$, approximating the fracture energy as:
\begin{equation}
    \Phi=\int_{\Gamma} G_c \, \text{d}S \approx \int_\Omega G_c\gamma(\phi,\nabla\phi) \, \text{d}V, \hspace{1cm} \text{for } \ell\rightarrow 0,
\end{equation}

\noindent where $G_c$ denotes the critical energy release rate for fracture, as established in classical fracture mechanics \citep{Griffith1920,kristensen2021assessment}. Using the principle of virtual work, the equations governing the coupled deformation-fracture-pore system are expressed as:
\begin{equation}\label{eq:energy}
 \int_\Omega \big\{ \bm{\sigma}:\delta\boldsymbol{\varepsilon}  - \mathbf{b} \cdot \delta \mathbf{u} +\omega\delta\phi + \bm{\upxi} \cdot \delta \nabla \phi + \dot{\zeta} \delta p- \mathbf{q} \cdot \delta \nabla p - q_m \delta p
    \big\} \, \text{d}V =  \int_{\partial \Omega} \left( \mathbf{T} \cdot \delta \mathbf{u} + q \delta p \right) \, \text{d}S ,
\end{equation}

\noindent where $\delta$ represents a virtual quantity, $\bm{\sigma}$ is the Cauchy stress tensor, $\mathbf{b}$ is the body force, and $\mathbf{T}$ denotes the traction on the boundary $\partial\Omega$. Also, the term $\omega$ refers to the micro-stress conjugate to the phase field $\phi$, while $\bm{\upxi}$ is the micro-stress vector conjugate to the gradient of the phase field $\nabla\phi$. In addition, $\dot{\zeta}$ denotes the rate of fluid mass content, corresponding to the mass of fluid per unit bulk volume during a unit of time, $\mathbf{q}$ is the fluid flux vector, $q_m$ is the fluid source, and $q$ is fluid flux per unit area applying on the boundary. Applying the Gauss divergence theorem to Eq. (\ref{eq:energy}) delivers the balance equations describing the coupled deformation-fracture-pore system:
\begin{equation}\label{eq:balance}
    \begin{split}
        &\nabla \cdot \bm{\sigma} + \mathbf{b} = 0  \\
        &\nabla \cdot \bm{\upxi} - \omega = 0 \\
        & \nabla \cdot \mathbf{q} + \Dot{\zeta}  = q_m
    \end{split} \hspace{2cm} \text{in } \,\, \Omega,
\end{equation}

\noindent along with appropriate boundary conditions,
\begin{equation}\label{eq:balanceBCs}
    \begin{split}
        &\bm{\sigma} \mathbf{n} = \mathbf{T}  \\
        &\bm{\upxi} \cdot \mathbf{n} = 0 \\
        & \mathbf{q} \cdot \mathbf{n} = - q
    \end{split} \hspace{2cm} \text{on } \,\, \partial \Omega.
\end{equation}

These equations represent the balance of linear momentum for the deformation field, the balance of microforces for the phase field, and mass conservation for fluid, respectively.

\subsection{Constitutive theory for phase field fracture}

The total potential energy density of the system for the coupled deformation-fracture-pore system is expressed as the sum of the elastic strain energy density $\psi$, fluid energy density $\psi_{fl}$\footnote{The fluid energy term accounts for the pressure $p$ and fluid volume fraction inside the domain, and can be expressed based on the storage coefficient $S$, defined in Section \ref{sec:Fluid flow equation}, as follows: $\psi_{fl}=S p^2 /2$.}, and the energy dissipated in creating new crack surfaces $\varphi$:
\begin{equation}\label{eq:TotalPotentialEnergy0}
W \left( \boldsymbol{\varepsilon} \left( \mathbf{u} \right), \, \phi, \,  \nabla \phi \right) = \psi \left( \boldsymbol{\varepsilon} \left( \mathbf{u} \right), \, g \left( \phi \right) \right)  +\psi_{fl}(\boldsymbol{\varepsilon} \left( \mathbf{u} \right),p)+  \varphi \left( \phi, \, \nabla \phi \right).
\end{equation}

The effect of the phase field on material stiffness is incorporated via the degradation function $g \left( \phi \right) = \left( 1 - \phi \right)^2 + \kappa \,$ with the conditions:
\begin{equation}
g \left( 0 \right) =1 , \,\,\,\,\,\,\,\,\,\, g \left( 1 \right) =0 , \,\,\,\,\,\,\,\,\,\, g' \left( \phi \right) \leq 0 \,\,\, \text{for} \,\,\, 0 \leq \phi \leq 1 \, .
\end{equation}

A small parameter $\kappa$ is included to prevent ill-conditioning as $\phi \to 1$. The fracture energy is approximated through the crack density function $\gamma(\phi, \nabla\phi)$:
\begin{align}
    \varphi \left( \phi, \, \nabla \phi \right) = G_c \gamma(\phi, \nabla\phi) = G_c \dfrac{1}{4c_w\ell}\left( w(\phi) + \ell^2 |\nabla\phi|^2\right) \, ,
\end{align}

\noindent where $\ell$ is the phase field length scale, $c_w$ is a scaling constant, and $w(\phi)$ is the geometric crack function. These variables are defined in Table \ref{tab:phase field-models} for the commonly used AT2 and AT1 models. See \cite{navidtehrani2021unified} for details.\\

\begin{table}[H]
    \caption{Geometric crack function $w(\phi)$, and scaling constant $c_w$ for the AT2, and AT1 models.}
    \label{tab:phase field-models}
    \centering
    \begin{tabular}{p{30mm} >{\centering\arraybackslash}p{20mm} >{\centering\arraybackslash}p{20mm}}
    \hline 
        \textbf{Model} &   $w(\phi)$ &  $c_w$ \\
        \hline \hline
         AT2  & $\phi^2$  & $1/2$ \\
         AT1   & $\phi$ & $2/3$ \\
        \hline
    \end{tabular}
\end{table}

In the evolution of the phase field order, the strain energy of the undamaged configuration $\psi_0$, drives fracture. For asymmetric stiffness degradation, the strain energy is split into a dissipative part $\psi_d$, and a stored part $\psi_s$, yielding the undamaged and damaged configurations:
\begin{equation}\label{eq:Split}
    \psi_0 \left( \boldsymbol{\varepsilon} \right) = \psi_d \left( \boldsymbol{\varepsilon} \right) + \psi_s \left( \boldsymbol{\varepsilon} \right) \, , \,\,\,\,\,\, \text{and} \,\,\,\,\,\, \psi \left( \boldsymbol{\varepsilon}, \phi \right) = g \left( \phi \right) \psi_d \left( \boldsymbol{\varepsilon} \right) + \psi_s \left( \boldsymbol{\varepsilon} \right) ,
\end{equation}

Thus, the total potential energy of the solid Eq. (\ref{eq:TotalPotentialEnergy0}), is expressed as:
\begin{equation}\label{eq:Free_energy}
     W =  g \left( \phi \right) \psi_d \left( \boldsymbol{\varepsilon} \right) + \psi_s \left( \boldsymbol{\varepsilon} \right) +\psi_{fl}(\boldsymbol{\varepsilon} \left( \mathbf{u} \right),p) + \frac{G_c}{4 c_w }  \left(\frac{1}{ \ell} {w(\phi)}+\ell |\nabla \phi|^2\right) .
\end{equation}

Hence, by considering the variation of energy with respect to the phase field variable, one can derive the fracture micro-stress variables $\omega$ and $\bm{\upxi}$ as \cite{navidtehrani2021unified},
\begin{equation}\label{eq:consOmega}
    \omega = \dfrac{\partial W}{\partial\phi} = {g^{\prime}(\phi)} \psi_d \left( \boldsymbol{\varepsilon} \right) +\frac{G_c}{4c_w \ell}  w^{\prime}(\phi) \, ,
\end{equation}
\begin{equation}\label{eq:consXi}
    \bm{\upxi} = \dfrac{\partial W}{\partial\nabla\phi} = \frac{\ell}{2c_w} G_{\mathrm{c}} \nabla \phi \, .
\end{equation}

\noindent Substituting these into Eq. (\ref{eq:balance}b), the phase field evolution equation reads:
\begin{equation}\label{eq:PhaseFieldStrongForm}
  \frac{G_c}{2c_w}  \left( \frac{w^{\prime}(\phi)}{2 \ell} - \ell \nabla^2 \phi \right) + {g^{\prime}(\phi)} \psi_d \left( \boldsymbol{\varepsilon} \right) = 0  .
\end{equation}

Finally, damage irreversibility is here enforced by defining a history variable: $\mathcal{H} = \text{max}_{t \in [0, \tau]} \psi_d \left( t \right)$.

\subsubsection{Strain energy decomposition as fracture driving force}
\label{seq:Drivnign-force}

The strain energy split as a fracture driving force was developed to prevent damage evolution under compression. Various phase field fracture driving forces can be found in the literature. In this study, we focus on the most widely used formulations and our novel generalised approach, particularised to the Drucker-Prager case. Alternative strain energy split approaches are discussed in \cite{haghighat2023efficient, HESAMMOKRI2023112080, vicentini2024energy}. Amor \textit{et al.} \cite{Amor2009} introduced the volumetric-deviatoric split to exclude energy associated with volumetric compaction. This split can be expressed in terms of the first invariant of the strain tensor \( I_1 (\boldsymbol{\varepsilon}) \) and the second invariant of the deviatoric part of the strain tensor \( J_2 (\boldsymbol{\varepsilon}) \) as follows:
\begin{equation}
\psi_d(\boldsymbol{\varepsilon})=\frac{1}{2} K\langle I_1 (\boldsymbol{\varepsilon})\rangle_{+}^2+ 2 \mu J_2 (\boldsymbol{\varepsilon})
\end{equation}
\begin{equation}
\psi_s(\boldsymbol{\varepsilon})=\frac{1}{2} K\langle I_1 (\boldsymbol{\varepsilon})\rangle_{-}^2 ,
\end{equation}

\noindent where \( K \) is the bulk modulus, \( \mu \) is the shear modulus, and the Macaulay brackets are defined as \( \langle a \rangle_{\pm} = {(a \pm |a|)}/{2} \). In this model, if the first invariant of strain tensor is negative (\( I_1 (\boldsymbol{\varepsilon}) < 0 \)), the fracture is driven by the distortion energy $\psi_d(\boldsymbol{\varepsilon})=2 \mu J_2 (\boldsymbol{\varepsilon}) $.\\

Miehe \textit{et al.} \cite{Miehe2010a} propose a split based on the decomposition of the principal strain tensor into positive and negative parts, defined as \( \bm{\epsilon}_{\pm} = \langle \bm{\epsilon} \rangle_{\pm} \). This spectral decomposition is given by
\begin{equation}
\psi_d(\boldsymbol{\varepsilon})=\frac{1}{2} \lambda\left(\left\langle I_1(\bm{\epsilon})\right\rangle_{+}\right)^2+\mu\left(\left(I_1(\bm{\epsilon}_{+})\right)^2-2 I_2(\bm{\epsilon}_{+})\right) 
\end{equation}
\begin{equation}
\psi_s(\boldsymbol{\varepsilon})=\frac{1}{2} \lambda\left(\left\langle I_1(\bm{\epsilon})\right\rangle_{-}\right)^2+\mu\left(\left(I_1(\bm{\epsilon}_{-})\right)^2-2 I_2(\bm{\epsilon}_{-})\right) ,
\end{equation}

\noindent where \( \lambda \) is the first Lamé constant and \( I_2(\bm{\epsilon}) \) is the second invariant of the strain tensor. \\

Later, Freddy and Royer-Carfagni \cite{Freddi2010} developed a decomposition approach known as the no-tension split, which was based on the work of Del Piero \cite{DelPiero1989} and aimed at masonry-like materials. This method can be expressed based on the principal strains (\( \epsilon_3 \geq \epsilon_2 \geq \epsilon_1 \)) as follows:
\begin{equation}
\psi_d(\boldsymbol{\varepsilon})=
\begin{cases}
\frac{E \nu}{2(1+\nu)(1-2 v)}\left(\epsilon_1+\epsilon_2+\epsilon_3\right)^2+\frac{E}{2(1+\nu)}\left(\epsilon_1^2+\epsilon_2^2+\epsilon_3^2\right) & \epsilon_1>0 \\
\frac{E v}{2(1+v)(1-2 v)}\left(\epsilon_3+\epsilon_2+2 v \epsilon_1\right)^2+\frac{E}{2(1+v)}\left[\left(\epsilon_3+v \epsilon_1\right)^2+\left(\epsilon_2+v \epsilon_1\right)^2\right] & \epsilon_2+\nu \epsilon_1>0 \\
\frac{E}{2\left(1-v^2\right)(1-2 v)}\left[(1-v) \epsilon_3+v \epsilon_1+v \epsilon_2\right]^2 & (1-\nu)\epsilon_3+\nu (\epsilon_1+\epsilon_2)>0 \\
0 & \text{else}
\end{cases}
\end{equation}
\begin{equation}
\psi_s(\boldsymbol{\varepsilon})=
\begin{cases}
0 & \epsilon_1>0 \\
\frac{E}{2} \epsilon_1^2 & \epsilon_2+\nu \epsilon_1>0 \\
\frac{E}{2\left(1-v^2\right)}\left(\epsilon_1^2+\epsilon_2^2+2 v \epsilon_1 \epsilon_2\right) & (1-\nu)\epsilon_3+\nu (\epsilon_1+\epsilon_2)>0 \\
\frac{E v}{2(1+\nu)(1-2 v)}\left(\epsilon_1+\epsilon_2+\epsilon_3\right)^2+\frac{E}{2(1+\nu)}\left(\epsilon_1^2+\epsilon_2^2+\epsilon_3^2\right) & \text{else}
\end{cases}
\end{equation}

\noindent where \( E \) is Young's modulus and \( \nu \) is Poisson's ratio. In this model, only positive principal stresses are considered for computing the fracture driving force. \\

However, there is growing interest in expanding the capabilities of phase field fracture models to incorporate arbitrary failure surfaces for crack nucleation and growth, so as to better represent the failure behaviour of rock-like materials \cite{molnar2020toughness,de2022nucleation,lopez2025classical}. This is of relevance in hydraulic fracture as shale rocks do not exhibit symmetric tension-compression fracture behaviour. While the injected fluid results in tractions normal to the crack surface, the stress state is often complex due to crack interaction, body forces and other boundary conditions. Therefore, an accurate simulation of hydraulic fracture under complex conditions necessitates a generalised phase field formulation capable of incorporating suitable failure surfaces. Recently, Navidtehrani \textit{et al.} \cite{Navidtehrani2022} developed a generalised approach to incorporate arbitrary failure surfaces into the phase field fracture driving force. The approach was demonstrated with the Drucker-Prager failure surface, which is relevant to shale rock and hence will be adopted here. Navidtehrani \textit{et al.} \cite{Navidtehrani2022} defined the material cohesion $c_f$ to be degraded by the phase field but a constant friction parameter $\beta_f$. Then, the strain energy split based on the Drucker-Prager model can be expressed as follows \cite{Navidtehrani2022}:
\begin{equation}
\psi_d=
\begin{cases}
\frac{1}{2}K I_1^2 (\boldsymbol\varepsilon)+2 \mu J_2 (\boldsymbol\varepsilon) & \text{for} \quad -6B\sqrt{J_2 (\boldsymbol\varepsilon)} < I_1 (\boldsymbol\varepsilon) \\
\frac{1}{18 B^2 K + 2 \mu} \left(-3BKI_1(\boldsymbol\varepsilon)+2 \mu \sqrt{J_2(\boldsymbol\varepsilon)} \right)^2 & \text{for} \quad -6B\sqrt{J_2 (\boldsymbol\varepsilon)} \geq I_1 (\boldsymbol\varepsilon) \,\, \& \,\,  2 \mu \sqrt{J_2 (\boldsymbol\varepsilon)} \geq 3 B K I_1 (\boldsymbol\varepsilon)\\
0 & \text{for} \quad 2 \mu \sqrt{J_2 (\boldsymbol\varepsilon)} < 3 B K I_1 (\boldsymbol\varepsilon) 
\end{cases}
\label{eq:psi_d3regimes}
\end{equation}
\begin{equation}
\psi_s=
\begin{cases}
0 & \text{for} \quad -6B\sqrt{J_2 (\boldsymbol\varepsilon)} < I_1 (\boldsymbol\varepsilon) \\
\frac{K \mu}{18 B^2 K + 2 \mu} \left(I_1(\boldsymbol\varepsilon)+6 B \sqrt{J_2(\boldsymbol\varepsilon)} \right)^2 & \text{for} \quad -6B\sqrt{J_2 (\boldsymbol\varepsilon)} \geq I_1 (\boldsymbol\varepsilon) \,\, \& \,\,  2 \mu \sqrt{J_2 (\boldsymbol\varepsilon)} \geq 3 B K I_1 (\boldsymbol\varepsilon) \\
\frac{1}{2}K I_1^2 (\boldsymbol\varepsilon)+2 \mu J_2 (\boldsymbol\varepsilon) & \text{for} \quad 2 \mu \sqrt{J_2 (\boldsymbol\varepsilon)} < 3 B K I_1 (\boldsymbol\varepsilon) 
\end{cases},
\label{eq:psi_s3regimes1}
\end{equation}

\noindent where $B$ is a material constant that is a function of the internal friction coefficient $\beta_f$, e.g. for Drucker–Prager failure surface middle circumscribes the Mohr–Coulomb surface:

\begin{equation}
B=\frac{2 \sin \beta_f}{\sqrt{3}(3+\sin \beta_f)} \,.
\end{equation}

\begin{figure}[H]
    \centering
    \begin{subfigure}[b]{0.4\textwidth}
         \centering
         \includegraphics[width=\textwidth]{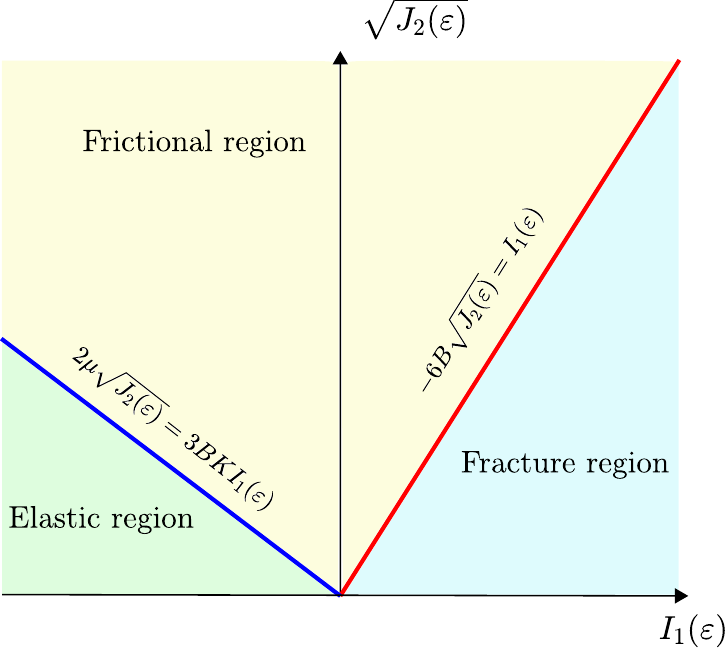}
         \caption{}
     \end{subfigure} \hspace{5mm}
     \begin{subfigure}[b]{0.4\textwidth}
         \centering
         \includegraphics[width=\textwidth]{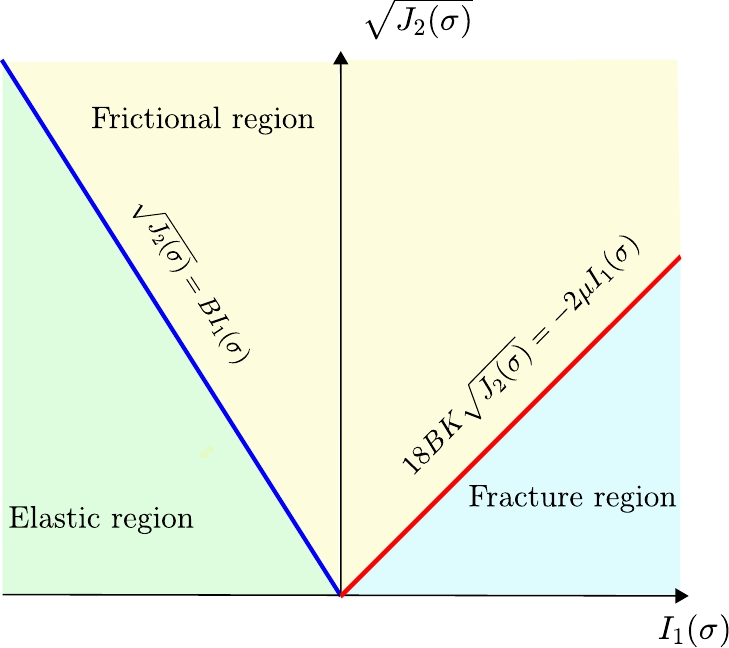}
         \caption{}
     \end{subfigure}
    \caption{Elastic, frictional, and fracture regions of Drucker-Prager based model in: (a) the strain space $(I_1(\boldsymbol{\varepsilon}), \sqrt{J_2 (\boldsymbol{\varepsilon})})$, and (b) the stress space $(I_1(\bm{\sigma}), \sqrt{J_2 (\bm{\sigma})})$.}
    \label{fig:Drucker-Prager-Space}
\end{figure}

Navidtehrani \textit{et al.} \cite{Navidtehrani2022} showed that with a Drucker-Prager based fracture driving force, different material behaviours, including confinement, frictional behaviour, and the dilatancy effect, can be captured. The strain and stress spaces in the Drucker-Prager model are illustrated in Fig. \ref{fig:Drucker-Prager-Space}. Both stress and strain spaces are divided into three different regions. In the elastic region, regardless of the value of the phase field variable \( \phi \), the material behaviour is completely elastic, with no loss of stiffness. On the opposite side, the fracture region, the entire stress and stiffness are degraded by the phase field, meaning that when \( \phi = 1 \) there is a traction-free crack. The material stiffness in the frictional region is anisotropic, indicating that only part of the stress and stiffness are degraded by the evolution of the phase field. Due to the frictional behaviour, applying more pressure results in higher shear stress. Finally, when \( \phi = 1 \), the stress lies on the failure line \( \sqrt{J_2 (\bm{\sigma})} = B I_1(\bm{\sigma}) \).

\subsection{Fluid flow equation through porous media}
\label{sec:Fluid flow equation}

To characterize the distribution of pore pressure \( p \) within a porous medium, a differential equation governing pore pressure must be defined. This can be achieved by examining the conservation of mass for the fluid, Eq. (\ref{eq:balance}c), in conjunction with a constitutive equation that relates fluid flux $\mathbf{q}$ and fluid mass  $\zeta$ to pore pressure. This is typically achieved by considering mass conservation and Darcy's law, whose principles are outlined here.\\

Darcy's law, developed by Henry Darcy \cite{darcy1856fontaines}, describes the relationship between pore pressure \( p \) and the flux vector \( \mathbf{q} \) under conditions of low flow rates, providing insights into fluid behaviour in porous media. For anisotropic cases, Darcy’s law is expressed as:
\begin{equation}\label{Eq:Darcy}
\mathbf{q} = -\rho_{fl} \frac{\bm{K}_{fl}}{\mu_{fl}} \nabla p,
\end{equation}

\noindent where gravity has been neglected, $\bm{K}_{fl}$ is the permeability tensor, and ${\mu_{fl}}$ represents the fluid dynamic viscosity.\\

\begin{figure}[H]
    \centering
    \includegraphics[width=0.45\textwidth]{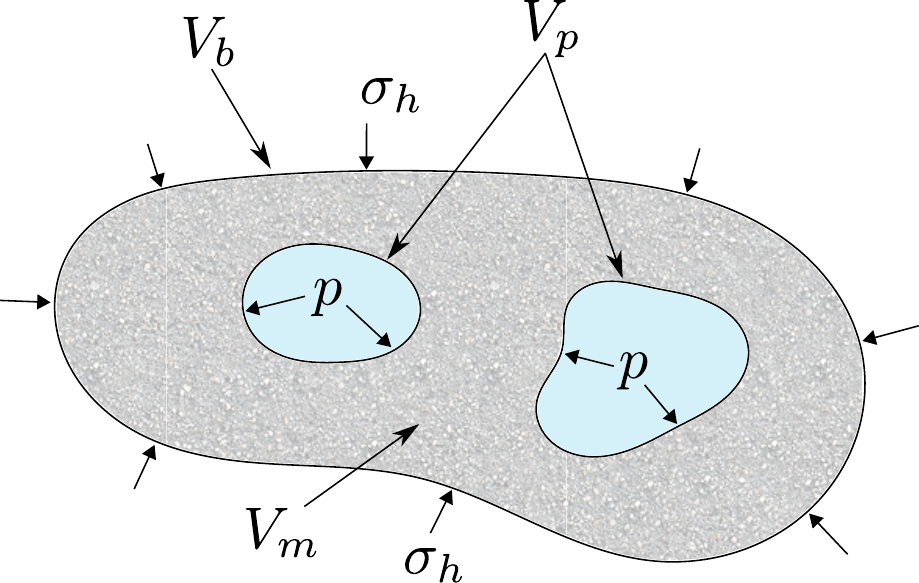}
    \caption{Porous material illustrating hydrostatic stress $\sigma_h$, pore pressure $p$, pore volume ($V_p$), bulk volume ($V_b$), and solid phase volume ($V_m$).}
    \label{fig:Porous rock config}
\end{figure}

The fluid mass content can be expressed using porosity \( n_p \) and fluid density \( \rho_{fl} \) as:
\begin{equation}\label{Eq:MassContent0}
\zeta_{fl} = \rho_{fl} n_p,
\end{equation}

\noindent where porosity is defined as the ratio of pore volume ($V_p$) to the bulk volume ($V_b$), i.e., $n_p = V_p / V_b$, as illustrated in Fig. \ref{fig:Porous rock config}. Changes in mass fluid content arise from alterations in porosity due to variations in pore pressure and the compression or expansion of fluid within the pores. This can be expressed as:
\begin{equation}\label{Eq:dMassContent00}
\text{d}\zeta_{fl} = \rho_{fl} \text{d}n_p + n_p \text{d}\rho_{fl},
\end{equation}

\noindent Here, fluid density changes due solely to pore pressure variations, while porosity changes result from both pore pressure \( p \) and hydrostatic stress \( \sigma_h \) variations. Consequently, Eq. (\ref{Eq:dMassContent00}) can be reformulated considering differential changes as:
\begin{equation}\label{Eq:dzeta}
\text{d}\zeta_{fl} = \rho_{fl} \left( \frac{\partial n_p}{\partial \sigma_h} \text{d}\sigma_h + \frac{\partial n_p}{\partial p} \text{d}p \right) + n_p \frac{\partial \rho_{fl}}{\partial p} \text{d}p.
\end{equation}

To expand Eq. (\ref{Eq:dzeta}), we briefly review the hydrostatic theory of poroelasticity. Consider a bulk volume of porous media containing voids and saturated with fluid subjected to hydrostatic stress \( \sigma_h \) and pore pressure \( p \), as shown in Fig. \ref{fig:Porous rock config}. Under static conditions, pore pressure is unable to sustain shear stress, and pore walls cannot transmit any shear traction. The solid is then subjected to two independent stresses, namely \( \sigma_h \) and \( p \), as well as two independent volumes (\( V_b \) and \( V_p \)). Therefore, four compressibilities can be defined:
\begin{equation}\label{Eq:Compresibilities}
C_{b\sigma_h} = \frac{1}{V_b} \frac{\partial V_b}{\partial \sigma_h}, \quad 
C_{bp} = \frac{1}{V_b} \frac{\partial V_b}{\partial p}, \quad 
C_{p\sigma_h} = \frac{1}{V_p} \frac{\partial V_p}{\partial \sigma_h}, \quad
C_{pp} = \frac{1}{V_p} \frac{\partial V_p}{\partial p}.
\end{equation}

Relationships between these compressibilities were established in Ref. \cite{Jaeger2009}, and are as follows:
\begin{equation}\label{eq:Cbp}
C_{bp} = C_{b\sigma_h} - C_{m} , \quad C_{p\sigma_h} = \frac{C_{b\sigma_h} - C_{m}}{n_p} , \quad C_{pp} = \frac{C_{b\sigma_h} - (1 + n_p) C_{m}}{n_p},
\end{equation}

\noindent where $C_{m}$ is the compressibility of the solid. Using Eq. (\ref{eq:Cbp}), we can determine changes in bulk modulus strain $\varepsilon_{b}$  (volumetric strain ($\varepsilon_{vol}$)) for a volume control $V_b$:
\begin{equation}\label{Eq:dEPSvol}
\text{d}\varepsilon_{b} = \text{d}\varepsilon_{vol} = \frac{\text{d}V_b}{V_b} = \frac{1}{V_b} \left(\frac{\partial V_b}{\partial \sigma_h} \text{d}\sigma_h + \frac{\partial V_b}{\partial p} \text{d}p\right) = C_{b\sigma_h} \text{d}\sigma_h + C_{bp} \text{d}p.
\end{equation}

This equation holds generally and does not assume a fixed bulk volume. Biot's coefficient ($\alpha$) is defined as:
\begin{equation}\label{Eq:Biot'sCoef}
\alpha = 1 - \frac{C_{{m}}}{C_{{b\sigma_h}}} = 1 - \frac{K_{{b\sigma_h}}}{K_{{m}}},
\end{equation}

\noindent where $K_{{b\sigma_h}}$ and $K_{{m}}$ represent the bulk moduli of the saturated porous media and solid phase, respectively. Then, Eq. (\ref{Eq:dEPSvol}) can be rewritten as:
\begin{equation}
\text{d}\varepsilon_{vol} = C_{b\sigma_h} (\text{d}\sigma_h + \alpha \text{d}p).
\end{equation}

Now, the variation of porosity \( n_p \) with respect to pore pressure and hydrostatic stress \( \sigma_h \) is given by:
\begin{equation}
\frac{\partial n_p}{\partial p} = \frac{1}{V_b} \frac{\partial V_p}{\partial p} - \frac{n_p}{V_b} \frac{\partial V_b}{\partial p}.
\end{equation}

While for the volume control (${V_b}$), we find:
\begin{equation}\label{Eq:dFidP}
\frac{\partial n_p}{\partial p} = n_p C_{pp} = C_{b\sigma_h} - (1 + n_p) C_{m} = (\alpha - n_p + n_p \alpha) C_{b\sigma_h}.
\end{equation}

Similarly,
\begin{equation}\label{Eq:dFidSh}
\frac{\partial n_p}{\partial \sigma_h} = n_p C_{p\sigma_h} = C_{b\sigma_h} - C_{m} = \alpha C_{b\sigma_h}.
\end{equation}

One can then reformulate Eq. (\ref{Eq:dzeta}) considering Eqs. (\ref{Eq:dFidP})-(\ref{Eq:dFidSh}), such that:
\begin{equation}\label{Eq:dFi1}
\text{d}\zeta_{fl} = \rho_{fl} \left((\alpha - n_p + n_p \alpha) C_{b\sigma_h} \text{d}p + \alpha C_{b\sigma_h} \left(\frac{1}{C_{b\sigma_h}} \text{d}\varepsilon_{vol} - \alpha \text{d}p \right) \right) + n_p \frac{\partial \rho_{fl}}{\partial p} \text{d}p.
\end{equation}

Rearranging yields:
\begin{equation}\label{Eq:dMassContent1}
\text{d}\zeta_{fl} = \rho_{fl} \left( (1 - \alpha)(\alpha - n_p) C_{b\sigma_h} + n_p \frac{1}{\rho_{fl}} \frac{\partial \rho_{fl}}{\partial p} \right) \text{d}p + \rho_{fl} \alpha \text{d}\varepsilon_{vol} = \rho_{fl} S \text{d}p + \rho_{fl} \alpha \text{d}\varepsilon_{vol},
\end{equation}

\noindent where $S$ is the storage coefficient, defined as:
\begin{equation}\label{Eq:dMassContent2}
S = (1 - \alpha)(\alpha - n_p) C_{b\sigma_h} + n_p C_{fl} = \frac{(1 - \alpha)(\alpha - n_p)}{K_{b\sigma_h}} + n_p C_{fl},
\end{equation}

\noindent where fluid compressibility is defined as $C_{fl}=\frac{1}{\rho_{fl}} \frac{\partial \rho_{fl}}{\partial p}$.

As the volumetric strain $\varepsilon_{vol}$ and pore pressure $p$ vary over time, the rate of change of fluid mass content can be expressed using Eq. (\ref{Eq:dMassContent1}) as:
\begin{equation}\label{Eq:dMassContent4}
\dot{\zeta}_{fl} = \rho_{fl} S \, \dot{p} + \rho_{fl} \alpha \,  \dot{\varepsilon}_{vol},
\end{equation}

Finally, substituting Eq. (\ref{Eq:dMassContent4}) into the fluid mass conservation equation, Eq. (\ref{eq:balance}c), yields:
\begin{equation}\label{Eq:dMassContent3}
\rho_{fl} \left(S \Dot{p} + \alpha \Dot{\varepsilon}_{vol} \right) + \nabla \cdot \left(-\rho_{fl} \frac{\bm{K}_{fl}}{\mu_{fl}} \nabla p \right) = q_m.
\end{equation}

\subsection{Constitutive Equations of Poroelasticity}
\label{sec:Constitutive equations of poroelasticity}

The strain-stress relationship for a material in the absence of pore pressure is given by:
\begin{equation}\label{Eq:complianceTensor}
\boldsymbol{\varepsilon} = \bm{C}^{-1} : \bm{\sigma},
\end{equation}

\noindent where $\bm{C}^{-1}$ denotes the compliance tensor of elasticity. Since the static pore pressure of fluid flow does not transmit shear stress to the solid structure and is negligible at low flow speeds, its effect is limited to volume changes within the domain, which can be modeled as follows:
\begin{equation}\label{Eq:PorePressureEffect}
\boldsymbol{\varepsilon} = \bm{C}^{-1} : \bm{\sigma} + \frac{1}{3}C_{bp}p \bm{I} = \bm{C}^{-1} : \bm{\sigma} + \frac{\alpha}{3 K_{b\sigma_h}} p \bm{I}.
\end{equation}

Using Biot's coefficient, as defined in Eq. (\ref{Eq:Biot'sCoef}) and considering an effective stress $\bm{\sigma}^{eff}$, which represents the stress carried by the solid skeleton, the total stress can be expressed as:
\begin{equation}\label{Eq:TotalStress}
 \bm{\sigma} = \bm{C} : \boldsymbol{\varepsilon} - \alpha p \bm{I} = \bm{\sigma}^{eff} - \alpha p \bm{I}.
\end{equation}

\subsection{Coupling phase field and fluid equation}
\label{sec:Phase field hydraulic fracture}

The microstructure of the solid comprises a porous matrix interspersed with microcracks. The fluid-filled pores constitute the material's intrinsic porosity, with the fluid pressure being governed by Biot's theory of poroelasticity through the principle of effective stress. These pores reside within the intact material and are incorporated into the continuum-scale balance equations. Microcracks can initiate or grow due to damage evolution, potentially coalescing into macroscopic fractures and serving as conduits for fluid transport. Their behaviour is captured by the phase field variable, which represents crack initiation and propagation. Physically, this scale separation assumes that microcracks are significantly smaller than the representative volume element (RVE) and interact with the surrounding pore network primarily by altering porosity, permeability, and fluid pressure distribution. The impact of microcracks on these properties may be negligible at low levels of damage (low $\phi$ values) but becomes more significant as the material approaches full fracture  ($\phi \to 1$). This assumption is prevalent in phase field models of hydraulic fracture, where the phase field impacts permeability and may also influence the effective stress and porosity. While more detailed models could explicitly resolve interactions between individual microcracks and pores at a finer scale based on micromechanics \cite{XIE2012919,jia2021experimental,Ulloa2022}, such approaches can be more complex in terms of implementation. Instead, our model adopts a homogenized approach, balancing physical accuracy with computational efficiency. In this section, we discuss three different methods for incorporating the effects of phase field evolution into the fluid equations.\\

The poroelastic theory can be combined with the phase field fracture framework by applying Biot’s theory of effective stress. Using Eq. (\ref{eq:Split}), the effective stress is defined as:
\begin{equation}\label{eq:EffStress}
\bm{\sigma}^{eff} = g(\phi) \frac{\partial \psi_d(\boldsymbol{\varepsilon})}{ \partial \boldsymbol{\varepsilon}} + \frac{\partial \psi_s(\boldsymbol{\varepsilon})}{ \partial \boldsymbol{\varepsilon}}.
\end{equation}

Substituting Eqs. (\ref{eq:EffStress}) and (\ref{Eq:TotalStress}) into the linear momentum equation (\ref{eq:balance}a), we obtain:
\begin{equation}
\nabla \cdot \left(\bm{\sigma}^{eff} - \alpha p \bm{I}\right) +\mathbf{b} = 0 .
\end{equation}

\begin{figure}[H]
    \centering
    \includegraphics[width=0.8\textwidth]{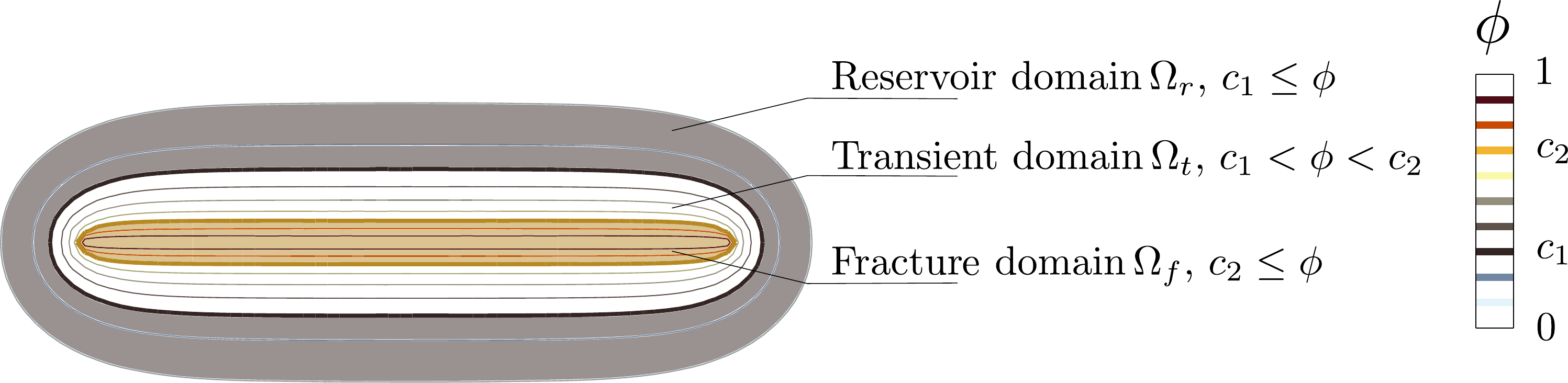}
    \caption{Reservoir ($\Omega_r$), transient ($\Omega_t$), and fracture ($\Omega_f$) domains identified by phase field variable $\phi$.}
    \label{fig:AxuFiled}
\end{figure}

There are several methods to couple fluid effects and the phase field equation. One approach, based on the work by Lee \textit{et al.} \cite{Lee2016}, is referred to as the domain decomposition method and has been utilised in other studies \cite{Zhou2018,Zhou2018c}. This method involves dividing the domain into three distinct regions: the reservoir (\( \Omega_r \)), fracture (\( \Omega_f \)), and transient (\( \Omega_t \)) domains, as illustrated in Fig. \ref{fig:AxuFiled}. These zones are identified using linear indicator functions \( \chi_{r} \) and \( \chi_{f} \), which depend on the phase field variable \( \phi \) and material constants \( c_1 \) and \( c_2 \):

\begin{equation}\label{Eq:Axu}
{\chi}_{r}\left({\phi}\right)=
\begin{cases}
    1 & {\phi} \le {c}_1 \\
    \frac{{c}_2 - {\phi}}{{c}_2 - {c}_1} & {c}_1 < {\phi} < {c}_2 \\
    0 & {c}_2 \le {\phi} 
\end{cases}, \quad
{\chi}_{f}\left({\phi}\right)=
\begin{cases}
    0 & {\phi} \le {c}_1 \\
    \frac{{{\phi} - {c}_1}}{{c}_2 - {c}_1} & {c}_1 < {\phi} < {c}_2 \\
    1 & {c}_2 \le {\phi} 
\end{cases},
\end{equation}

\noindent where the material constants \( c_1 \) and \( c_2 \) determine whether a given point in the domain belongs to the reservoir, fracture, or transient zones. The continuity equation (Eq. \eqref{Eq:dMassContent3}) describes the fluid flow in the reservoir domain. This equation can also be applied to the fracture domain by setting \( S = C_{fl} \) and neglecting the volumetric strain rate term, \( \dot{\varepsilon}_{vol} \). In the transient zone, the fluid equation is formulated to ensure mass conservation is continuous across all domains and at their boundaries. This approach can be interpreted as an implicit method for capturing the influence of microcracks on fluid behaviour. Thus, fluid and solid parameters between the reservoir and fracture domains are then interpolated as follows:
\begin{equation}
\alpha = {\chi}_{r} {\alpha}_{r} + {\chi}_{f}
\end{equation}
\begin{equation}
{n}_{p} = {\chi}_{r} n_{pr} + {\chi}_{f}
\end{equation} 
\begin{equation}\label{Eq:Kzhou}
\bm{K}_{fl} = \chi_{r} \bm{K}_{r} + {\chi}_{f} \bm{K}_{f},
\end{equation}

\noindent where $\alpha_{r}$ and $n_{pr}$ are Biot’s coefficient and porosity of the reservoir. In the fracture domain, $\alpha=1$ and $n_{p}=1$, while $\bm{K}_{r}$ and $\bm{K}_{f}$ represent the permeability tensors of the reservoir and fracture domains, respectively. The assumption of linear interpolation of properties between the reservoir and the fracture zone is arguably the simplest one but dedicated experiments are needed to establish more physically-based interpolation functions, with particular attention to complex scenarios, such as highly confined states where grain crushing and compaction can occur.\\

An alternative method, proposed by Miehe \textit{et al.} \cite{miehe2015minimization}, considers Poiseuille-type flow within the crack by modifying Darcy's law to define fracture permeability as follows:
\begin{equation}\label{Eq:K2ndmethod}
K_{\mathrm{f}} = \frac{w_h^2}{12}\left(\bm{I} - \mathbf{n}_{\phi} \otimes \mathbf{n}_{\phi}\right),
\end{equation}

\noindent where $\mathbf{n}_{\phi}$ is the crack normal vector defined by the phase field gradient ($\mathbf{n}_{\phi} = \nabla \phi / |\nabla \phi|$), and $w_h$ is the crack opening calculated using the element size $h_e$:
\begin{equation}
w_h = \langle \left\| h_e \left( 1 + \mathbf{n}_{\phi} \cdot \boldsymbol{\varepsilon} \cdot \mathbf{n}_{\phi} \right) \right\| \rangle_+ .
\end{equation}

The permeability tensor for the modified Darcy approach is then:
\begin{equation}\label{Eq:Kzhou1}
\bm{K}_{fl} = \bm{K}_{r} + \phi^{b} \bm{K}_{f},
\end{equation}

\noindent where $b$ is a permeability transient indicator. \\

We here propose a third method, combining these two approaches, which is henceforth referred to as the \emph{hybrid permeability method}. In this hybrid method, we use the domain decomposition approach but adopt the definition of $\bm{K}_{f}$ from Eq. (\ref{Eq:K2ndmethod}) as follows:
\begin{equation}\label{Eq:Kzhou3}
\bm{K}_{fl} = \chi_{r} \bm{K}_{r} + \phi^{b} {\chi}_{f} \bm{K}_{f}.
\end{equation}

The hybrid method leverages the advantages of both the domain decomposition and modified Darcy methods while addressing their respective limitations. As demonstrated in Section \ref{sec:Coupling_Method}, the domain decomposition method does not account for the effect of crack opening on permeability. Additionally, it assumes a uniform permeability raise across the fracture region in all directions, whereas, in reality, permeability enhancement occurs primarily along the crack direction. In contrast, the modified Darcy method incorporates an anisotropic permeability tensor, effectively capturing directional permeability changes. However, this effect becomes significant at an unrealistic distance from the crack region. To mitigate this issue, the characteristic length scale must be chosen sufficiently small, but this, in turn, increases computational costs by constraining the element size. The proposed hybrid method addresses these limitations by combining the strengths of both approaches. A hybrid approach offers greater flexibility in calibrating parameters with experimental data. Additionally, by leveraging the advantages of the domain decomposition method, the influence of microcracks can be accounted for through the phase field value while also enabling a more precise representation of permeability through cracks. Microcracks can significantly influence permeability and other material properties, such as Biot's coefficient and porosity. Their effects, along with their evolution, can be captured through the phase field variable by appropriately selecting the parameters $c_1$ and $c_2$. These parameters can be calibrated experimentally, for example, through permeability testing of fractured rock samples. In such laboratory core tests, specimens are subjected to controlled fluid flow while measuring the resulting pressure drop and flow rate. These measurements are then used to estimate the effective permeability and infer suitable values for $c_1$ and $c_2$ through numerical modelling.\\

The general form of the fluid flow equation, applicable to all methods discussed, is as follows:
\begin{equation}\label{Eq:dMassContent5}
\rho_{fl} \left( S(\alpha(\phi), n_p(\phi)) \Dot{p} + \alpha(\phi) \chi_{r}(\phi) \Dot{\varepsilon}_{vol} \right) + \nabla \cdot \left( -\rho_{fl} \frac{\bm{K}_{fl}(\phi)}{\mu_{fl}} \nabla p \right) = q_m.
\end{equation}

\noindent where the permeability can be defined based on mentioned method as follows:
\begin{equation}\label{Eq:PermeabilityAll}
\bm{K}_{fl}=
\begin{cases}
\chi_{r}{K}_{r}\bm{I}+{\chi}_{f}{K}_{f}\bm{I} & \text{Domain decomposition method (Lee \textit{et al.} \cite{Lee2016})} \\

{K}_{r}\bm{I}+\phi^{b} \underbrace{ \left [\frac{w_h^2}{12}\left(\bm{I}-\mathbf{n}_{\phi} \otimes \mathbf{n}_{\phi}\right) \right]}_{\bm{K}_{f}} & \text{Modified Darcy method (Miehe \textit{et al.} \cite{miehe2015minimization})} \\

\chi_{r}{K}_{r}\bm{I}+{\chi}_{f} \phi^{b} \underbrace{ \left [\frac{w_h^2}{12}\left(\bm{I}-\mathbf{n}_{\phi} \otimes \mathbf{n}_{\phi}\right) \right]}_{\bm{K}_{f}} & \text{Hybrid method (Present work)} \\
    
\end{cases}.
\end{equation}

A challenging aspect common to the modified Darcy method and our hybrid formulation is the complexity of estimating $\mathbf{n}_{\phi}$ accurately \cite{CHUKWUDOZIE2019957}, particularly at points near the crack tip or where the phase field gradient vanishes. Various approaches have been presented to overcome this (see, e.g., Refs. \cite{Bryant2018,ZIAEIRAD2016304}). Here, a new protocol is established, whereby $\mathbf{n}_{\phi}$ is estimated in those complicated regions using the phase field gradient of a neighbouring integration point. The detailed procedure is presented in Algorithm~\ref{alg:gradient}. The first step is to determine, for each integration point, whether we are near a crack ($\phi > 0.5$) or if the phase field gradient is zero. If one of these conditions is met, the second step involves identifying the closest neighbouring Gauss point where $\phi=1$ and $| \nabla \phi | \neq 0$. 
In the third step, the cosine of the angle between the phase field gradient vectors of the current point and the neighbouring integration point is computed. If the cosine exceeds 0.866 (i.e., the angle between the vectors is less than 30$^\circ$, implying that the point is near to the crack, but not close to the crack tip), the gradient direction at the current point is considered reliable and used for computing $\mathbf{n}_{\phi}$. Otherwise, if the cosine is below this threshold, indicating proximity to the crack tip, the gradient at the neighbouring Gauss point is adopted for the current point.\\

\begin{algorithm}[H]
\caption {Determination of crack direction near the crack tip or at points with zero phase field gradient}
\label{alg:gradient}
\begin{algorithmic}[1]
\State Check if $\phi > 0.5$ or $|\nabla \phi| = 0$ at the current integration point.
\If $\, \phi > 0.5$ or $|\nabla \phi| = 0$
    \State Identify the nearest neighbouring point with $\phi_{\text{neighbour}}=1$ and $|\nabla \phi_{\text{neighbour}}| \neq 0$.
    \State Compute $\cos\theta = \dfrac{\nabla \phi \cdot \nabla \phi_{\text{neighbour}}}{|\nabla \phi| |\nabla \phi_{\text{neighbour}}|}$.
    \If $\cos\theta < 0.866$
        \State  $\mathbf{n}_{\phi}=\nabla \phi_{\text{neighbour}}/ |\nabla \phi_{\text{neighbour}}|$.
    \Else
        \State  $\mathbf{n}_{\phi}={\nabla \phi}/{|\nabla \phi|}$.
    \EndIf
\EndIf
\end{algorithmic}
\end{algorithm}

\section{Numerical implementation}
\label{Sec:NumImplementation}

We proceed to describe a general implementation of our model, considering the weak and discretised versions of the balance equations (Section \ref{Sec:FEM-Formulation}), the computation of the material Jacobian (Section \ref{Sec:Jacobian}) and the solution scheme (Section \ref{Sec:Solution}).

\subsection{Weak formulation and finite element implementation}
\label{Sec:FEM-Formulation}

To implement the formulation presented in Section \ref{Sec:Theory} within a finite element framework, the weak form of the coupled governing equations in Eq. (\ref{eq:balance}) is constructed, using the test functions $\delta \mathbf{u}$, $\delta \phi$, and $\delta p$:
\begin{equation}
  \int_\Omega \Big\{ \left(\bm{\sigma}^{eff} - \alpha p \bm{I}\right) : \delta \boldsymbol{\varepsilon} - \mathbf{b} \cdot \delta \mathbf{u} \Big\} \, \text{d}V 
  = \int_{\partial \Omega} \left( \mathbf{T} \cdot \delta \mathbf{u} \right) \, \text{d}S,
\end{equation}
\begin{equation}
\int_{\Omega} \left\{ {g^{\prime}(\phi)\delta \phi} \, \mathcal{H} +
        \frac{1}{2c_w} G_c \left[ \frac{1}{2 \ell} {w^{\prime}(\phi)} \delta \phi - \ell \nabla \phi  \cdot \nabla \delta \phi \right] \right\}  \, \text{d}V = 0,
\end{equation}
\begin{equation}
\int_{\Omega} \left\{\left(\rho_{fl} \left(S\Dot{p} + \alpha\chi_r\Dot{\varepsilon}_{vol}\right) - q_m \right) \delta p  + \frac{\rho_{fl}}{\mu_{fl}}  (\bm{K}_{fl} {\nabla p}) \cdot \nabla \delta p  \right\} \, \text{d}V 
+ \int_{\partial\Omega} \delta p \mathbf{q} \cdot \mathbf{n} \, \text{d}S = 0,
\end{equation}

The primary variables $\mathbf{u}$, $\phi$, and $p$ are approximated using the shape functions $N_i$ corresponding to node $i$ as follows:
\begin{equation}\label{Eq:ShapeFun}
\mathbf{u} = \sum_i^n \mathbf{N}_i \mathbf{u}_i, \quad \phi = \sum_i^n N_i \phi_i, \quad p = \sum_i^n N_i p_i.
\end{equation}

The gradients of these variables are computed by differentiating the shape functions with respect to the spatial coordinates, resulting in the following $\bm{B}$-matrices:
\begin{equation}\label{Eq:BMatrices}
\boldsymbol{\varepsilon} = \sum_i^n \bm{B}_i^u \mathbf{u}_i, \quad \nabla \phi = \sum_i^n \bm{B}_i \phi_i, \quad \nabla p = \sum_i^n \bm{B}_i p_i.
\end{equation}

Using the approximations in Eqs.~\eqref{Eq:ShapeFun} and \eqref{Eq:BMatrices}, the nodal residuals are expressed as:
\begin{align}
    & \mathbf{R}_i^\mathbf{u} = \int_\Omega \left\{ \left(\bm{B}^\mathbf{u}_i\right)^T (\bm{\sigma}^{eff} - \alpha p \bm{I}) - \mathbf{N}_i^T \mathbf{b} - \mathbf{N}_i^T \mathbf{T} \right\} \, \text{d}V, \\
    & R_i^\phi = \int_\Omega \left\{ {g^{\prime}(\phi)}N_i \mathcal{H} + \frac{G_c}{2c_w \ell}  \left[\frac{w^{\prime}(\phi)}{2 } N_i + \ell^2 \,  \left( \bm{B}_i \right)^T \nabla\phi\right]\right\} \, \text{d}V, \\
    & R_i^{p} = \int_{\Omega} \left[ \left(\rho_{fl} \left(S\Dot{p}+\alpha\chi_r\Dot{\varepsilon}_{vol}\right) - q_m \right) N_i+ \bm{B}_i^T \left(\rho_{fl}\frac{\bm{K}_{fl}}{\mu_{fl}}{\nabla p} \right)  \right] \, \text{d}V - \int_{\partial\Omega} N_i q\, \text{d}S.
\end{align}

The stiffness matrix is obtained by taking the variation of the residual with respect to each relevant primary variable:
\begin{align}\label{Eq:stiffness}
    & \bm{K}_{ij}^{\mathbf{u}} = \frac{\partial \mathbf{R}_{i}^{\bm{u}} }{\partial \bm{u}_{j} } = 
        \int_{\Omega} \left\{  {(\bm{B}_{i}^{\mathbf{u}})}^{T} \bm{C} \, \bm{B}_{j}^{\mathbf{u}} \right\} \, \text{d}V, \\
    & \bm{K}_{ij}^{\phi} = \frac{\partial R_{i}^{\phi} }{ \partial \phi_{j} } =  \int_{\Omega} \left\{ \left( {g''(\phi)} \mathcal{H} + \frac{G_{c}}{4 c_w \ell}  {w''(\phi)} \right) N_{i} N_{j} + \frac{G_{c} \ell}{2 c_w}   \, \bm{B}_i^T\bm{B}_j \right\} \, \text{d}V, \\
    & \bm{K}_{ij}^{p} = \frac{\partial {R}_i^{p}}{\partial p_j} = 
    \int_{\Omega} \left\{ \frac{1}{\delta t} N_i \left(\rho_{fl} S \right) N_j + (\bm{B}_i)^T \cdot \left(\rho_{fl} \frac{\bm{K}_{fl}}{\mu_{fl}}\right) \cdot \bm{B}_j - N_i \frac{\partial q_m}{\partial p} N_j \right\} \, \text{d}V 
    - \int_{\partial \Omega} N_i \frac{\partial q}{\partial p} N_j \, \text{d}S,
\end{align}

\noindent where $\bm{C}$ represents the Jacobian, obtained by taking the second variation of the strain energy with respect to the strain tensor.

\subsection{Computation of the material Jacobian}
\label{Sec:Jacobian}

The computation of the material Jacobian is intrinsically linked to the choice of strain energy decomposition. Let us start by expressing the strain energy density as a function of the undamaged strain energy $\psi_0 \left( \boldsymbol{\varepsilon} \right)$ and the stored strain energy $\psi_s \left( \boldsymbol{\varepsilon} \right)$, such that
\begin{equation}
    \psi \left( \boldsymbol{\varepsilon}, \phi \right) = g \left( \phi \right) \psi_0 \left( \boldsymbol{\varepsilon} \right) + \left( 1 - g \left( \phi \right) \right) \psi_s \left( \boldsymbol{\varepsilon} \right),
\end{equation}

\noindent where the tangential stiffness tensor is given by:
\begin{equation}\label{Eq:Stiffness}
\bm{C}=g \left( \phi \right)\frac{\partial^2 \psi_0}{\partial \boldsymbol\varepsilon^2}+ \left( 1 - g \left( \phi \right) \right)\frac{\partial^2 \psi_s}{\partial \boldsymbol\varepsilon^2}=g \left( \phi \right)\bm{C}_0+ \left( 1 - g \left( \phi \right) \right)\bm{C}_s,
\end{equation}

\noindent Here, $\bm{C}_0$ and $\bm{C}_s$ are the tangential stiffness tensors for the undamaged and fully cracked configurations, respectively. Calculating $\bm{C}_s$ provides the anisotropic tangential stiffness tensor $\bm{C}$.\\

The strain energy splits defined in Section \ref{seq:Drivnign-force} can be divided into two main groups. The first group includes those based on the strain tensor in its original form (i.e., without rotations), such as the volumetric-deviatoric split \cite{Amor2009} and the Drucker-Prager model \cite{de2022nucleation,Navidtehrani2022}. The second group is based on principal strains, such as spectral decomposition \cite{Miehe2010a} and the no-tension model \cite{Freddi2010}. The first group can be directly obtained by differentiation with respect to the strain tensor, whereas for the second group, the Jacobian is first determined for the principal directions and subsequently rotated to the original coordinate system.\\

For the first group, it can be shown that the volumetric-deviatoric split is a special case of the Drucker-Prager model when $B=0$. Hence, let us derive $\bm{C}_s$ for the Drucker-Prager model and particularise later. Thus,
\begin{equation} \label{eq:Cs1}
\bm{C}_s=\frac{\partial^2 \psi_s}{\partial \boldsymbol\varepsilon^2}=
\begin{cases}
0 & \text{if} \quad -6B\sqrt{J_2 (\boldsymbol\varepsilon)} < I_1 (\boldsymbol\varepsilon), \\
\bm{C}_s^{DP} & \text{if} \quad -6B\sqrt{J_2 (\boldsymbol\varepsilon)} \geq I_1 (\boldsymbol\varepsilon) \,\, \& \,\,  2 \mu \sqrt{J_2 (\boldsymbol\varepsilon)} \geq 3 B K I_1 (\boldsymbol\varepsilon), \\
\bm{C}_0 & \text{if} \quad 2 \mu \sqrt{J_2 (\boldsymbol\varepsilon)} < 3 B K I_1 (\boldsymbol\varepsilon).
\end{cases}
\end{equation}

\noindent where $\bm{C}_s^{DP}$ is defined as:
\begin{equation}\label{eq:CsDP}
\begin{aligned}
(C_s^{DP})_{ijkl}=\frac{K \mu}{9 B^2 K + \mu}\left(\frac{\partial I_{1}}{\partial \varepsilon_{i j}}+\frac{3 B}{\sqrt{J_{2}}} \frac{\partial J_{2}}{\partial \varepsilon_{i j}}\right)\left(\frac{\partial I_{1}}{\partial \varepsilon_{k l}}+\frac{3 B}{\sqrt{J_{2}}} \frac{\partial J_{2}}{\partial \varepsilon_{k l}}\right)+ \\ \left(\frac{6 B a_{1}\left(I_{1}+6 B \sqrt{J_{2}}\right)}{\sqrt{J_{2}}}\right)\left(\frac{\partial^{2} J_{2}}{\partial \varepsilon_{i j} \partial \varepsilon_{k l}}-\frac{1}{2 J_{2}} \frac{\partial J_{2}}{\partial \varepsilon_{i j}} \frac{\partial J_{2}}{\partial \varepsilon_{k l}}\right).
\end{aligned}
\end{equation}

Considering $B=0$ in Eqs. (\ref{eq:Cs1})-(\ref{eq:CsDP}), renders the material Jacobian for the volumetric-deviatoric split. \\

For the second group, the Jacobian in the principal direction $\bm{C}_s^{'}$ is calculated. For the spectral decomposition, the fully cracked stiffness tensor in the principal direction is given by:
\begin{equation}
(C_s^{'})_{ijkl}=\frac{1-\operatorname{sgn}\left({I_1(\bm{\epsilon})}\right)}{2}  \delta_{ij} \delta_{kl} \lambda+  2 \mu \left(\delta_{ij} \delta_{kl}  -
\frac{\partial^2 I_2(\bm{\epsilon}_{-}) }{\partial \mathrm{\epsilon_{ij}^{-}} \partial \mathrm{\epsilon_{kl}^{-}}} \right) \frac{\partial {\epsilon_{ij}^{-}} }{\partial \mathrm{\epsilon_{ij}}} \frac{\partial {\epsilon_{kl}^{-}} }{\partial \mathrm{\epsilon_{kl}}},
\end{equation}

\noindent where $\delta_{ij}$ is the Kronecker delta. The variation ${\partial {\epsilon}_{ij}^{-}}/{\partial \mathrm{\epsilon_{ij}}}$ is defined as:
\begin{equation}
\frac{\partial {\epsilon_{ij}^{-}} }{\partial \mathrm{\epsilon_{ij}}}=\left\{\begin{array}{cc}
0 & \epsilon_{ij}>0, \\
\frac{1}{2} & \epsilon_{ij}=0, \\
1 & \epsilon_{ij}<0.
\end{array}\right. ,
\quad
\operatorname{sgn}(x)=
\left\{\begin{array} {cc} 
1 & x>0, \\
0 & x=0, \\
-1 & x<0.
\end{array}\right.
\end{equation}

For the no-tension model, the material Jacobian in the principal direction is:
\begin{equation}
(C_s^{'})_{ijkl}=\frac{\partial \psi_s}{\partial \epsilon_{ij} \partial \epsilon_{kl} }=
\begin{cases}
0 & \epsilon_1>0, \\
\delta_{i1} \delta_{j1} \delta_{k1} \delta_{l1} \, E &  \epsilon_2 + \nu \epsilon_1 > 0, \\
\delta_{ij} \delta_{kl} (1-\delta_{i3}) \big(\delta_{ik}+(1-\delta_{k3}) \, \nu \big) \, \frac{E}{1 - \nu^2} & (1-\nu)\epsilon_3 + \nu (\epsilon_1 + \epsilon_2) > 0, \\
\bm{C}_0 & else.
\end{cases}
\end{equation}

The tangential stiffness matrix in the original direction $\bm{C}$ is obtained by rotating $\bm{C}^{'}$ using:
\begin{equation}\label{Eq:Rotation}
C_{qrst} = a_{qi} a_{rj} a_{sk} a_{tl} C_{ijkl}^{'} \, ,
\end{equation}

\noindent where $\bm{a}$ is the transpose of the direction cosines matrix for the principal directions, $\bm{a}^{'} = [\boldsymbol{v}_1, \boldsymbol{v}_2, \boldsymbol{v}_3]$, with $\boldsymbol{v}_1$, $\boldsymbol{v}_2$, and $\boldsymbol{v}_3$ as the principal vectors of the strain tensor, satisfying:
\begin{equation}
(\boldsymbol{\varepsilon} - \epsilon_{ii} \bm{I}) \cdot \boldsymbol{v}_i = 0,
\end{equation}
\noindent for $i=1,2,3$, and $\bm{I}$ as the identity matrix.

\subsection{Solution scheme}
\label{Sec:Solution}

After computing all necessary components of residual and stiffness matrices, we can solve the nonlinear coupled equations using an iterative procedure based on the Newton-Raphson method. The algorithm is detailed in Algorithm \ref{alg:Solution Scheme}. As shown, the coupled stiffness matrices are omitted ($\bm{K}^{\mathbf{u}, \phi} = 0$, $\bm{K}^{\mathbf{u}, p} = 0$, $\bm{K}^{\phi, p} = 0$). While these stiffness matrices can enhance the convergence rate for strongly coupled equations, solving the equations separately reduces the size of subproblem, thereby saving computational time and storage per iteration, resulting in less computational effort overall.

\begin{algorithm}[H]
\caption{Solution algorithm for phase field hydraulic fracture in $[t_n, t_{n+1}]$}
\label{alg:Solution Scheme}
\begin{algorithmic}[1]
\State \textbf{Input:} Displacement field $\mathbf{u}_n$, phase field $\phi_n$, history field $\mathcal{H}_n$, and fluid pressure field $p_n$ at time $t_n$.
\State \textbf{Initialization:} Set the initial guess for Newton-Raphson iterations at $t_{n+1}$: $\mathbf{u}_{n+1}^0$, $\phi_{n+1}^0$, $p_{n+1}^0$. Initialize the iteration counter $i = 0$.
\Repeat
    \State Compute $\mathbf{R}^{\mathbf{u}}$ and $\bm{K}^{\mathbf{u}}$ for the variables $\mathbf{u}$, $\phi$, $p$\textsuperscript{*}.
    \State Compute $\mathbf{R}^{\phi}$ and $\bm{K}^{\phi}$ for the variables $\mathcal{H}$, $\phi$\textsuperscript{*}.
    \State Compute $\mathbf{R}^{p}$ and $\bm{K}^{p}$ for the variables $\dot{\varepsilon}_{vol}$, $\phi$, $p$\textsuperscript{*}.
    \State Solve the coupled system of equations for $\mathbf{u}_{n+1}^{i+1}$, $\phi_{n+1}^{i+1}$, $p_{n+1}^{i+1}$ using:
    \begin{equation}\label{eq:sysEq}
    \begin{aligned}
    \begin{bmatrix}
    \mathbf{u}^{i+1} \\
    \phi^{i+1} \\
    p^{i+1}
    \end{bmatrix}_{t_{n+1}}
    = 
    \begin{bmatrix}
    \mathbf{u}^{i} \\
    \phi^{i} \\
    p^{i}
    \end{bmatrix}_{t_{n+1}}
    -
    \begin{bmatrix}
    \bm{K}^{\mathbf{u}} & 0 & 0 \\
    0 & \bm{K}^{\phi} & 0 \\
    0 & 0 & \bm{K}^{p}
    \end{bmatrix}_t^{-1}
    \begin{bmatrix}
    \mathbf{R}^{\mathbf{u}} \\
    \mathbf{R}^\phi \\
    \mathbf{R}^p
    \end{bmatrix}_t.
    \end{aligned}
    \end{equation}
    \State Compute the norm of the residual for the updated variables, $||\mathbf{R}(\mathbf{u}_{n+1}^{i+1}, \phi_{n+1}^{i+1}, p_{n+1}^{i+1})||$.
    \If{$||\mathbf{R}|| < \text{TOL}$}
        \State Converged. Proceed to the next time increment $t_{n+2}$.
    \Else
        \State Increment the iteration counter $i \gets i+1$.
    \EndIf
\Until Convergence is achieved.
\end{algorithmic}
\vspace{0.5em}
\textbf{*} \footnotesize{The variables are selected based on the solution scheme described in Table \ref{tab:Soluetion_scheme}.}
\end{algorithm}

Various solution schemes exist for coupled equations, such as the monolithic and staggered schemes \cite{gerasimov2019penalization,kristensen2020phase}. In the monolithic scheme, all equations are solved simultaneously, updating all variables in each equation. In contrast, the staggered method updates only the primary variable of an equation while using variables of other equations from the previous increment (single-pass staggered) or the last iteration (multi-pass staggered). The monolithic scheme is unconditionally stable, allowing for larger time increments, but it often requires more iterations to achieve convergence due to the highly nonlinear behaviour. On the other hand, the staggered scheme converges with fewer iterations but requires smaller time increments for accurate results.\\ 

The required variables in steps 4 to 6 of Algorithm \ref{alg:Solution Scheme} for different solution schemes are shown in Table \ref{tab:Soluetion_scheme}. A combination of monolithic and staggered approaches can be used for systems with more than two coupled equations. For example in the mixed monolithic scheme, the linear momentum equation and phase field evolution equation are solved using the monolithic scheme, while the fluid equation is solved with the other two in a multi-pass staggered manner. In the mixed staggered scheme, the linear momentum and phase field evolution equations are solved using a single-pass staggered scheme, while the fluid equation is solved with a multi-pass staggered approach.

\begin{table}[H]
\caption{Selection of variables for steps 4 to 6 of Algorithm \ref{alg:Solution Scheme} based on the solution scheme. In $f_n^i$, the subscript $n$ represents the time increment number, while the superscript $i$ denotes the iteration number.}
\centering
\begin{tabular}{l @{\hskip 20pt} lll @{\hskip 30pt} ll @{\hskip 30pt} lll}
\hline
\hline
\multicolumn{1}{c@{\hskip 20pt}}{\textbf{Solution Scheme}} & 
\multicolumn{3}{c@{\hskip 30pt}}{\textbf{Step 4}} & 
\multicolumn{2}{c@{\hskip 30pt}}{\textbf{Step 5}} & 
\multicolumn{3}{c}{\textbf{Step 6}} \\
& $\mathbf{u}$ & $\phi$ & $p$ & $\mathcal{H}$ & $\phi$ & $\dot{\varepsilon}_{vol}$ & $\phi$ & $p$ \\
\hline
\hline
Monolithic & $\mathbf{u}_{n+1}^i$  & $\phi_{n+1}^i$  & $p_{n+1}^i$  & $\mathcal{H}_{n+1}^i$  & $\phi_{n+1}^i$  & $(\dot{\varepsilon}_{vol})_{n+1}^i$ & $\phi_{n+1}^i$  & $p_{n+1}^i$ \\
Single-pass staggered & $\mathbf{u}_{n+1}^i$  & $\phi_{n}$  & $p_{n}$   & $\mathcal{H}_{n}$ & $\phi_{n+1}^i$ & $(\dot{\varepsilon}_{vol})_{n}$ & $\phi_{n}$  & $p_{n+1}^i$ \\
Multi-pass staggered & $\mathbf{u}_{n+1}^i$  & $\phi_{n+1}^{i-1}$  & $p_{n+1}^{i-1}$  & $\mathcal{H}_{n+1}^{i-1}$ & $\phi_{n+1}^i$ & $(\dot{\varepsilon}_{vol})_{n+1}^{i-1}$ & $\phi_{n+1}^{i-1}$  & $p_{n+1}^i$ \\
Mixed monolithic & $\mathbf{u}_{n+1}^i$  & $\phi_{n+1}^i$ & $p_{n+1}^{i-1}$ & $\mathcal{H}_{n+1}^i$ & $\phi_{n+1}^i$ & $(\dot{\varepsilon}_{vol})_{n+1}^i$ & $\phi_{n+1}^i$  & $p_{n+1}^i$ \\
Mixed staggered & $\mathbf{u}_{n+1}^i$  & $\phi_{n+1}^i$  & $p_{n+1}^{i-1}$ & $\mathcal{H}_{n}$ & $\phi_{n+1}^i$ & $(\dot{\varepsilon}_{vol})_{n+1}^i$ & $\phi_{n+1}^i$  & $p_{n+1}^i$ \\
\hline
\end{tabular}
\label{tab:Soluetion_scheme}
\end{table}

As described in \ref{Sec:Abaqus}, we implement this framework, and the solution schemes provided in Table \ref{tab:Soluetion_scheme}, within the commercial finite element package Abaqus. A novel procedure is exploited to carry out the numerical implementation at the integration point level, without the need to define residuals and stiffness matrices, which are here provided for the sake of generality. 

\section{Numerical experiments}
\label{Sec:Examples}

Four case studies are extensively investigated to evaluate the proposed methods and highlight the relevance of the two novel ingredients proposed: the Drucker-Prager-based split and the hybrid permeability approach. In the first case study (Section \ref{sec:Coupling_Method}), we analysed a rectangular domain with a central vertical crack to examine the coupling effects between the phase field variable and the permeability tensor. This configuration allowed us to assess how the phase field influences permeability in fractured regions, demonstrating the efficacy of the hybrid permeability approach presented. The second case study, presented in Section \ref{Sec:StickSlip}, focused on a stick-slip problem, illustrating the capability of the Drucker-Prager-based split method to model stick-slip behaviour accurately. This example highlights the suitability of the Drucker-Prager-based split in simulating stress redistribution and frictional resistance in geotechnical applications. The third case study (Section \ref{Sec:FractureInter}) investigated the influence of different fracture-fluid coupling methods and strain energy decompositions as the driving force for fracture propagation in a crack interaction problem. By considering different decomposition approaches, we evaluated how different fracture-driving mechanisms affect crack growth and interaction. Finally, the fourth case study, presented in Section \ref{Sec:Axisym} involved modelling an axisymmetric domain with initial stress, subjected to fluid injection to simulate multiaxial conditions. This scenario allowed us to assess the applicability of the proposed framework under complex loading conditions, relevant to subsurface applications involving fluid-driven fracture under multiaxial stress states. Unless otherwise stated, the AT2 model is employed.

\subsection{Influence of the approach adopted to model the coupling between permeability and phase field}
\label{sec:Coupling_Method}

\begin{figure}[H]
    \centering
    \includegraphics[width=0.6\textwidth]{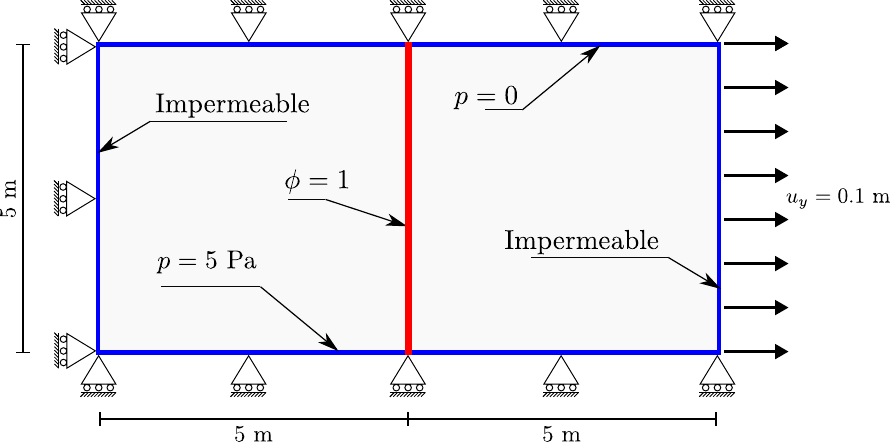}
    \caption{Geometry, dimensions, and boundary conditions of a rectangular domain with a central vertical fracture subjected to pressure at the bottom.}
    \label{fig:Darcy-Poiseuille-Config}
\end{figure}

We begin by investigating the impact of various coupling methods between permeability and phase field. This study examines three distinct coupling strategies, as discussed in Section \ref{sec:Phase field hydraulic fracture}. The problem setup involves a rectangular domain with a vertical crack located at the centre, see Fig. \ref{fig:Darcy-Poiseuille-Config}. The focus of this analysis is on fluid behaviour within the crack rather than crack propagation. To this end, a pre-existing vertical fracture is introduced at the centre of the domain. Fluid pressure is applied with the following boundary conditions: (i) a \( p = 0 \) Pa pressure at the top, maintained constant throughout the analysis, and (ii) a linearly increasing pressure going from 0 to \( p = 5 \) Pa over 100 seconds. Both lateral boundaries are considered impermeable. Following the application of pressure, a horizontal displacement of \( u_x = 0.1 \) m is imposed on the left boundary over an additional 100 seconds to investigate the effect of crack opening under the different coupling methods. The material parameters, as outlined in Table \ref{tab:ParametersCase1}, are chosen for illustrative purposes and are not intended to represent realistic values. For example, the critical fracture energy release rate, \( G_c \), is set to a very high value (\( 10^6 \) J/m\(^2\)) to prevent crack propagation during the pressure loading phase. The domain is discretized using a uniform mesh of bilinear quadrilateral elements, each with a size of 10 cm. This analysis primarily focuses on the phase field fracture AT2 model. However, for the sake of completeness, the effect of the coupling method on the AT1 model is also investigated.

\begin{table}[H]
\begin{center}
\caption{Material and model parameters for the first case study, aimed at investigating the permeability-phase field coupling.}
\label{tab:ParametersCase1}
\begin{tabular}{m{7cm} c c c }
\hline
Parameter & Symbol & Value & Unit\\
\hline \hline
Young’s modulus & \( E \) & 50 & GPa \\
Poisson’s ratio & \( \nu \) & 0.3 & -- \\
Characteristic length scale & \( \ell \) & 0.5 & m \\
Critical fracture energy & \( G_c \) & \( 10^6 \) & J/m\(^2\) \\
Biot’s coefficient of reservoir domain & \( \alpha_r \) & 0.002 & -- \\
Porosity of reservoir domain & \( \varepsilon_{pr} \) & 0.002 & -- \\
Density of the fluid & \( \rho_{fl} \) & 1000 & kg/m\(^3\) \\
Dynamic viscosity of the fluid & \( \mu_{fl} \) & 0.001 & Pa·s \\
Compressibility of fluid & \( c_{fl} \) & \( 10^{-8} \) & Pa\(^{-1}\) \\
Permeability of reservoir domain & \( K_{r} \) & 0 & m\(^2\) \\
Permeability of fracture domain & \( K_{f} \) & 1 & m\(^2\) \\
\hline
\end{tabular}
\end{center}
\end{table}

\begin{figure}[H]
    \centering
    \begin{subfigure}[b]{0.4\textwidth}
         \centering
         \includegraphics[width=\textwidth]{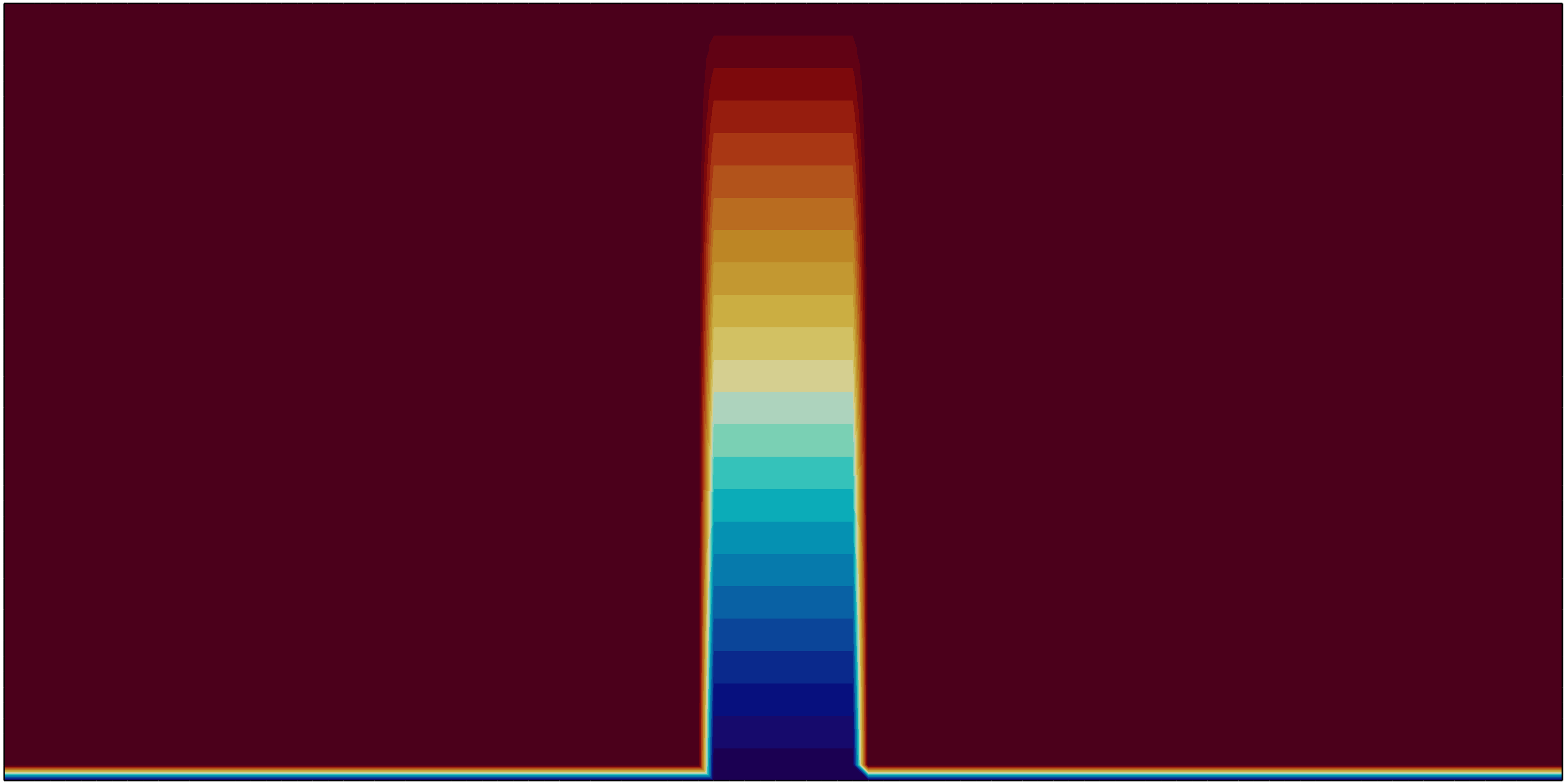}
         \caption{}
     \end{subfigure}
    \begin{subfigure}[b]{0.4\textwidth}
         \centering
         \includegraphics[width=\textwidth]{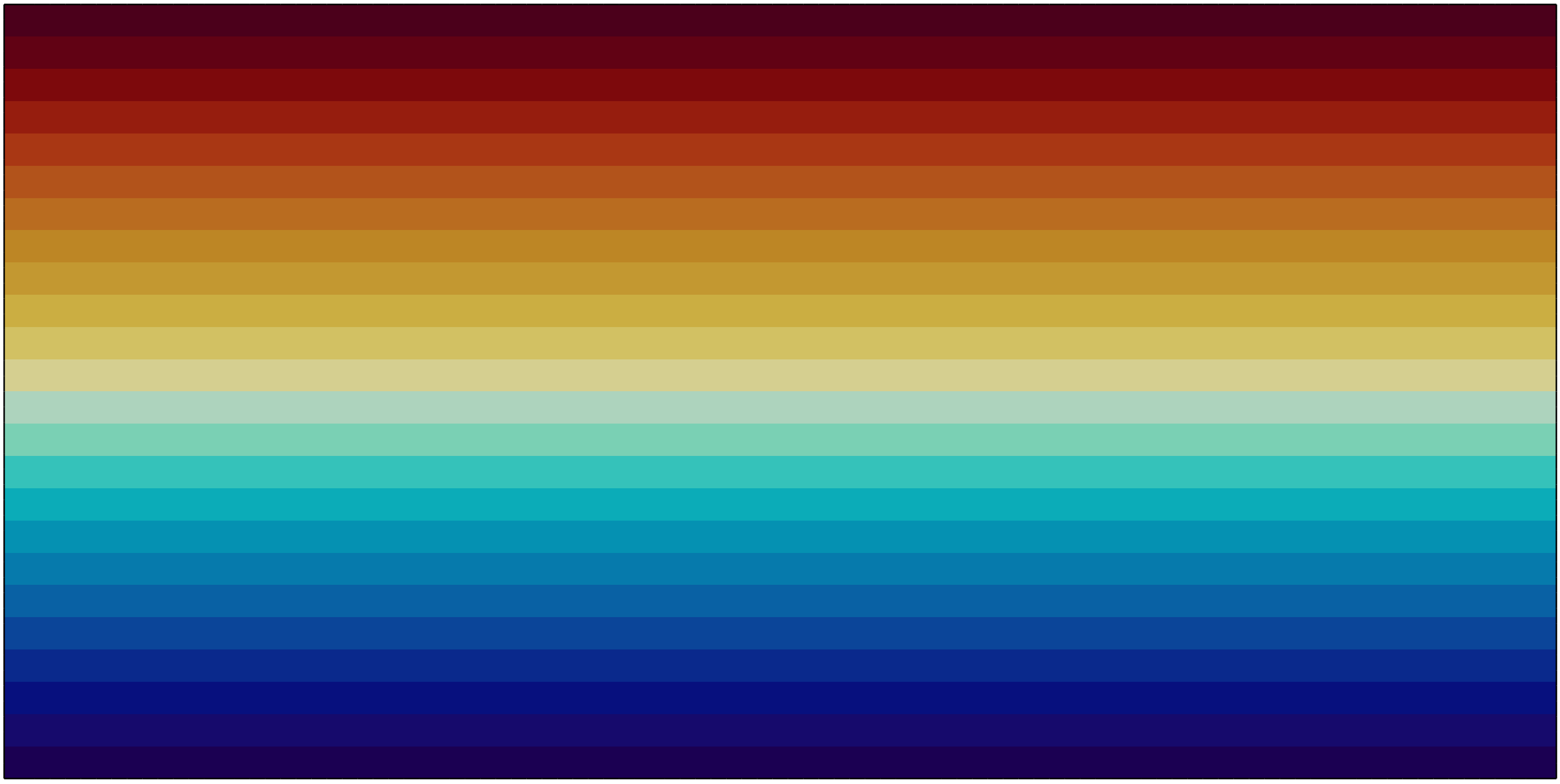}
         \caption{}
     \end{subfigure}
     \begin{subfigure}[b]{0.1\textwidth}
         \centering
         \includegraphics[width=\textwidth]{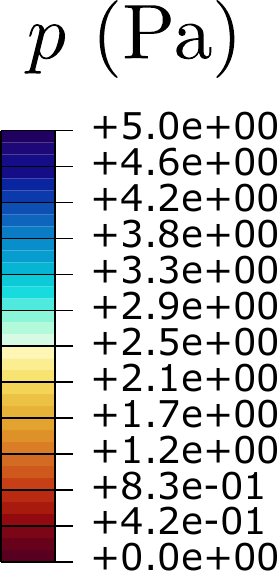}
         \vspace{0 mm}
     \end{subfigure}
    \caption{ Fluid pressure contours for the rectangular domain with a centered crack: (a) Permeability determined using the domain decomposition and hybrid methods, (b) Permeability determined using the modified Darcy method.}
    \label{fig:Perm-P}
\end{figure}

Before analysing the effects of different coupling methods on fluid flux, we first examine their impact on pressure distribution within the domain. Fig. \ref{fig:Perm-P} compares the pressure distribution for the three coupling methods considered to simulate the interplay between permeability and phase field. In Fig. \ref{fig:Perm-P}a, the pressure distribution is shown for the domain decomposition and hybrid methods. With reservoir permeability \( K_r = 0 \), the pressure in that region is zero and is only distributed across the transient and fracture domains. In contrast, Fig. \ref{fig:Perm-P}b displays a uniform pressure distribution for the modified Darcy method due to element size contributions to crack width \( (w_h = \langle | h_e (1 + \mathbf{n}_{\phi} \cdot \boldsymbol{\varepsilon} \cdot \mathbf{n}_{\phi}) | \rangle_+) \), influencing permeability in such a way that there is no region in the domain with zero permeability. Thus, the modified Darcy method introduces artificial permeability in the undamaged region, which does not accurately reflect physical behaviour. In contrast, the domain decomposition method and the proposed hybrid method preserve the physical permeability of the undamaged region.

\begin{figure}[H]
    \centering
    \begin{subfigure}[b]{0.45\textwidth}
         \centering
         \includegraphics[width=\textwidth]{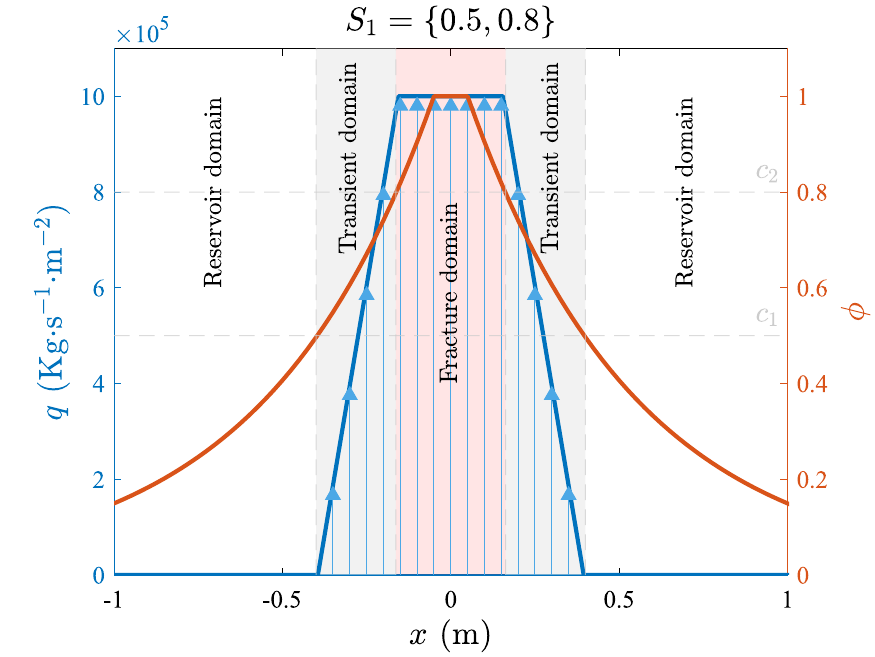}
        \caption{}
     \end{subfigure}
    \begin{subfigure}[b]{0.45\textwidth}
         \centering
         \includegraphics[width=\textwidth]{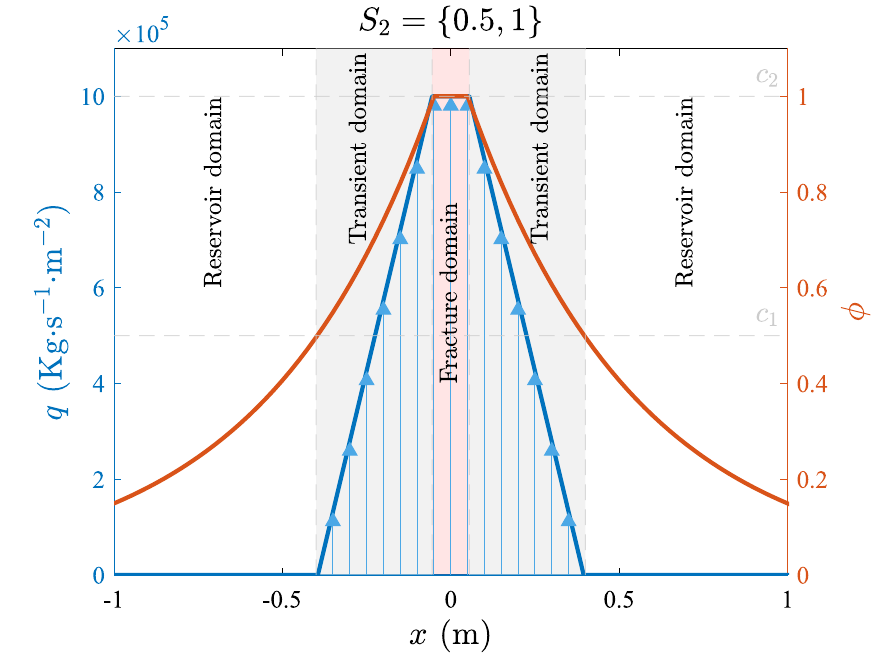}
         \caption{}
     \end{subfigure} 
     \begin{subfigure}[b]{0.45\textwidth}
         \centering
         \vspace{3 mm}
         \includegraphics[width=\textwidth]{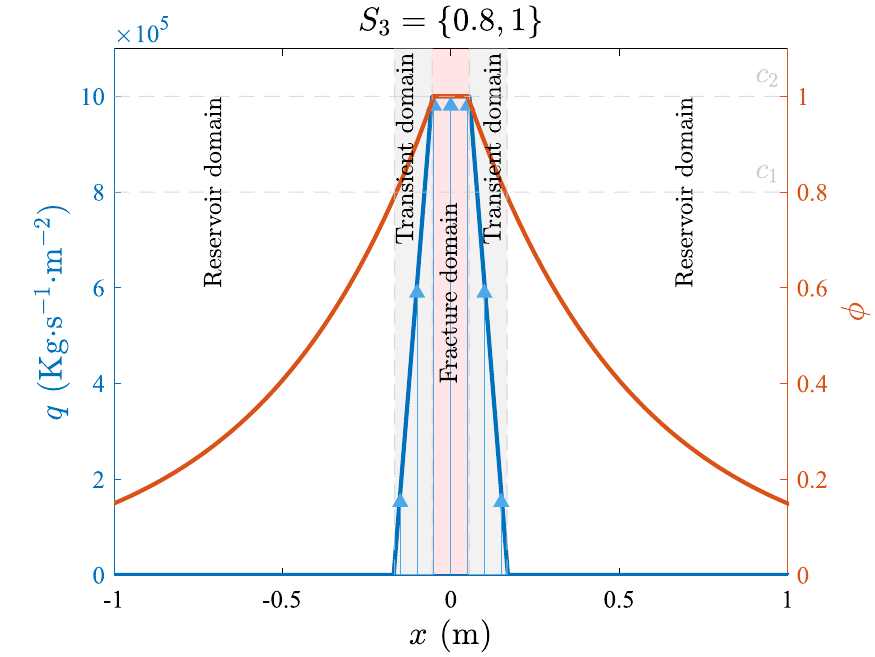}
         \caption{}
     \end{subfigure}
     \begin{subfigure}[b]{0.45\textwidth}
         \centering
         \includegraphics[width=\textwidth]{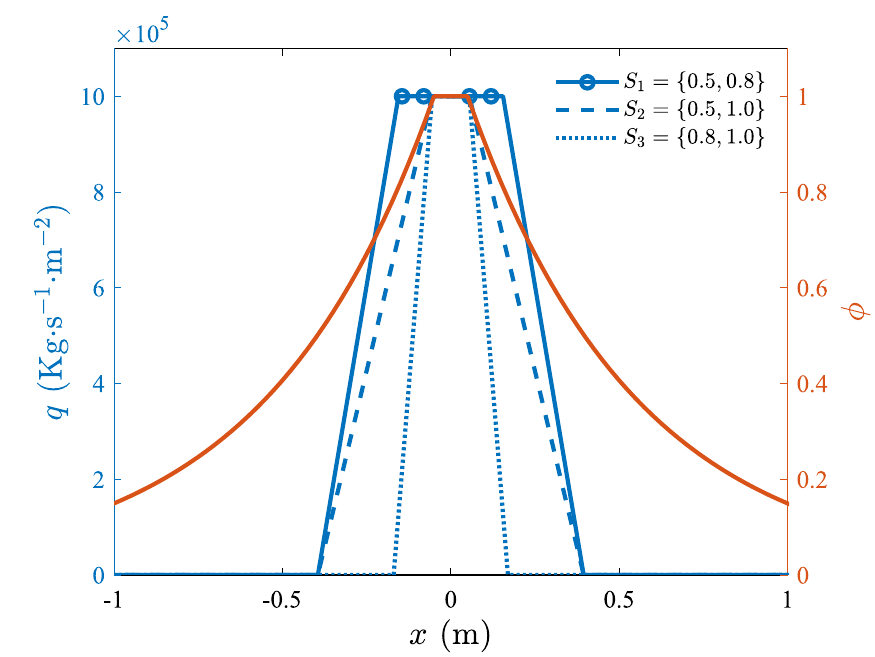}
         \caption{}
     \end{subfigure}
    \caption{ Phase field profile and fluid flux distribution along the $x$-direction at the top of the domain for the domain decomposition method \cite{Lee2016} and: (a) \( S_1 = \{0.5, 0.8\} \), (b) \( S_2 = \{0.5, 1\} \), (c) \( S_3 = \{0.8, 1\} \), while (d) shows the comparison of fluid flux distribution for all sets.}
    \label{fig:Darcy-Poiseuille-P-K1-Flux}
\end{figure}

Additional, quantitative insight can be gained by plotting the flux distribution, as shown in Fig. \ref{fig:Darcy-Poiseuille-P-K1-Flux} for the case of the domain decomposition method \cite{Lee2016}. The results are obtained using domain indicator variables, see Eq. (\ref{Eq:Axu}), with three material constant sets, \( S_n = \{c_1, c_2\} \): \( S_1 = \{0.5, 0.8\} \), \( S_2 = \{0.5, 1\} \), and \( S_3 = \{0.8, 1\} \). With no phase field evolution (constant \( \phi \)), the permeability tensor \( \bm{K}_{fl} \) is constant across the domain and time. The fluid flux for each set is shown in Fig. \ref{fig:Darcy-Poiseuille-P-K1-Flux}a-c, with a comparison of all sets being given in Fig. \ref{fig:Darcy-Poiseuille-P-K1-Flux}d. The division of the domain into three regions based on \( \phi \) values and material constants \( c_1 \) and \( c_2 \) is illustrated in Figs. \ref{fig:Darcy-Poiseuille-P-K1-Flux}a-c. Reservoir permeability is equal where \( \phi < c_1 \), while fracture domain permeability, \( K_f \), applies where \( \phi > c_2 \). The transient domain’s permeability varies linearly, affecting fluid flux along the x-direction (Fig. \ref{fig:Darcy-Poiseuille-P-K1-Flux}a-c).\\

The effect of the three constant sets on fluid flux is compared in Fig. \ref{fig:Darcy-Poiseuille-P-K1-Flux}d, showing equal flux in the fracture domain due to consistent permeability. The transient zone width varies with the selected \( c_1 \) and \( c_2 \) values, impacting mass flow rate \( Q \), which takes values of \( Q = 550 \) tons/s for \( S_1 \), \( Q = 450 \) tons/s for \( S_2 \), and \( Q = 225 \) tons/s for \( S_3 \). This emphasizes the importance of carefully calibrating the values of \( c_1 \) and \( c_2 \) for accurate modelling.

\begin{figure}[H]
    \centering
    \begin{subfigure}[b]{0.45\textwidth}
         \centering
         \includegraphics[width=\textwidth]{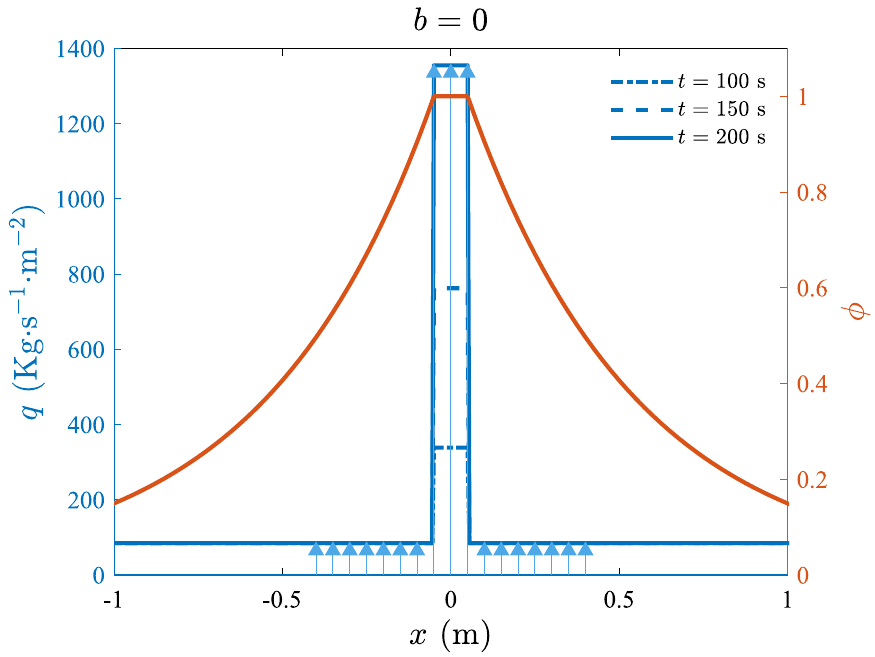}
        \caption{}
     \end{subfigure}
    \begin{subfigure}[b]{0.45\textwidth}
         \centering
         \includegraphics[width=\textwidth]{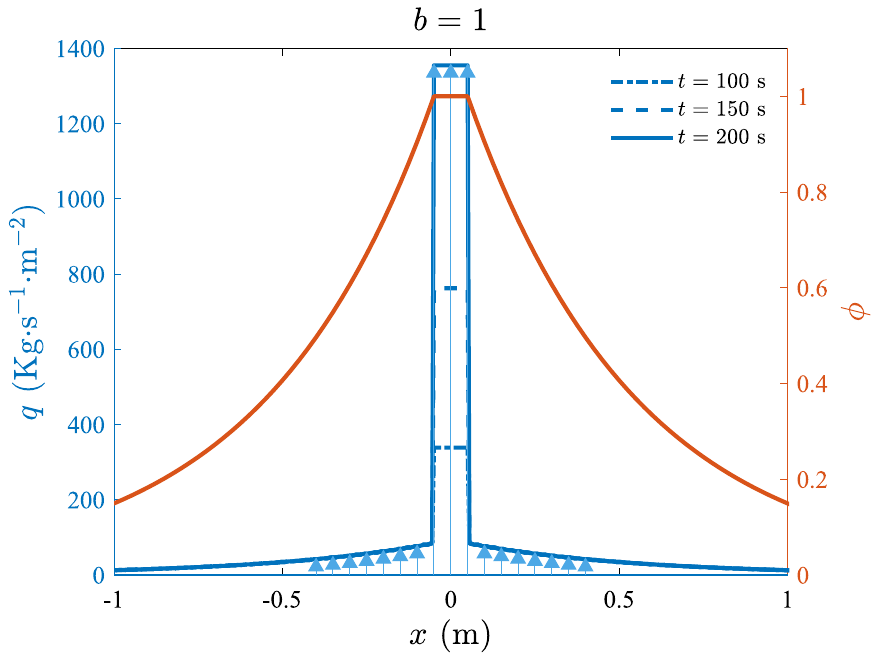}
         \caption{}
     \end{subfigure} 
     \begin{subfigure}[b]{0.45\textwidth}
         \centering
         \vspace{3 mm}
         \includegraphics[width=\textwidth]{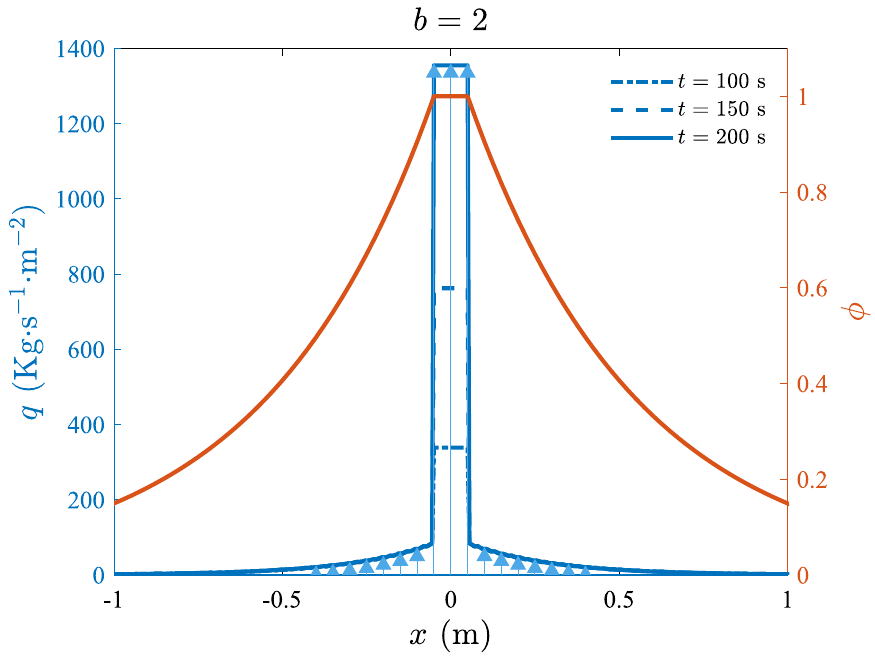}
         \caption{}
     \end{subfigure}
     \begin{subfigure}[b]{0.45\textwidth}
         \centering
         \includegraphics[width=\textwidth]{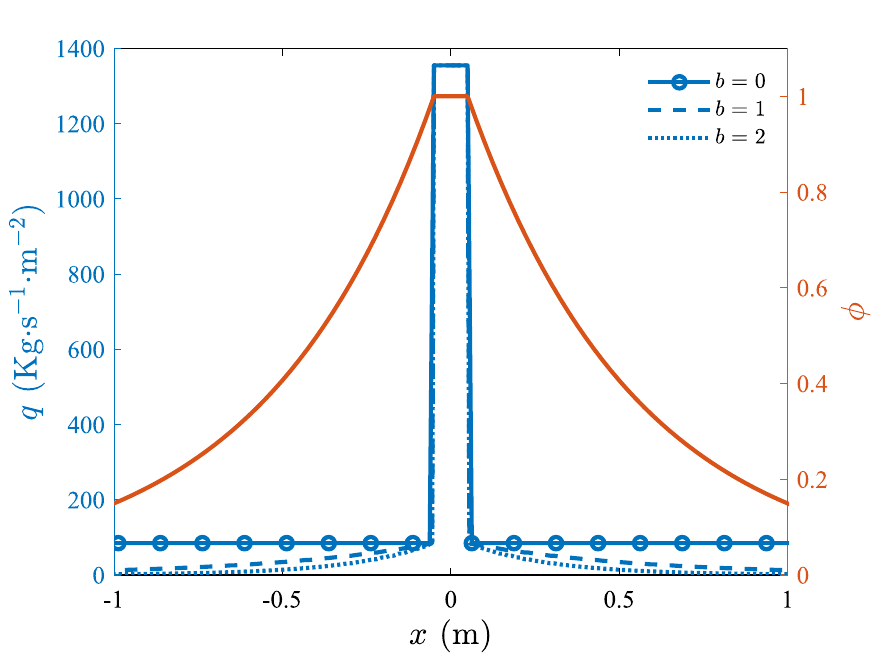}
         \caption{}
     \end{subfigure}
    \caption{ Phase field profile and fluid flux distribution along the $x$-direction at the top of the domain for the modified Darcy method \cite{miehe2015minimization} and: (a) \( b=0 \), (b) \( b=1 \), (c) \( b=2 \), while (d) shows the comparison of fluid flux distribution for all \( b \) values.}
    \label{fig:Darcy-Poiseuille-P-K2-Flux}
\end{figure}

For the modified Darcy method \cite{miehe2015minimization}, permeability is modeled as an anisotropic tensor to represent Poiseuille-type flow in cracks. Three values of the transition parameter \( b = \{0, 1, 2\} \) were considered. Fig. \ref{fig:Darcy-Poiseuille-P-K2-Flux}a-c shows fluid flux for each \( b \) value at times \( t = 100 \) s, \( t = 150 \) s, and \( t = 200 \) s. For \( b = 0 \) (no transition), flux remains uniform across regions with \( \phi < 1 \), though permeability in zero-\( \phi \) areas is non-zero due to element size contributions. This changes with linear (\( b = 1 \)) and quadratic (\( b = 2 \)) transitions, where permeability in low-\( \phi \) regions decreases as \( b \) increases, see Fig. \ref{fig:Darcy-Poiseuille-P-K2-Flux}d. Employing a large transient parameter $b$ results in a narrower flux profile.

\begin{figure}[H]
    \centering
    \begin{subfigure}[b]{0.45\textwidth}
         \centering
         \includegraphics[width=\textwidth]{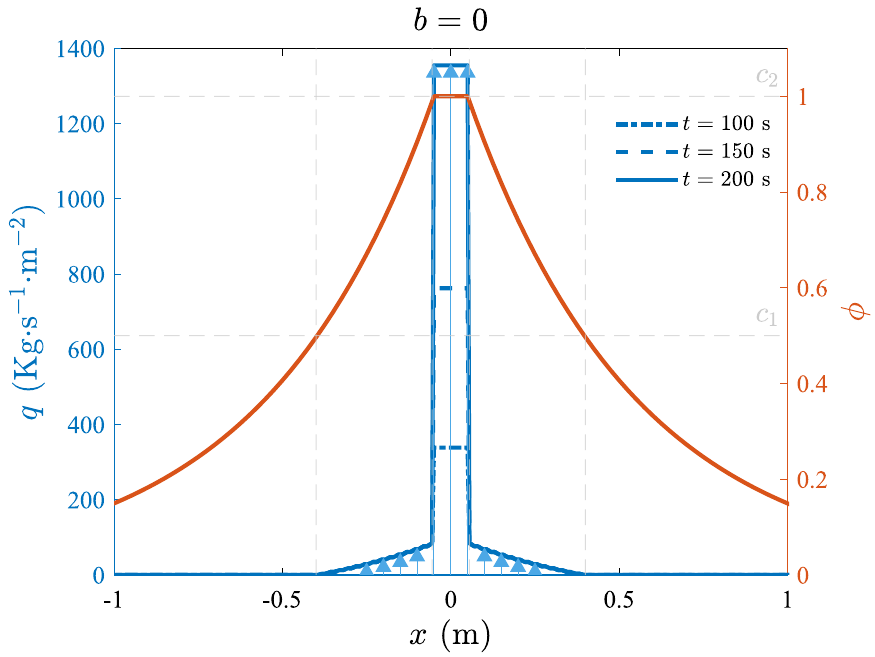}
        \caption{}
     \end{subfigure}
    \begin{subfigure}[b]{0.45\textwidth}
         \centering
         \includegraphics[width=\textwidth]{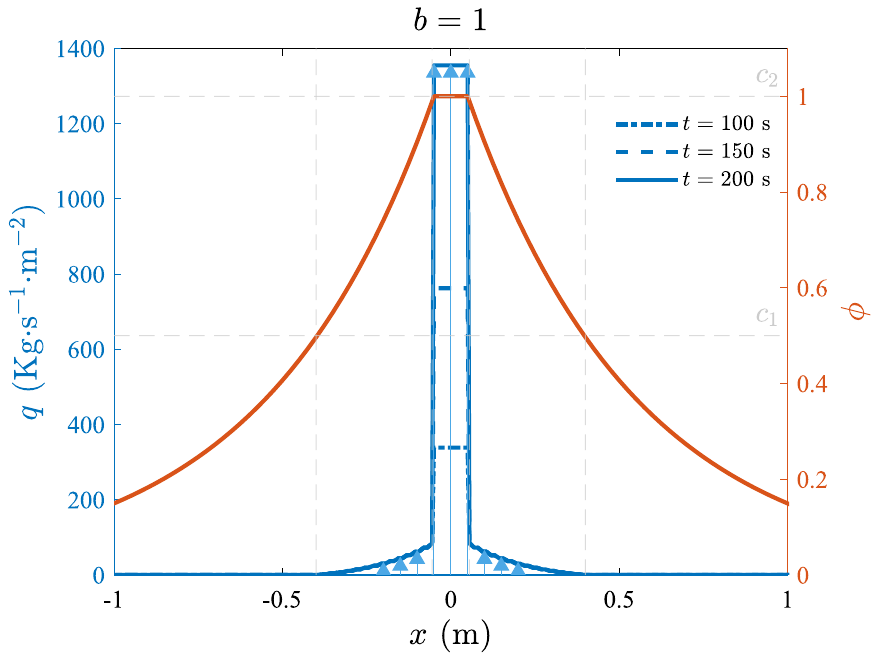}
         \caption{}
     \end{subfigure} 
     \begin{subfigure}[b]{0.45\textwidth}
         \centering
         \vspace{3 mm}
         \includegraphics[width=\textwidth]{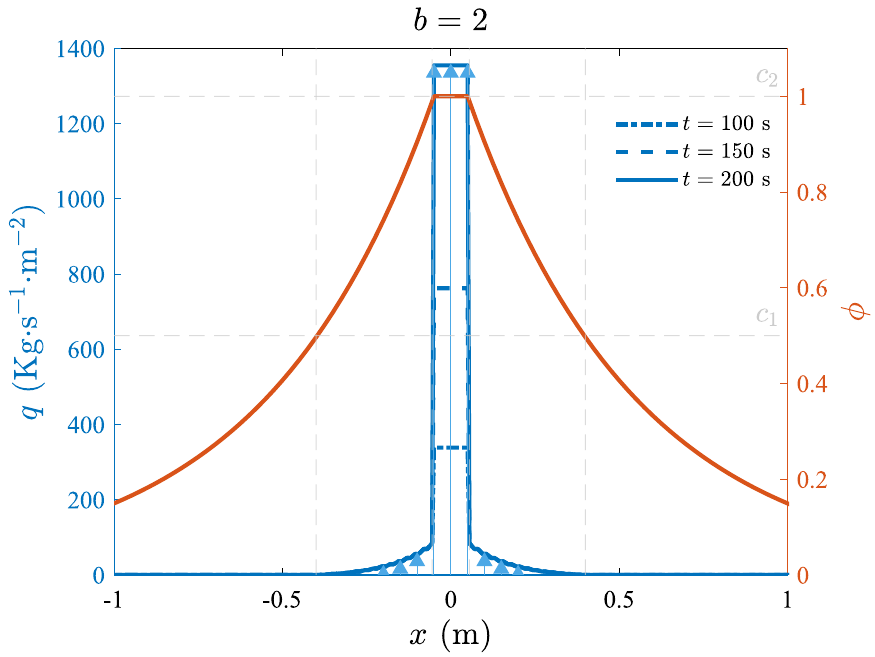}
         \caption{}
     \end{subfigure}
     \begin{subfigure}[b]{0.45\textwidth}
         \centering
         \includegraphics[width=\textwidth]{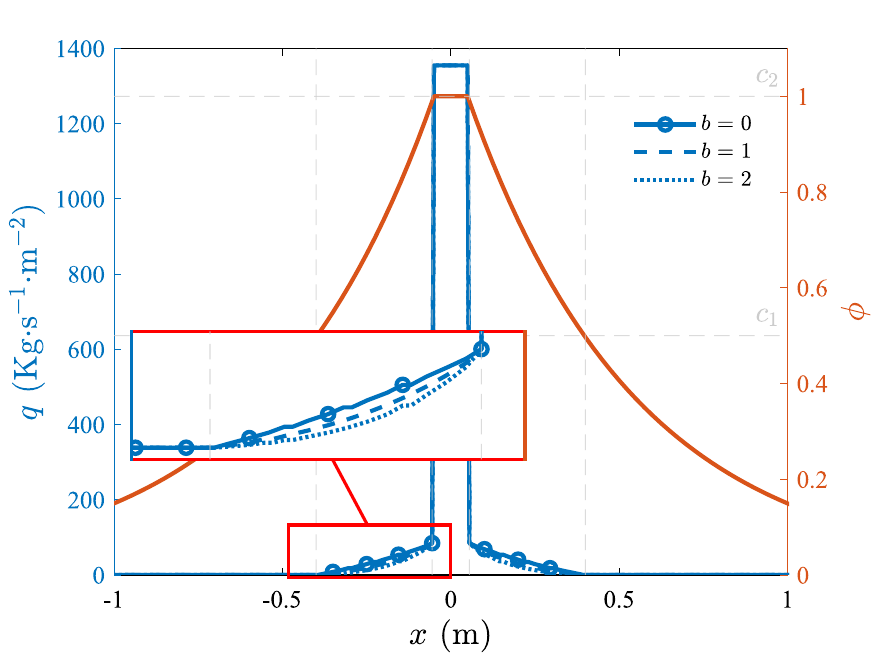}
         \caption{}
     \end{subfigure}
    \caption{ Phase field profile and fluid flux distribution along the $x$-direction at the top of the domain for the proposed hybrid method for the following choices of parameters: \( c_1=0.5 \), \( c_2=1 \) and: (a) \( b=0 \), (b) \( b=1 \), (c) \( b=2 \), while (d) shows the comparison of fluid flux distribution for all \( b \) values.}
    \label{fig:Darcy-Poiseuille-P-K3-Flux}
\end{figure}

The hybrid method combines the domain decomposition and modified Darcy methods. The results obtained for material constants \( c_1 = 0.5 \), \( c_2 = 1 \), and \( b = \{0, 1, 2\} \) are shown in Figs. \ref{fig:Darcy-Poiseuille-P-K3-Flux}a-c, where the fluid flux is plotted at \( t = 100 \) s, \( t = 150 \) s, and \( t = 200 \) s, for each \( b \) value. As observed, fluid flux is zero for \( \phi < c_1 \), given the zero permeability of the reservoir domain. Comparison of flux profiles in the transient zone (\( c_1 < \phi < c_2 \)) in Fig. \ref{fig:Darcy-Poiseuille-P-K3-Flux}d reveals minimal effect from \( b \) due to domain indicator variables \( \chi_r \) and \( \chi_f \) varying linearly in the transition zone.

\begin{figure}[H]
    \centering
    \begin{subfigure}[b]{0.45\textwidth}
         \centering
         \includegraphics[width=\textwidth]{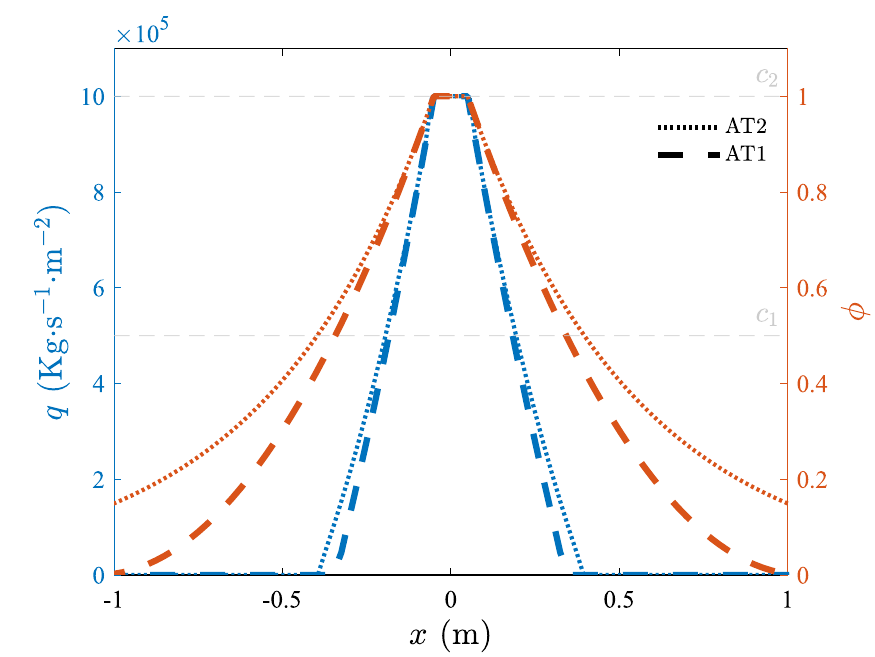}
         \caption{Domain decomposition method}
     \end{subfigure}
     \begin{subfigure}[b]{0.45\textwidth}
         \centering
         \includegraphics[width=\textwidth]{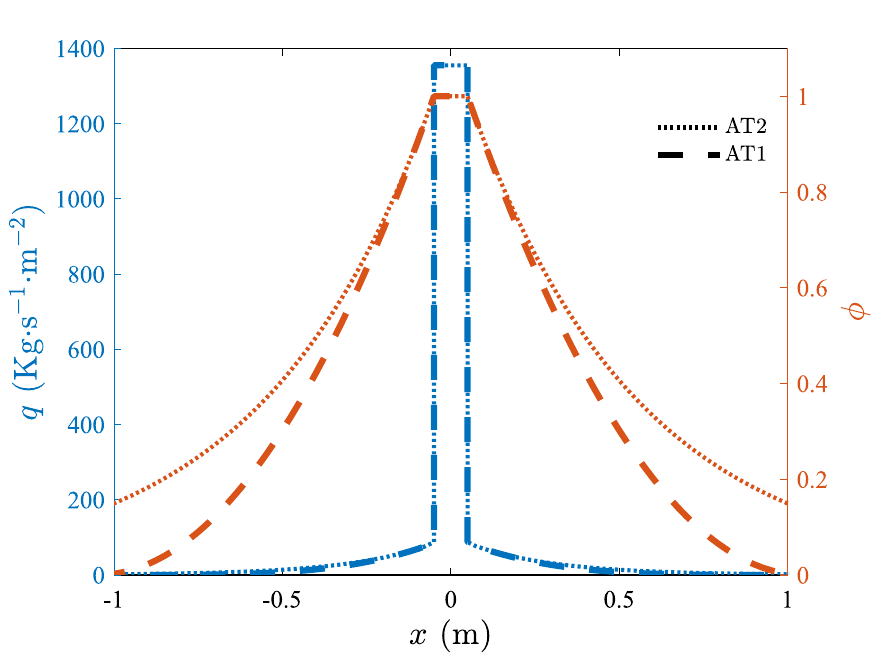}
         \caption{Modified Darcy method}
     \end{subfigure}
     \begin{subfigure}[b]{0.45\textwidth}
         \centering
         \includegraphics[width=\textwidth]{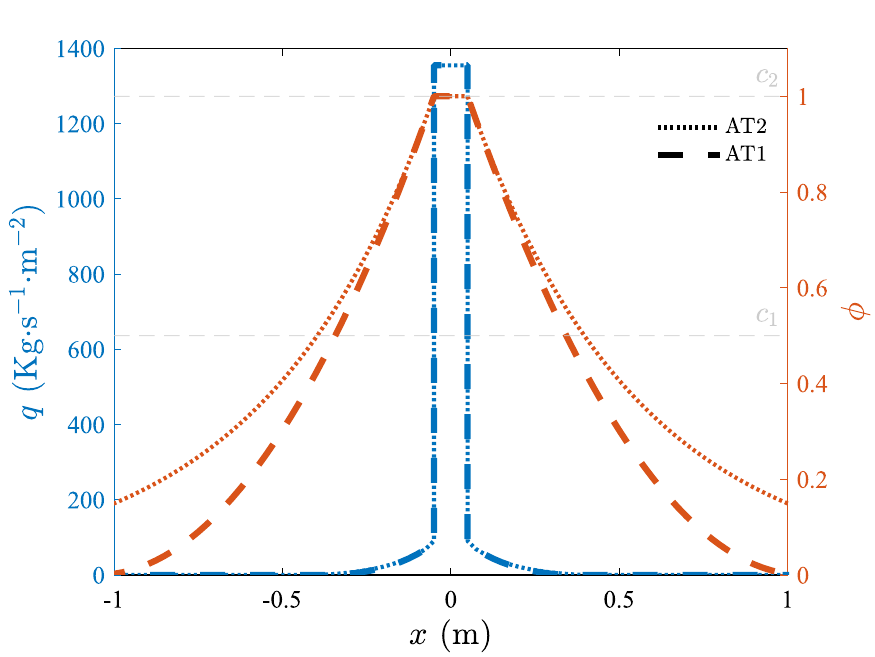}
         \caption{Hybrid method}
     \end{subfigure}
    \caption{Phase field profile and fluid flux distribution along the $x$-direction at the top of the domain for AT2 and AT1 models, (a) domain decomposition method with parameter $c_1=0.5$ and $c_2=1$, (b) modified Darcy method with transient parameter $b=2$, and (c) hybrid methods for selecting parameters $c_1=0.5$, $c_2=1$, and $b=2$.}
    \label{fig:Darcy-Poiseuille-AT1}
\end{figure}

We proceed to assess the influence of the specific phase field model adopted (AT2 vs AT1), as the AT2 model produces a broader damage zone, relative to the AT1 model. Consequently, less sensitivity to the coupling method is expected for the latter case. Their comparison is shown in Fig. \ref{fig:Darcy-Poiseuille-AT1}, where both the phase field profile and the fluid flux distribution are shown.  For the domain decomposition method, the difference is not significant if \( c_1 \) is selected to be sufficiently large (e.g., \( c_1 \geq 0.5 \)), as depicted in Fig. \ref{fig:Darcy-Poiseuille-AT1}a. This is because the difference in the phase field profiles of the AT1 and AT2 models becomes negligible for values of \( \phi \geq 0.5 \). In the case of the modified Darcy method, if the transient parameter \( b \) is sufficiently large (\( b \geq 2 \)), the difference becomes negligible, as shown in Fig. \ref{fig:Darcy-Poiseuille-AT1}b. Finally, as shown in Fig. \ref{fig:Darcy-Poiseuille-AT1}c, the hybrid method is largely insensitive to the choice of phase field model for the aforementioned choices of parameters. It must be noted that, since the history field method does not yield an optimal phase field fracture profile for the AT1 model, the penalty method is here employed to enforce the irreversibility condition. For further details, see Ref. \cite{gerasimov2019penalization}.

The comparison of Figs. \ref{fig:Darcy-Poiseuille-AT1}a and \ref{fig:Darcy-Poiseuille-AT1}b shows that the fluid flux obtained from the modified Darcy method approaches that of the hybrid method when the transient parameter $b$ is sufficiently large. However, this conclusion is intrinsic to the benchmark considered here, where a constant phase field is assumed. In cases involving phase field evolution (i.e., crack propagation) the results of these two methods can differ significantly. This is because in the modified Darcy method only the permeability is a function of the phase field variable, while Biot's coefficient and porosity remain constant. In contrast, the hybrid method accounts for variations in permeability, Biot coefficient, and porosity due to phase field evolution.\\

In summary, this study highlights the importance of appropriate parameter selection for accurate results. The domain decomposition method benefits from isotropic permeability but lacks sensitivity to crack opening changes. The modified Darcy method, which models anisotropic permeability and Poiseuille-type flow, suggests \( b \geq 2 \) for an effective transition. The hybrid method combines the strengths of both, providing distinct permeability domains while accounting for crack opening effects.

\subsection{Stick–slip modelling using a Drucker-Prager-based split}
\label{Sec:StickSlip}

\begin{figure}[H]
    \centering
    \begin{subfigure}[b]{0.45\textwidth}
         \centering
         \includegraphics[width=\textwidth]{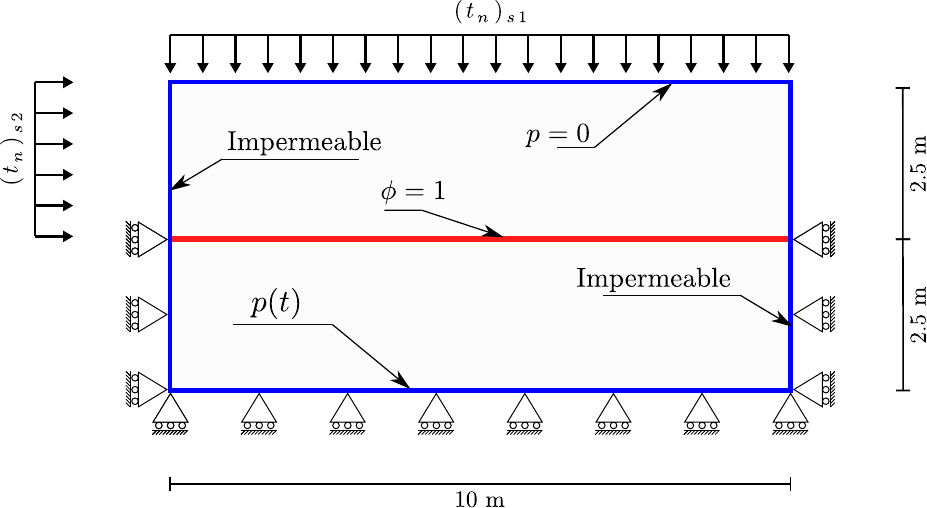}
         \caption{Biaxial initial state}
     \end{subfigure}  \hspace{2mm}
    \begin{subfigure}[b]{0.45\textwidth} 
         \centering
         \includegraphics[width=\textwidth]{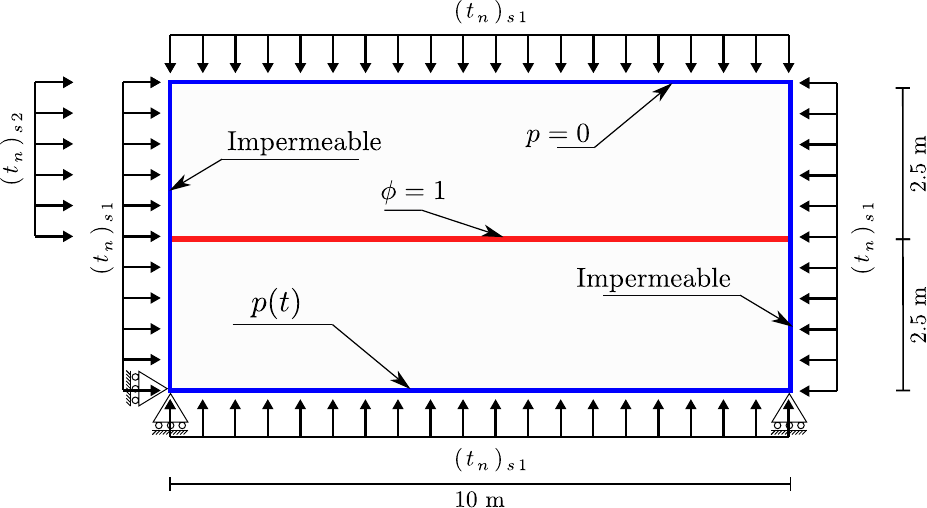}
         \caption{Triaxial initial state }
     \end{subfigure} 
    \caption{Geometry and boundary conditions of stick–slip problems: (a) Biaxial initial state, and (b) Triaxial initial state.}
    \label{fig:Drucker-Prager-Config}
\end{figure}

In this case study, we demonstrate the capability of the proposed Drucker-Prager-based split of the strain energy density to model stick-slip within hydraulic fracturing. As shown in Fig. \ref{fig:Drucker-Prager-Config}, a rectangular domain with a central horizontal crack is considered. To comprehensively investigate different regions of the strain space within the Drucker-Prager-based split (discussed in Section \ref{seq:Drivnign-force}), we consider two loading configurations that result in a path in the strain space that begins either in the elastic region or in the frictional region. Each boundary condition configuration is named according to the stress state at the end of the first step. The boundary conditions are applied in three steps, each lasting \(10^6\) seconds.\\

The first configuration is referred to as \emph{biaxial initial state} (Fig. \ref{fig:Drucker-Prager-Config}a) and involves fixing the lower half of the domain. Both domain sides are impermeable. A traction \((t_n)_{s1} = 8.86\) MPa is applied linearly over time to the top boundary during the first step. As the plane strain is adopted, the model experiences biaxial stress in this step. In the second step, a traction \((t_n)_{s2} = 5\) MPa is applied to the upper half of the left side, ramped over time. In the third step, the fluid pressure \(p\) is increased linearly over time to reach \(p = 15.55\) MPa at the domain's bottom.\\

The second configuration, referred to as the \emph{triaxial initial state}, is designed to ensure that the strain invariant path includes the elastic region (see Fig. \ref{fig:Drucker-Prager-Space}). In this setup, the left bottom corner has its displacements constrained in both directions, while the right bottom corner restricts only vertical displacement. As in the first configuration, both domain sides are impermeable. A uniform traction \((t_n)_{s1} = 8.86\) MPa is applied to the entire boundary during the first step, which leads to a triaxial stress state. In the second step, a traction \((t_n)_{s2} = 5\) MPa is applied linearly to the upper half of the left side. Finally, in the third step, fluid pressure is applied to the bottom side until it reaches \(p = 15.55\) MPa.\\

In both configurations, a horizontal crack is introduced by setting \( \phi = 1 \). A uniform mesh of bilinear quadrilateral elements is used, with a characteristic finite element length of 10 cm. The material properties adopted correspond to those employed in the previous case study (see Table \ref{tab:ParametersCase1}), unless specified otherwise. Poisson’s ratio equals \(\nu=0.2\), the characteristic length scale is \(\ell=0.2\) m, the permeability of the reservoir domain is assumed to be \(K_{r}=10^{-15}\) m\(^2\), and the permeability of the fracture domain equals \(K_{f}=1.333 \times \, 10^{-6}\) m\(^2\). A high value of the material toughness \( G_c \) is set to prevent crack propagation. The domain decomposition method with constants \(c_1=0.5\) and \(c_2=1\) is used for the permeability coupling, and a monolithic solution scheme is employed. 

\begin{figure}[H]
    \centering
        \begin{subfigure}[b]{0.45\textwidth} 
         \centering
         \includegraphics[width=\textwidth]{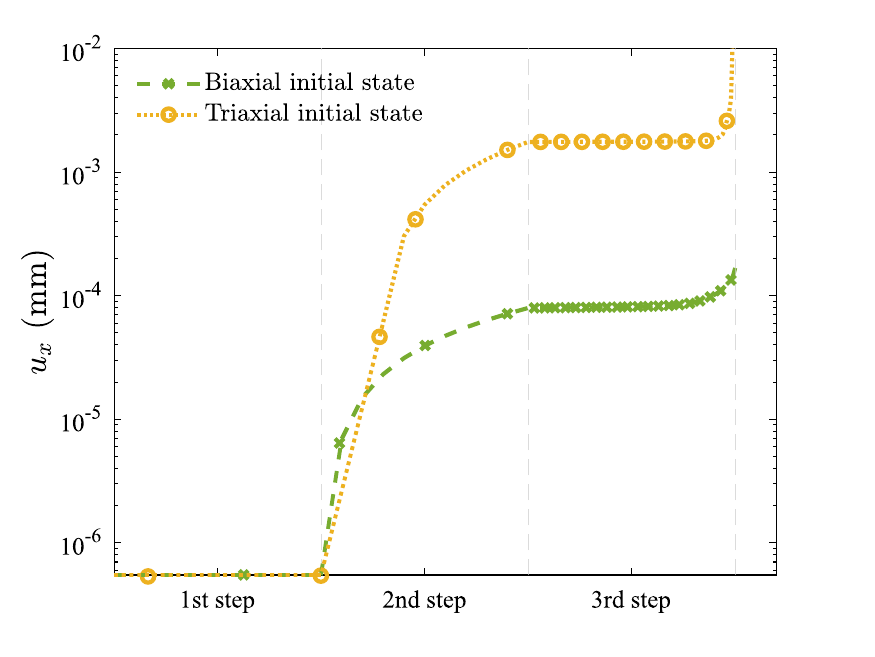}
         \caption{}
     \end{subfigure}  \hspace{2mm}
          \begin{subfigure}[b]{0.45\textwidth}
         \centering
         \includegraphics[width=\textwidth]{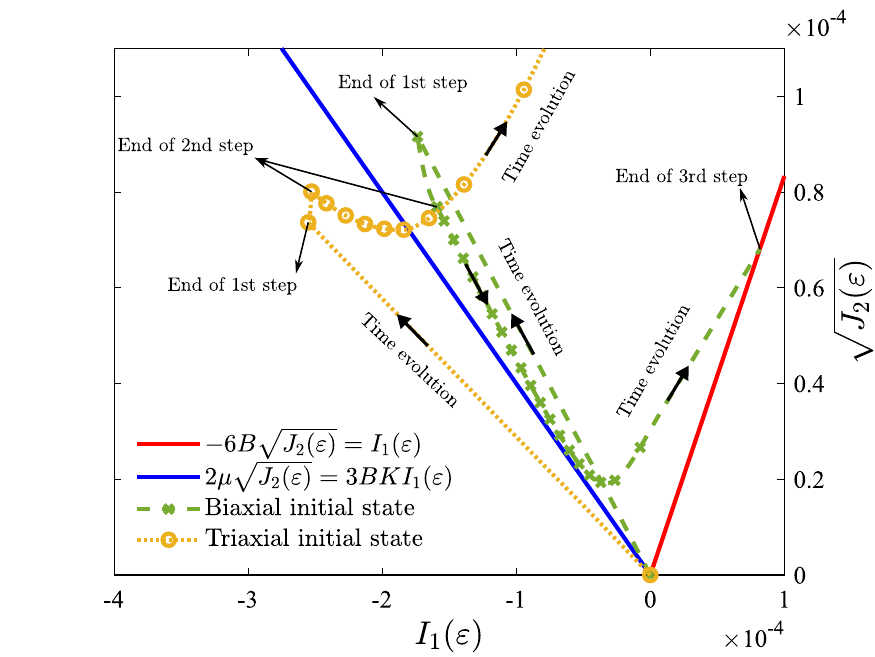}
         \caption{}
     \end{subfigure} 
    \begin{subfigure}[b]{0.45\textwidth}
         \centering
         \includegraphics[width=\textwidth]{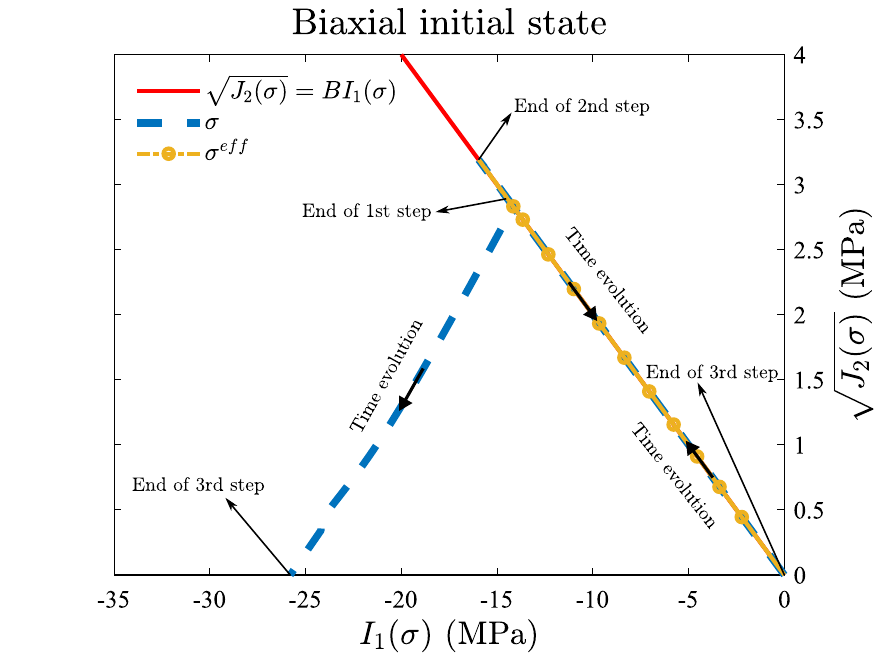}
         \caption{}
     \end{subfigure}  \hspace{2mm}
    \begin{subfigure}[b]{0.45\textwidth} 
         \centering
         \includegraphics[width=\textwidth]{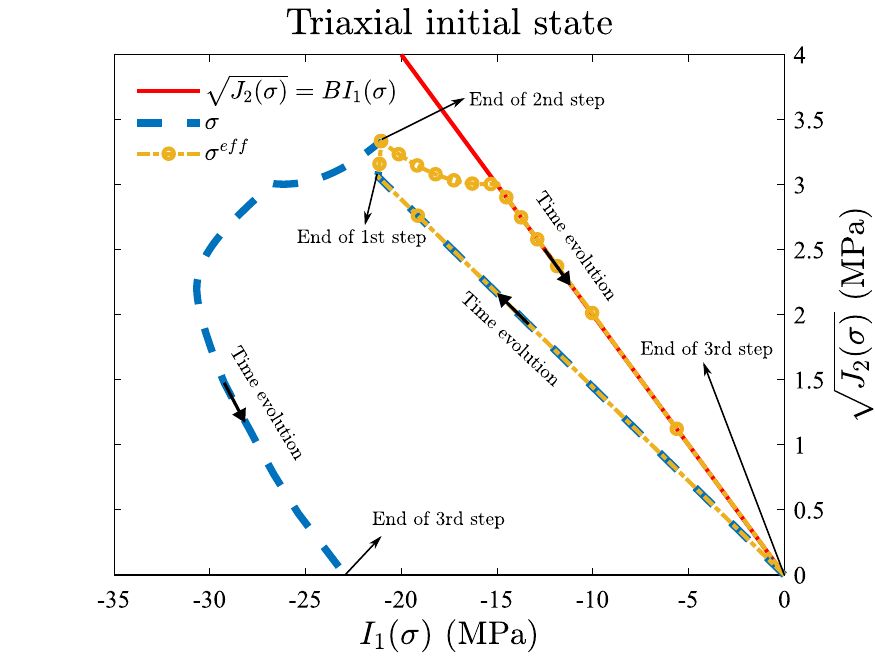}
         \caption{}
     \end{subfigure}  
    \caption{Stick–slip problem: (a) Horizontal displacement \( u_x \) of top side at different steps, and (b) comparison of the strain path for biaxial and triaxial initial states. The bottom half of the figure shows the total (\( \mathbf{\sigma} \)) and effective (\( \mathbf{\sigma}^{\text{eff}} \)) stress paths of integration point with \( \phi=1 \) for (c) the biaxial initial state, and (d) the triaxial initial state.}
    \label{fig:Drucker-Prager-Result}
\end{figure}

To analyse the results, we first examine the displacement at the top of the domain for each problem. Fig. \ref{fig:Drucker-Prager-Result}a compares the horizontal displacement \( u_x \) for both configurations. Initially, \( u_x \) remains nearly zero during the first step but increases when traction \( (t_n)_{s2} \) is applied. After reaching equilibrium at the end of the second step, fluid pressure increases during the third step. Initially, there is no displacement change, but as the fluid pressure rises, the displacement increases until elements along the horizontal crack lose stiffness, causing the displacement solution to diverge. The horizontal displacement \( u_x \) at the top provides insight into the stress and strain paths. We examine the strain space \( (I_1 (\boldsymbol{\varepsilon}), \sqrt{J_2 (\boldsymbol{\varepsilon})}) \) (Fig. \ref{fig:Drucker-Prager-Result}b) and the stress space \( (I_1 (\mathbf{\sigma}), \sqrt{J_2 (\mathbf{\sigma})}) \) (Figs. \ref{fig:Drucker-Prager-Result}c,d) for an element along the crack line in each problem, analysing both the total stress \( \mathbf{\sigma} \) and the effective stress \( \mathbf{\sigma}^{\text{eff}} \).\\

Let us start with the biaxial initial state, where the strain is in the frictional region (see Fig. \ref{fig:Drucker-Prager-Space}a). Since \( \phi = 1 \) at this integration point, the material follows the Drucker-Prager failure criterion, as seen in Fig. \ref{fig:Drucker-Prager-Result}c. Applying traction \( (t_n)_{s1} \) initiates the effective stress from \( (I_1 (\mathbf{\sigma}^{\text{eff}}) = 0, \sqrt{J_2 (\mathbf{\sigma}^{\text{eff}})} = 0) \) along the criterion line \( \sqrt{J_2 (\mathbf{\sigma})} = B I_1 (\mathbf{\sigma}) \). In the second step, traction \( (t_n)_{s2} \) increases \( J_2 (\mathbf{\sigma}^{\text{eff}}) \) while \( I_1 (\mathbf{\sigma}^{\text{eff}}) \) also increases due to frictional behaviour. \( I_1 (\mathbf{\sigma}) \) decreases with fluid pressure \( p \) until reaching zero stress, as indicated by the strain line intersecting the line \( -6 B  \sqrt{J_2 (\boldsymbol{\varepsilon})} = I_1 (\boldsymbol{\varepsilon}) \). Consequently, the total stress loses its deviatoric part, transitioning to a hydrostatic state.\\

The triaxial initial state, shown in Figs. \ref{fig:Drucker-Prager-Result}b and d, starts in the elastic region, meaning that the material remains elastic without phase field influence. In the first step, the application of a traction \( (t_n)_{s1} \) increases strains and stresses elastically. In the second step, the application of a traction \( (t_n)_{s2} \) changes the stress and strain paths, which still lie within the elastic region. With fluid pressure being applied in the third step, \( I_1 (\mathbf{\sigma}^{\text{eff}}) \) decreases, intersecting the failure criterion \( \sqrt{J_2 (\mathbf{\sigma})} = B I_1 (\mathbf{\sigma}) \). Here, the material behaviour follows the Drucker-Prager criterion, and with \( \phi = 1 \), the effective stress lies on the failure line \( \sqrt{J_2 (\mathbf{\sigma})} = B I_1 (\mathbf{\sigma}) \). A continued rise in fluid pressure causes the effective stress to return toward the origin, with the strain path intersecting the line \( -6 B  \sqrt{J_2 (\boldsymbol{\varepsilon})} = I_1 (\boldsymbol{\varepsilon}) \), indicating zero effective stress and complete stiffness loss, leading to a hydrostatic stress state.

This case study demonstrates that the stick-slip behaviour of rock joints can be effectively modelled by incorporating pore pressure variations using a Drucker-Prager-based split model. Stick-slip behaviour refers to the alternating phases of sticking and sudden slipping along a rock joint, driven by the accumulation and abrupt release of effective stress. As illustrated in Figures \ref{fig:Drucker-Prager-Result}c,d, the stress path captures these transitions. This modelling approach accounts for variable field conditions, including in-situ stress, and elucidates how pore pressure fluctuations influence the frictional behaviour of rock joints, determining whether they remain stationary (stick) or undergo slip, as relevant to many applications such as reducing seismic hazard during mining process \cite{GONZALEZ2022104975}.

\subsection{Influence of the fracture driving force on fluid-driven crack interactions}
\label{Sec:FractureInter}

\begin{figure}[H]
    \centering
    \begin{subfigure}[b]{0.45\textwidth}
         \centering
         \includegraphics[width=\textwidth]{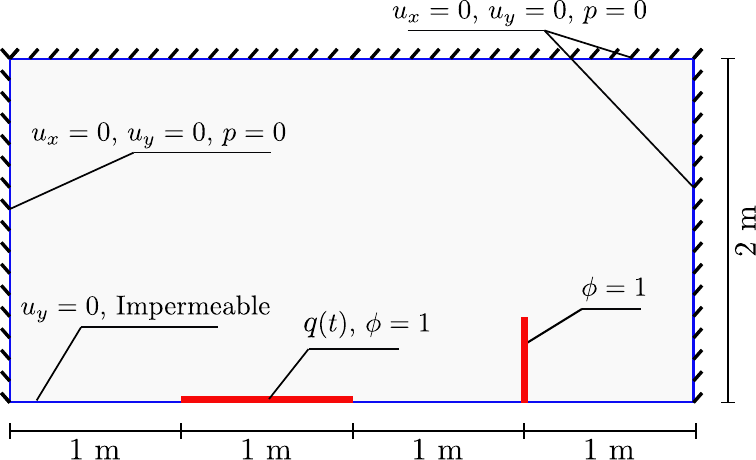}
         \caption{}
     \end{subfigure} 
    \begin{subfigure}[b]{0.41\textwidth} 
         \centering
         \includegraphics[width=\textwidth]{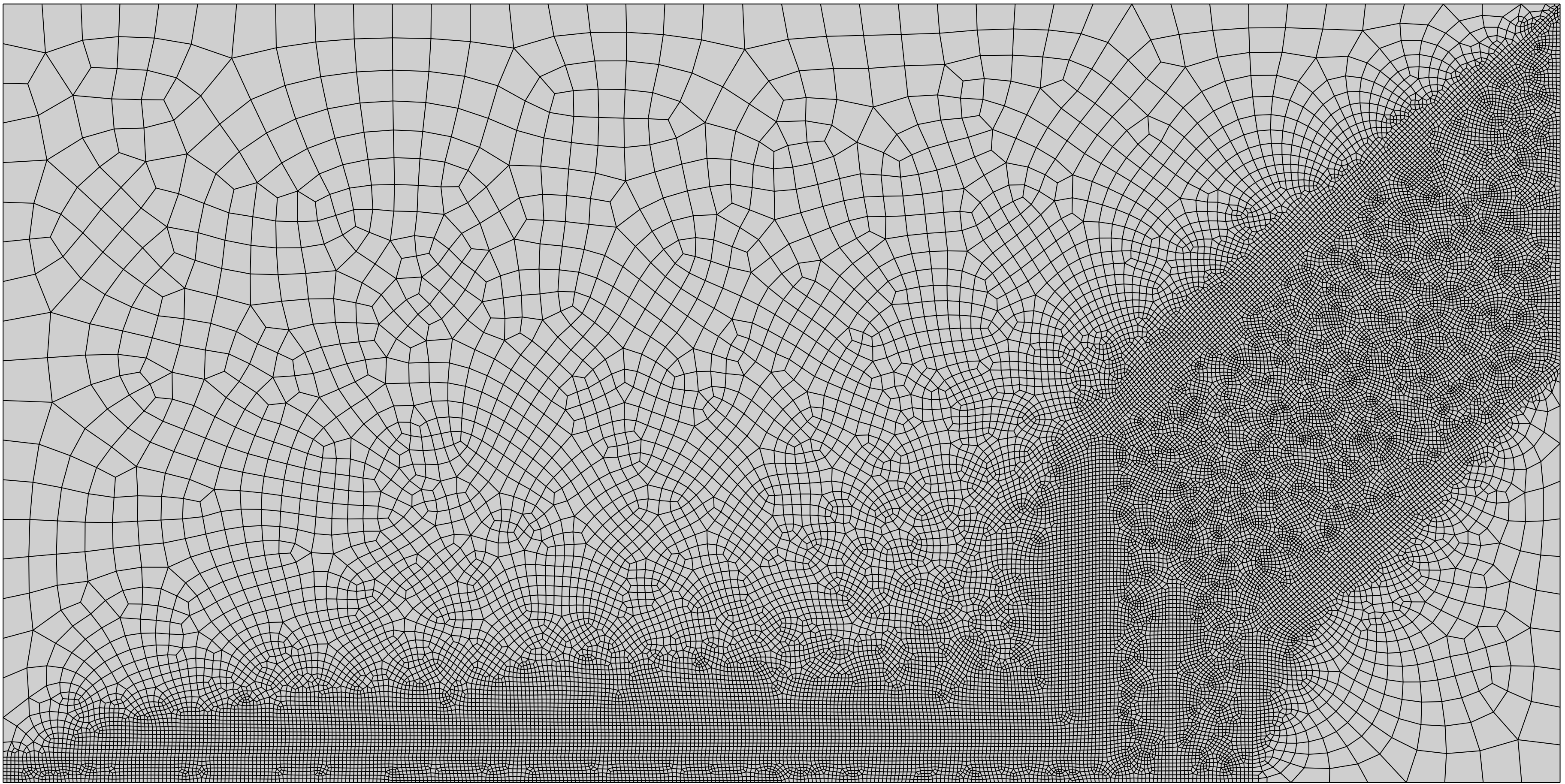}
         \vspace{0.8mm}
         \caption{}
     \end{subfigure} 
    \caption{Crack-interaction problem: (a) Geometry and boundary conditions, and (b) finite element mesh discretization.}
    \label{fig:Crack-Interaction-Geo}
\end{figure}

In this case study, we investigate the influence of the adopted strain energy decomposition and of the fracture-fluid coupling method (discussed in Section \ref{sec:Phase field hydraulic fracture}) on the crack propagation behaviour of hydraulic fractures. A square domain with two pre-existing cracks is considered, arranged is such a way so as to examine both tensile and shear contributions of the strain energy density in crack propagation (Fig. \ref{fig:Crack-Interaction-Geo}a). Due to symmetry, only half of the boundary value problem is simulated. The displacement and pressure at the domain perimeter are fixed, and each crack has a width of \(1\) cm ($\phi=1$ prescribed over a row of elements). A fluid flux of \( q_m = 80 \, \text{kg} \cdot \text{s}^{-1} \cdot \text{m}^{-2} \) is applied to the horizontal crack at $t=0$ and held constant throughout the analysis. Material properties follow those used in the second case study, but considering a Young's modulus of \(E=210\) GPa, a Poisson's ratio equal to \(\nu=0.3\), a characteristic length scale of \( \ell = 0.02 \) m, and a critical fracture energy release rate of \( G_c=2700 \) J/m\(^2\). The domain decomposition method is used for permeability coupling with constants \(c_1=0.4\) and \(c_2=1\). The model is discretized using bilinear quadrilateral elements. A total of 31,277 elements are used, with the mesh being refined along the anticipated crack propagation region, giving a minimum element size of \(0.01\) m (Fig. \ref{fig:Crack-Interaction-Geo}b). The mixed staggered method is used, with a time increment of \(0.05\) s over a total duration of \(500\) s.\\

As described in Section \ref{seq:Drivnign-force}, the present phase field framework for hydraulic fracture includes five different treatments of the fracture driving force: no split, volumetric-deviatoric, spectral, no tension and Drucker-Prager. All five are considered in this case study. For the Drucker-Prager-based split, the parameter \( B=-0.2 \) is used.

\begin{figure}[H]
    \centering
         \begin{subfigure}[b]{0.45\textwidth}
         \centering
         \includegraphics[width=\textwidth]{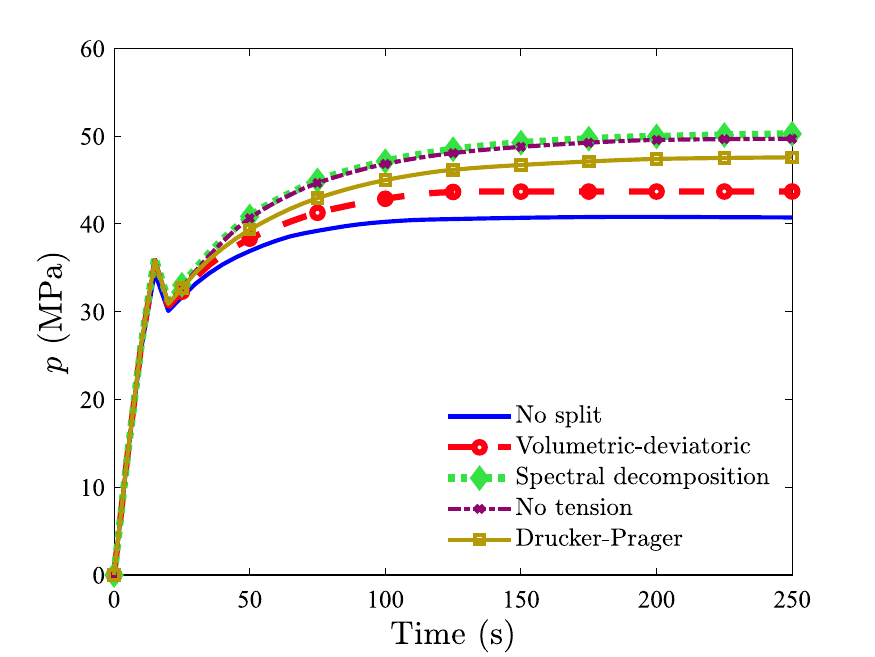}
         \vspace{0.0001 mm}
         \caption{}
     \end{subfigure}
     \begin{subfigure}[b]{0.45\textwidth}
         \centering
         \includegraphics[width=\textwidth]{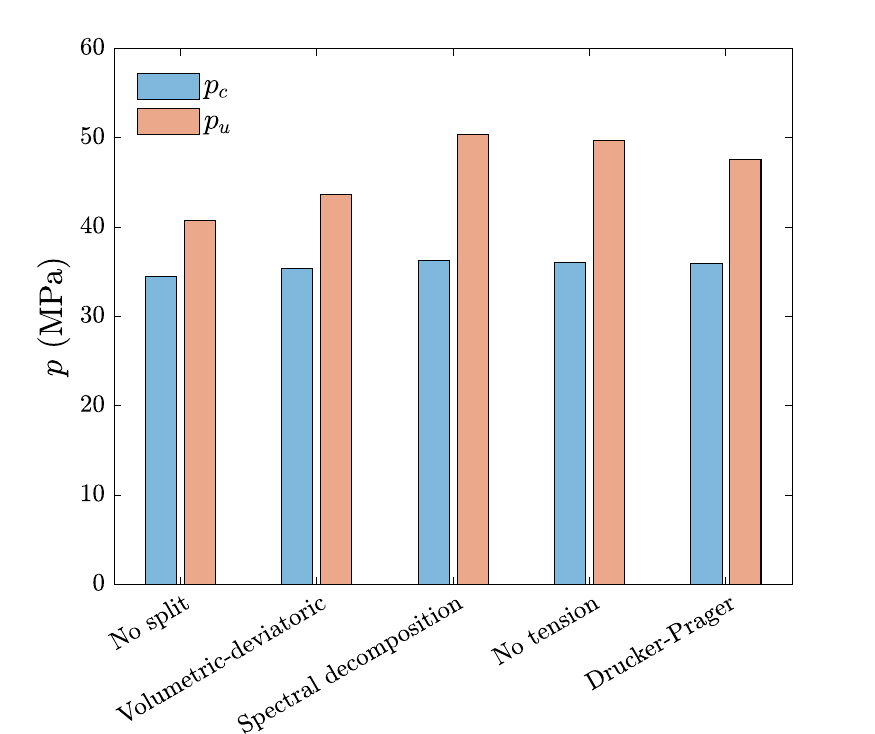}
         \caption{}
     \end{subfigure} 
    \begin{subfigure}[b]{0.5\textwidth}
         \centering
         \vspace{5mm}
         \includegraphics[width=\textwidth]{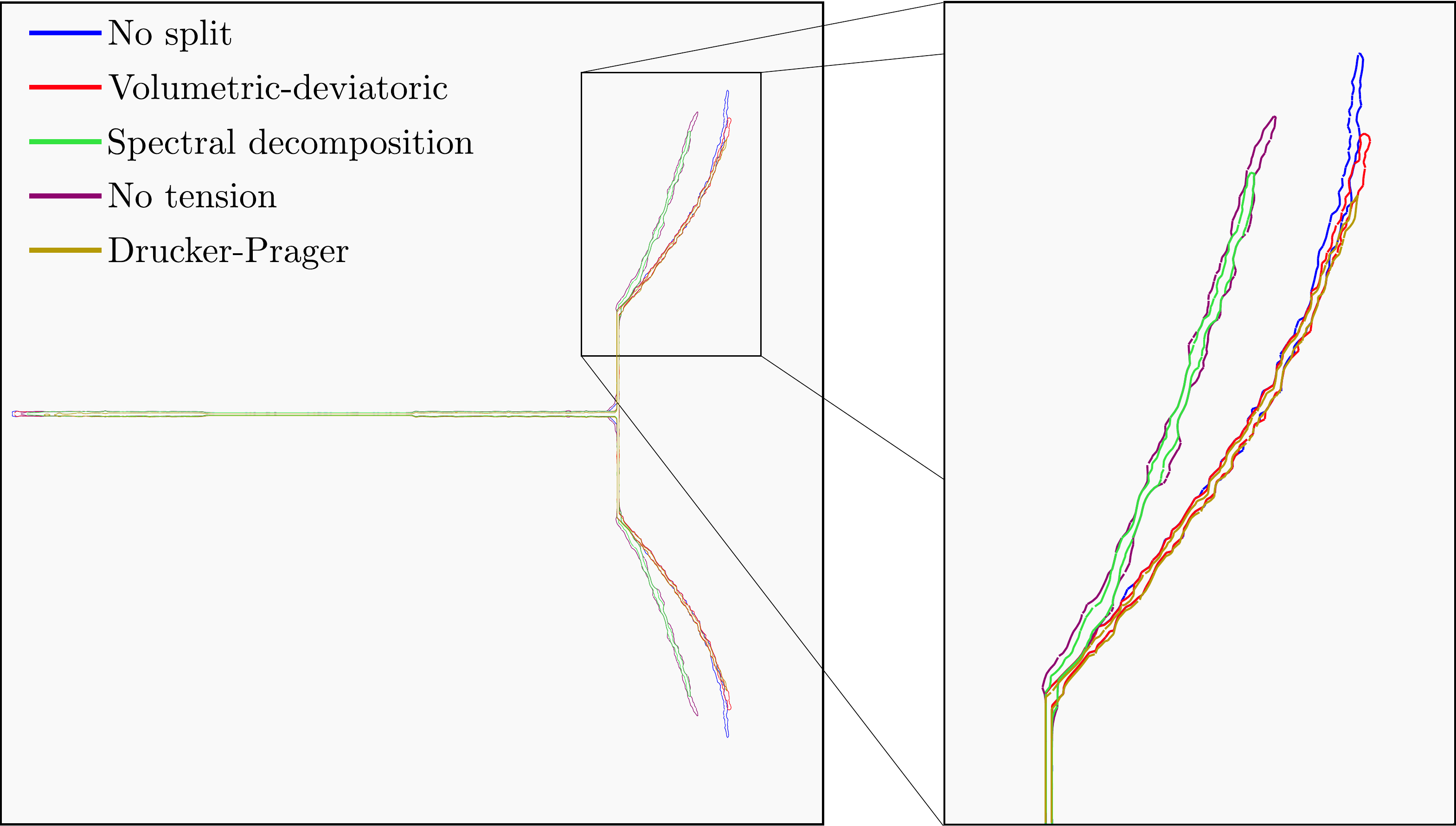}
         \caption{}
     \end{subfigure} \hspace{5mm}
     \begin{subfigure}[b]{0.28\textwidth}
         \centering
         \includegraphics[width=\textwidth]{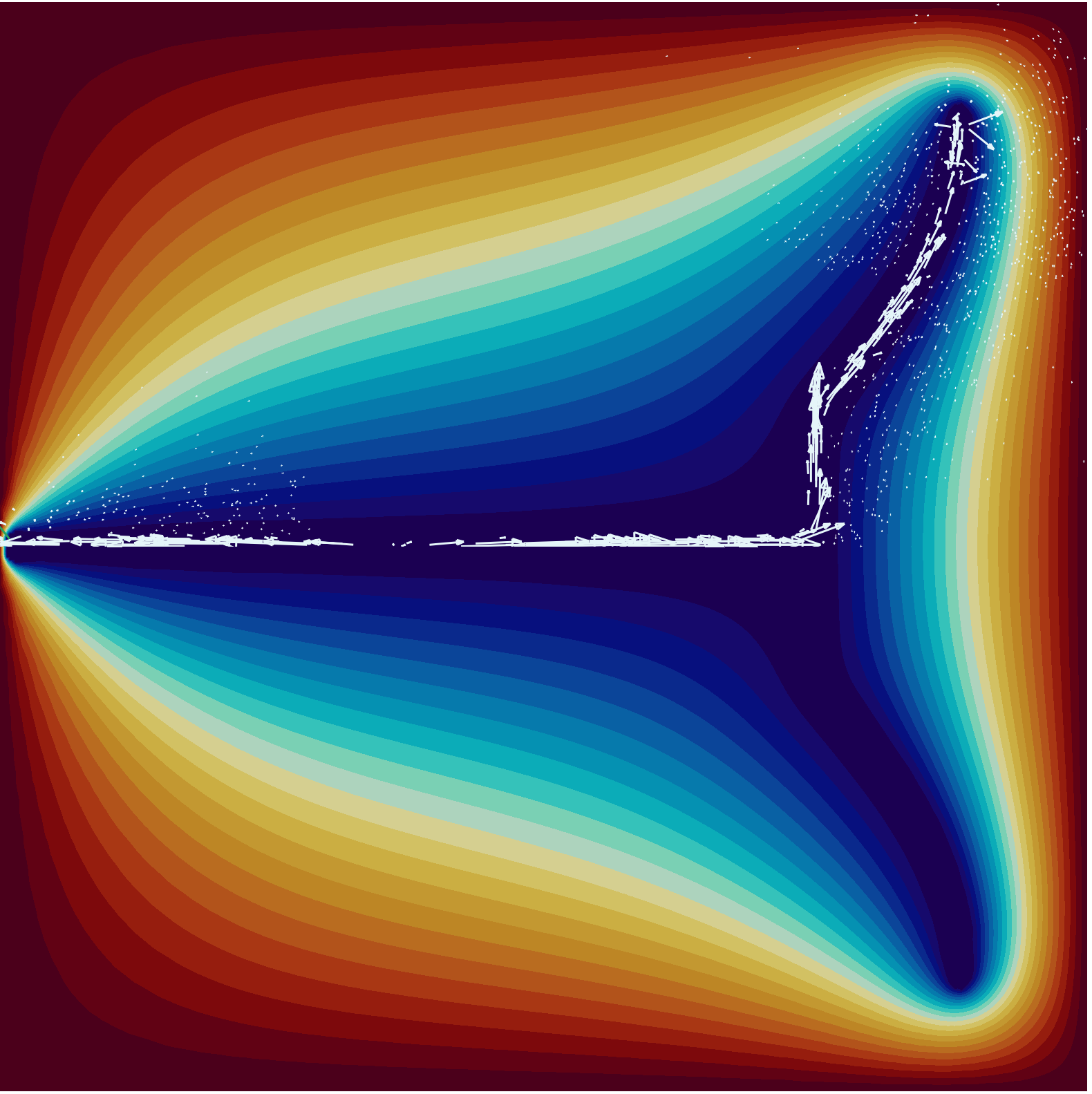}
         \caption{}
     \end{subfigure} \hspace{1mm}
     \begin{subfigure}[b]{0.11\textwidth}
         \centering
         \includegraphics[width=\textwidth]{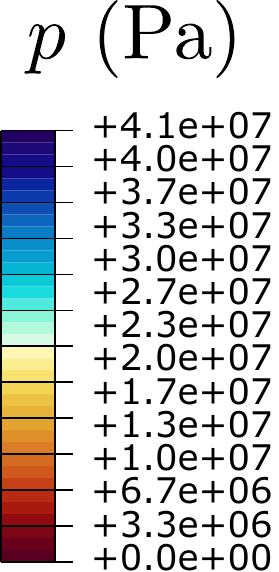}
         \vspace{0mm}
     \end{subfigure}
    \caption{Crack-interaction problem: (a) Evolution of fluid pressure \( p \) versus time at a point on the horizontal crack for different strain energy decompositions, (b) critical \( p_c \) and ultimate \( p_u \) fluid pressures, (c) crack paths for different fracture driving forces, and (d) fluid pressure \( p \) contour and fluid flux vector at the steady state for the case without strain energy decomposition.}
    \label{fig:Crack-Interaction-result}
\end{figure}

The results obtained are shown in Fig. \ref{fig:Crack-Interaction-result}. In all cases, applying volumetric fluid flux to the horizontal crack causes fluid pressure to increase until it reaches a critical value \( p_c \), as shown in Fig. \ref{fig:Crack-Interaction-result}a. When the fluid pressure reaches \( p_c \), crack propagation initiates. The critical pressure varies only slightly depending on the strain energy split, as the fracture of the initial crack is primarily driven by tensile stresses. The highest critical pressure \( p_c \) is attained with the spectral decomposition. The fluid pressure \( p \) decreases with crack propagation until the two cracks merge, causing the pressure in the second crack to rise. The fluid pressure then continues to increase, with subsequent crack propagation being gradual and influenced by shear stresses. Eventually, a steady state in crack growth is reached, indicated by the lack of further pressure increases after \(\text{Time} = 200\) s in Fig. \ref{fig:Crack-Interaction-result}a.\\

The critical pressure \( p_c \) and ultimate pressure \( p_u \) (steady-state pressure) are reported in Fig. \ref{fig:Crack-Interaction-result}b. The highest ultimate pressure is attained with the spectral decomposition, where the material is weaker in shear compared to tension. Conversely, the lowest ultimate pressure corresponds to the no split case, where all the strain energy drives fracture. The impact of the shear contributions of the strain energy density to the crack trajectory is shown in Fig. \ref{fig:Crack-Interaction-result}c: decompositions incorporating shear strain energy contributions (no split, volumetric-deviatoric, and Drucker-Prager) exhibit a higher degree of deflection from the vertical crack tip, while tensile-based decompositions (spectral and no tension models) show straighter paths. Fig. \ref{fig:Crack-Interaction-result}d shows the fluid pressure contour and flux vectors, where the pressure in the fractured area is uniform due to low permeability. These results are qualitatively the same for all the fracture driving forces.
The fluid flux begins at the horizontal crack and follows the crack path, facilitating propagation towards the vertical crack.

The influence of the property coupling method was extensively assessed in Section \ref{sec:Coupling_Method} for the case of a uniform phase field (stationary crack), with a focus on permeability. The analysis is extended here to consider their effect on crack growth. To this end, the crack-interaction boundary value problem illustrated Fig. \ref{fig:Crack-Interaction-Geo} is evaluated, with the no tension strain energy decomposition approach and the three different coupling models: modified Darcy (power-law), domain decomposition and the presently proposed hybrid one. In the modified Darcy method, only the permeability tensor is dependent on the phase field variable. In contrast, the domain decomposition and hybrid methods incorporate the phase field dependency into the permeability tensor, Biot's coefficient, and porosity. As a result, during fracture propagation, the critical pressure is expected to be higher in the modified Darcy method due to the insensitivity of Biot's coefficient to phase field evolution. Conversely, in the domain decomposition and hybrid methods, Biot's coefficient approaches unity in the crack region and at the crack tip, thereby enhancing the influence of pore pressure on the deformation process.\\

\begin{figure}[H]
    \centering
    \begin{subfigure}[b]{0.45\textwidth}
    \centering
    \includegraphics[width=\linewidth]{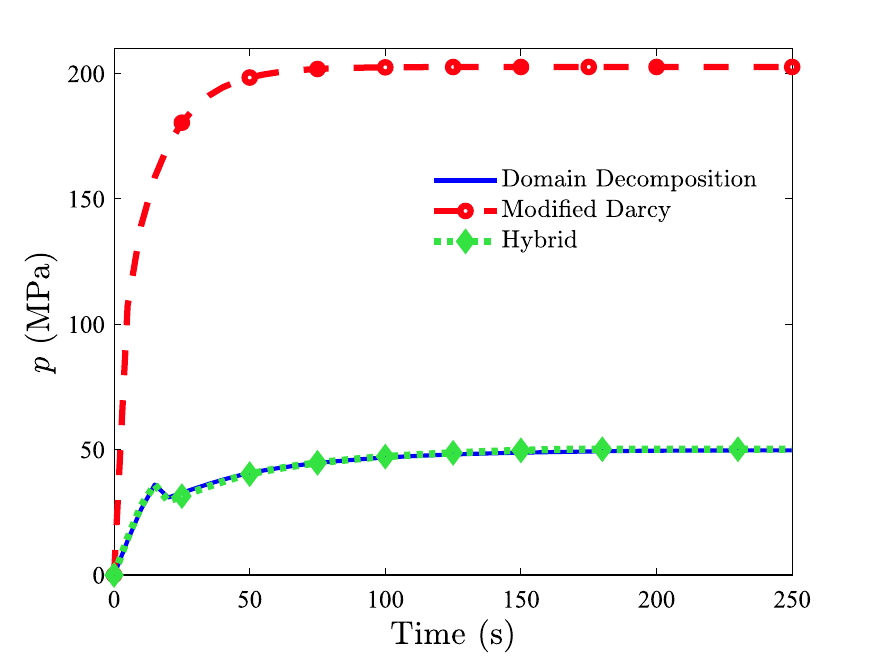}
    \caption{}
    \end{subfigure} \hspace{5mm}
    \begin{subfigure}[b]{0.28\textwidth}
    \centering
    \includegraphics[width=\linewidth]{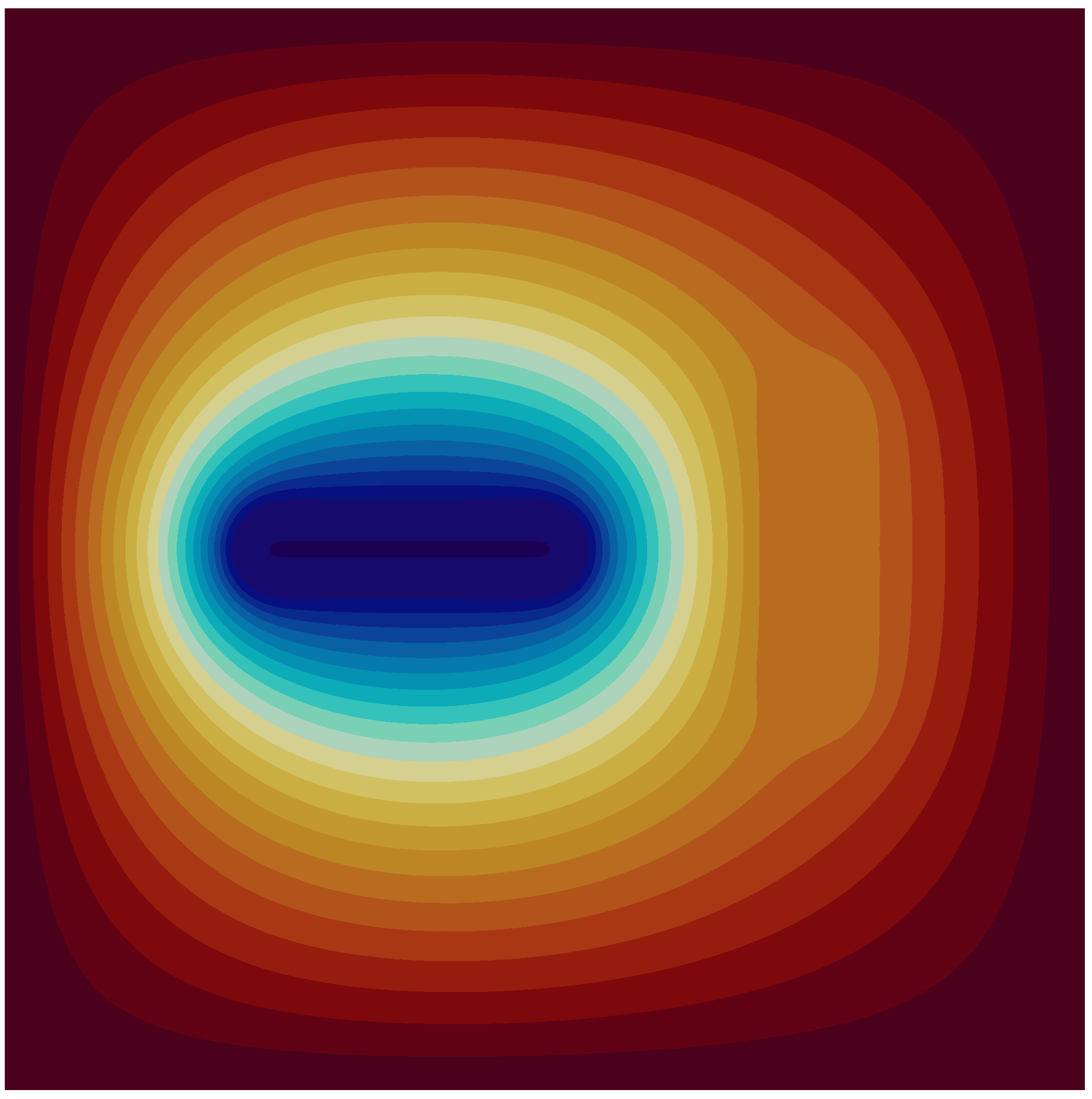}
    \vspace{-1.5mm}
    \caption{Modified Darcy}
    \end{subfigure} \hspace{1mm}
    \begin{subfigure}[b]{0.11\textwidth}
    \centering
    \includegraphics[width=\linewidth]{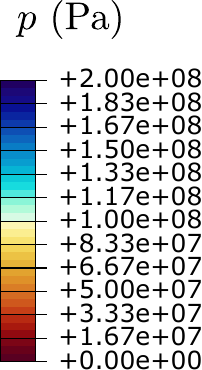}
    \vspace{6mm}
    \end{subfigure}  \\ \vspace{5mm} \hspace{15 mm}
    \begin{subfigure}[b]{0.28\textwidth}
    \centering
    \includegraphics[width=\linewidth]{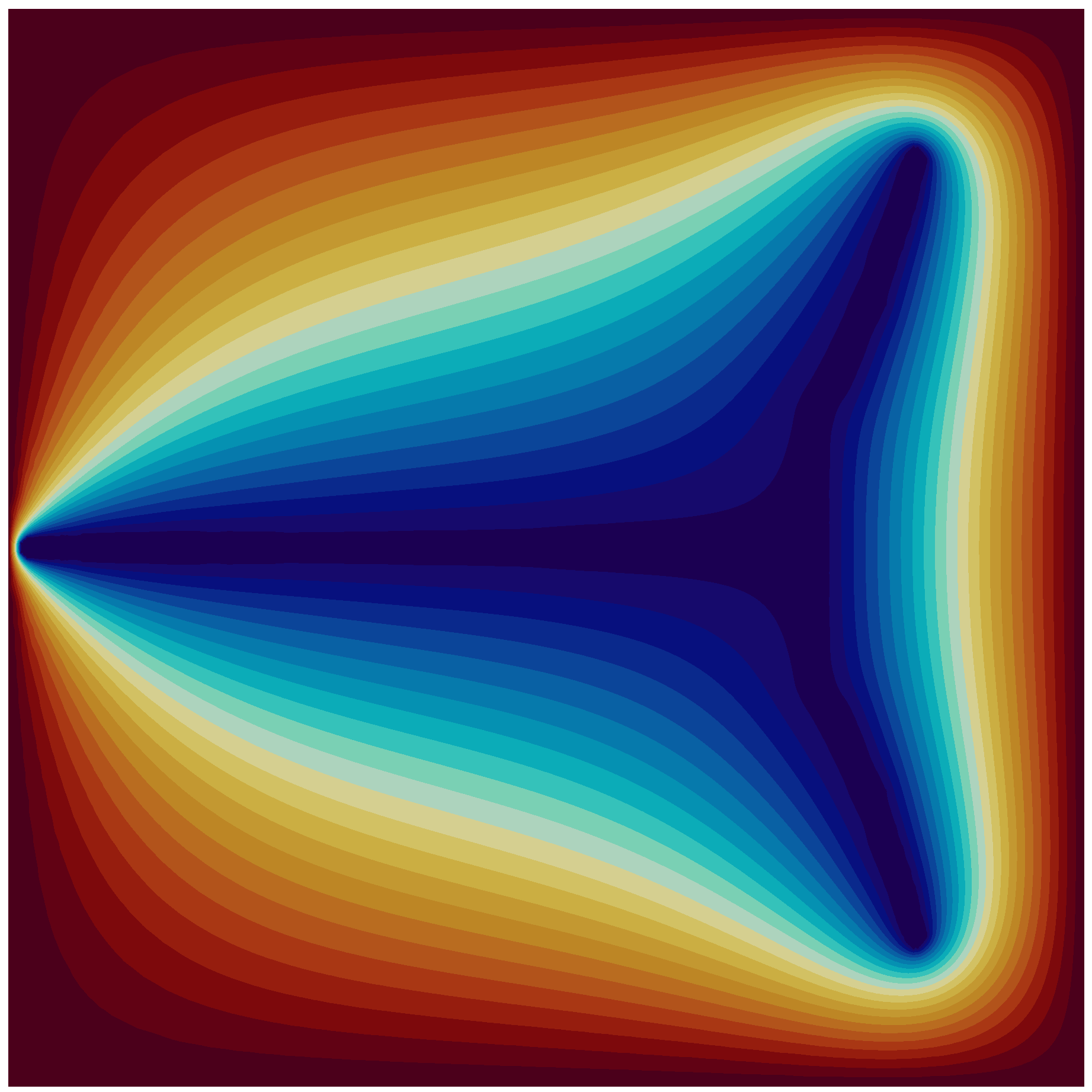}
    \caption{Domain decomposition}
    \end{subfigure} \hspace{18mm}
    \begin{subfigure}[b]{0.28\textwidth}
    \centering
    \includegraphics[width=\linewidth]{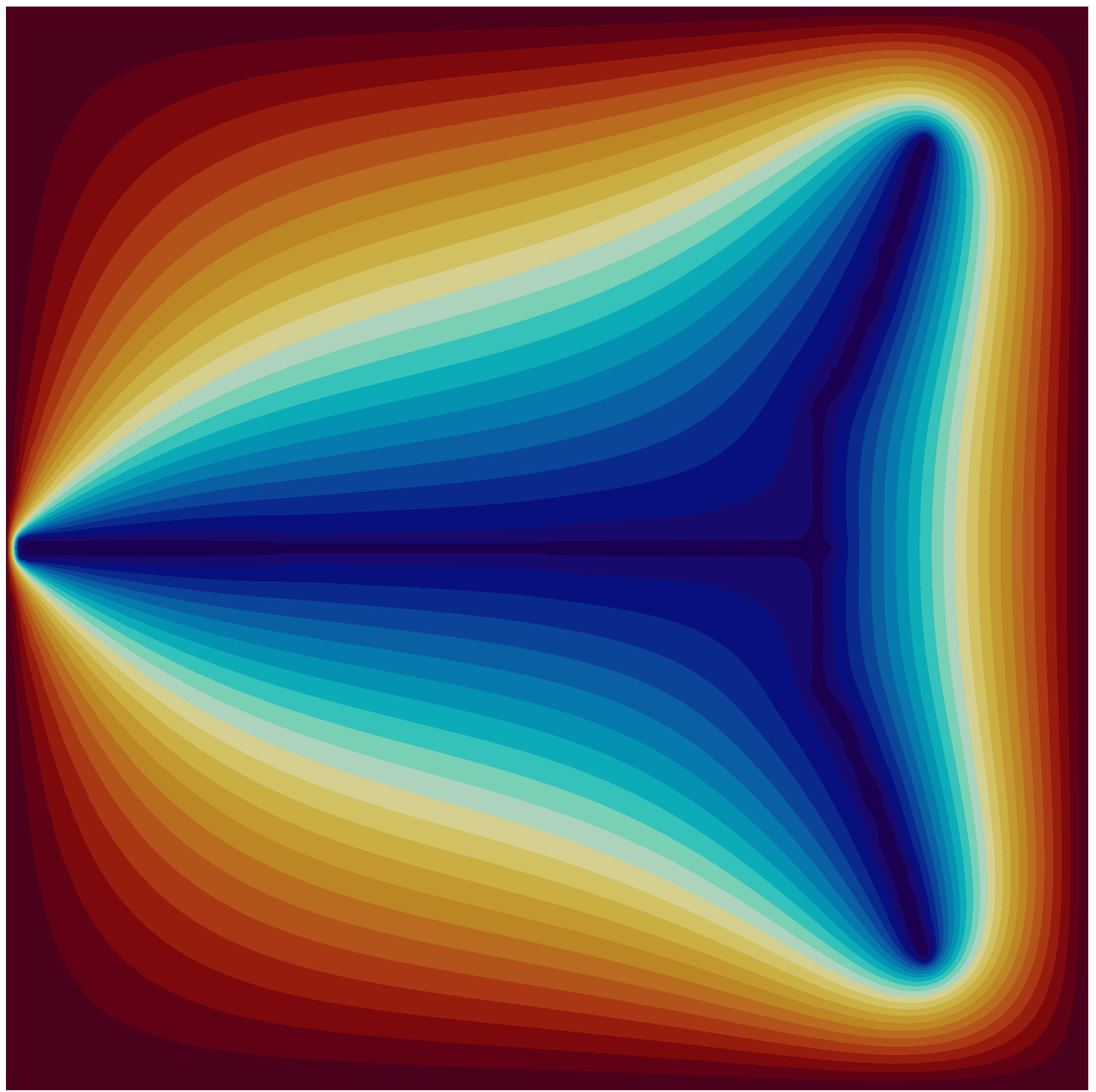}
    \caption{Hybrid}
    \end{subfigure} \hspace{1mm}
    \begin{subfigure}[b]{0.11\textwidth}
    \centering
    \includegraphics[width=\linewidth]{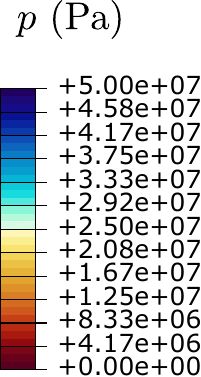}
    \vspace{-0.8mm}
    \end{subfigure}
    \caption{Crack interaction analysis under various coupling strategies employing a no-tension model for a fluid flux of \( q_m = 80 \, \text{kg} \cdot \text{s}^{-1} \cdot \text{m}^{-2} \). (a) Temporal evolution of fluid pressure \( p \), and fluid pressure \( p \) contour distributions at $t=150$ s for: (b) the modified Darcy method, (c) the domain decomposition method, and (d) the present hybrid method.}
    \label{fig:Crack-Interaction-Coupling}
\end{figure}

The results obtained are shown in Fig. \ref{fig:Crack-Interaction-Coupling} for a fluid flux \( q_m = 80 \, \text{kg} \cdot \text{s}^{-1} \cdot \text{m}^{-2} \) and a transient parameter of $b=2$. In the modified Darcy method, these conditions do not result in crack growth. The fluid pressure rises to 200 MPa, and the system reaches steady state without crack propagation, as shown in the pressure contours provided in Fig. \ref{fig:Crack-Interaction-Coupling}b. This result aligns with the assumption that Biot's coefficient remains constant and is not influenced by the phase field. To determine the critical pressure, a significantly higher fluid flux of \( q_m = 70{,}000 \, \text{kg} \cdot \text{s}^{-1} \cdot \text{m}^{-2} \) is applied, leading to crack propagation at a pressure of $p = 12{,}350$ MPa. Conversely, in the hybrid method the evolution of Biot's coefficient and porosity with $\phi$ promotes crack propagation at a lower pressure. The observed critical pore pressure is comparable to that of the domain decomposition method, approximately $p = 37.3$ MPa. As depicted in Fig. \ref{fig:Crack-Interaction-Coupling}a, the time evolution of pore pressure in the hybrid method closely follows that of the domain decomposition method. However, a comparison of the fluid pressure distributions for the domain decomposition (Fig.~\ref{fig:Crack-Interaction-Coupling}c) and hybrid (Fig.~\ref{fig:Crack-Interaction-Coupling}d) approaches reveals that, due to the consideration of anisotropic permeability in the hybrid method, the pressure is more concentrated near the crack in the hybrid model compared to the domain decomposition method, which employs an isotropic permeability tensor.

In this case study, we observe that the choice of fracture driving force significantly affects both the pressure field and the crack path. However, this effect can be negligible if the fracture is driven primarily by tensile stress. On the other hand, shear stress can play a significant role in determining the crack trajectory and pressure distribution. Fracture driving forces that mainly consider tensile stresses, such as the spectral decomposition and no tension models, result in higher pressures for crack propagation. In contrast, methods like the volumetric-deviatoric and Drucker-Prager-based splits incorporate shear stress effects, influencing both crack path and pressure distribution. The lowest pressure is observed in the original formulation (no split), where the entire strain energy density contributes to fracture propagation. Moreover, this crack-interaction boundary value problem is used to assess the influence of the coupling strategy under crack growth conditions, extending the analysis in Section \ref{sec:Coupling_Method}. While in Section \ref{sec:Coupling_Method}, focused on permeability and considering a uniform and stationary phase field, the hybrid approach was closer to the modified Darcy model, under crack growth conditions our hybrid approach aligns better with the domain decomposition method. This is primarily because in the domain decomposition and hybrid methods, the Biot coefficient evolves with the phase field. The consideration of anisotropic permeability in the hybrid method shows a more concentrated distribution of pressure near the crack relative to the domain decomposition method, showcasing its greater modelling flexibility, as it allows for the evolution of material behaviour via the phase field while also leveraging an anisotropic permeability tensor based on computed crack openings.

\subsection{Simultaneous Injection into an axisymmetric boundary with initial stress}
\label{Sec:Axisym}

\begin{figure}[H]
    \centering
    \begin{subfigure}[b]{0.7\textwidth}
         \centering
         \includegraphics[width=\textwidth]{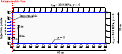}
         \caption{}
         \label{fig:Shadow-config}
     \end{subfigure} \\ \hspace{12mm}
    \begin{subfigure}[b]{0.55\textwidth} 
         \centering
         \includegraphics[width=\textwidth]{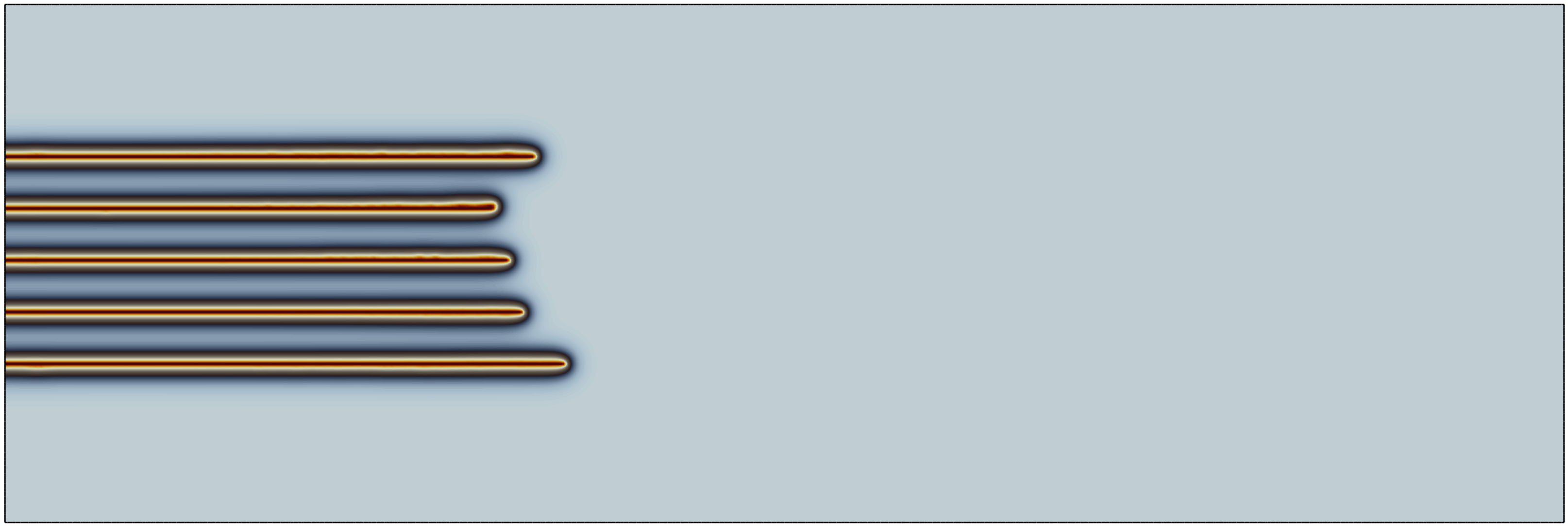}
         \caption{Time= 500 s}
         \label{fig:Shadow-500-Phi}
     \end{subfigure} \hspace{3mm}
     \begin{subfigure}[b]{0.06\textwidth}
         \includegraphics[width=\textwidth]{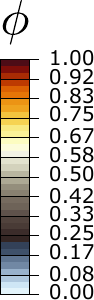}
         \vspace{0mm}
     \end{subfigure} \\ \hspace{19mm}
    \begin{subfigure}[b]{0.55\textwidth}
         \centering
         \includegraphics[width=\textwidth]{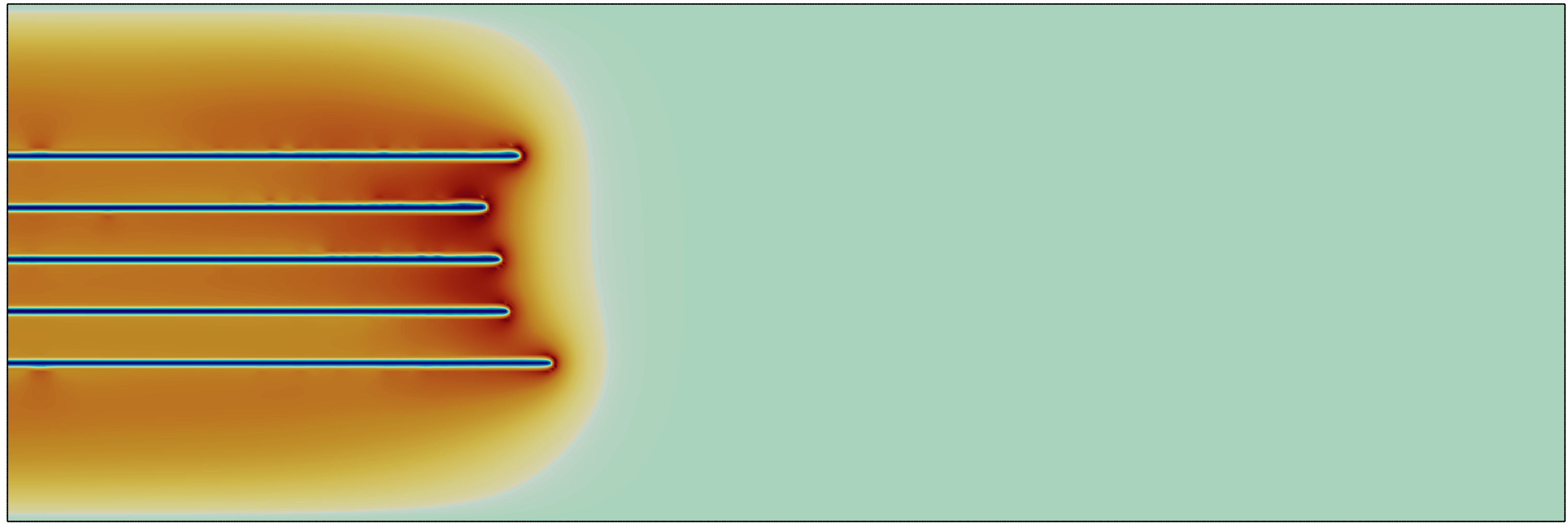}
         \caption{Time= 500 s}
         \label{fig:Shadow-500-P}
     \end{subfigure} \hspace{3mm}
     \begin{subfigure}[b]{0.09\textwidth}
         \includegraphics[width=\textwidth]{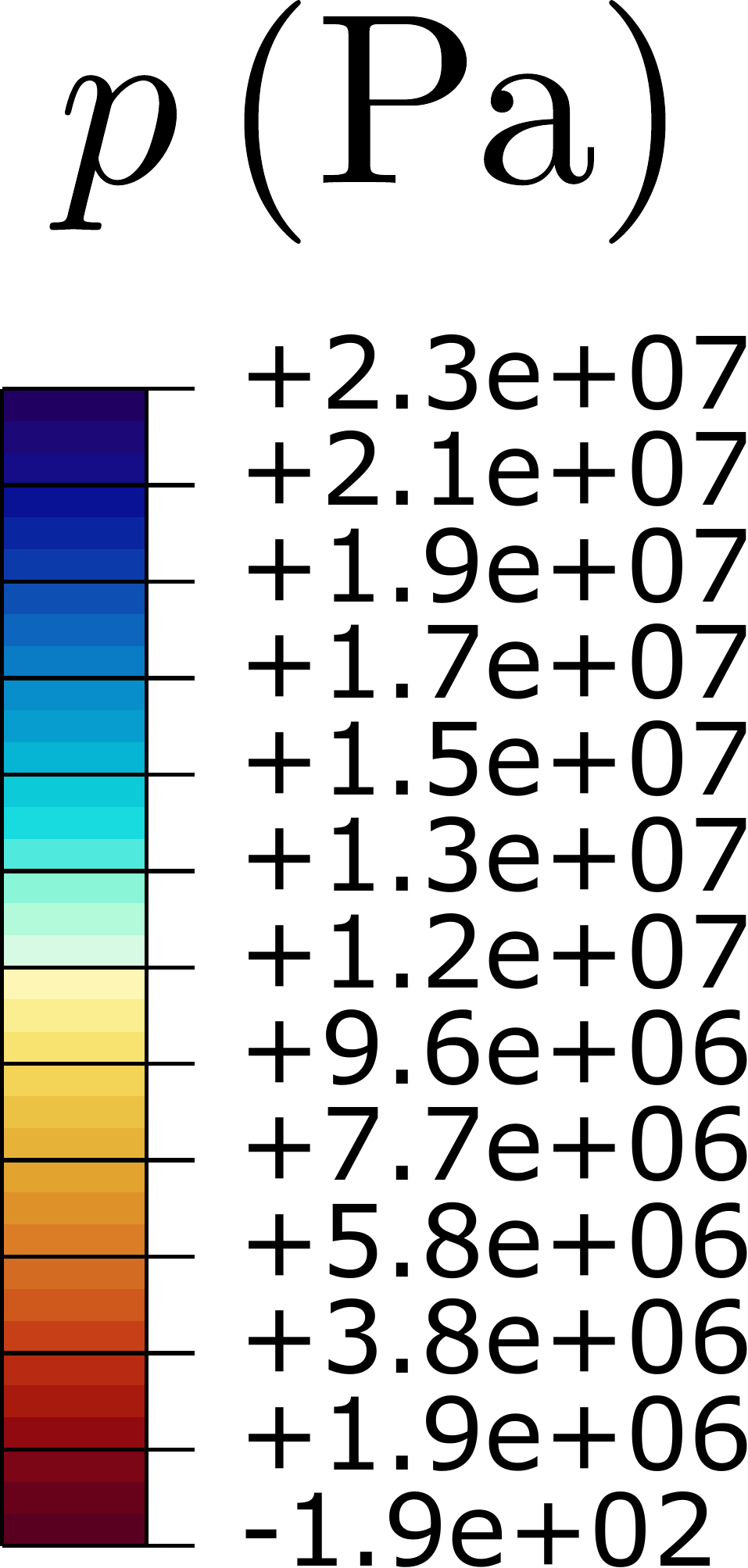}
         \vspace{0mm}
     \end{subfigure} \\ \hspace{12mm}
     \begin{subfigure}[b]{0.55\textwidth}
         \centering
         \includegraphics[width=\textwidth]{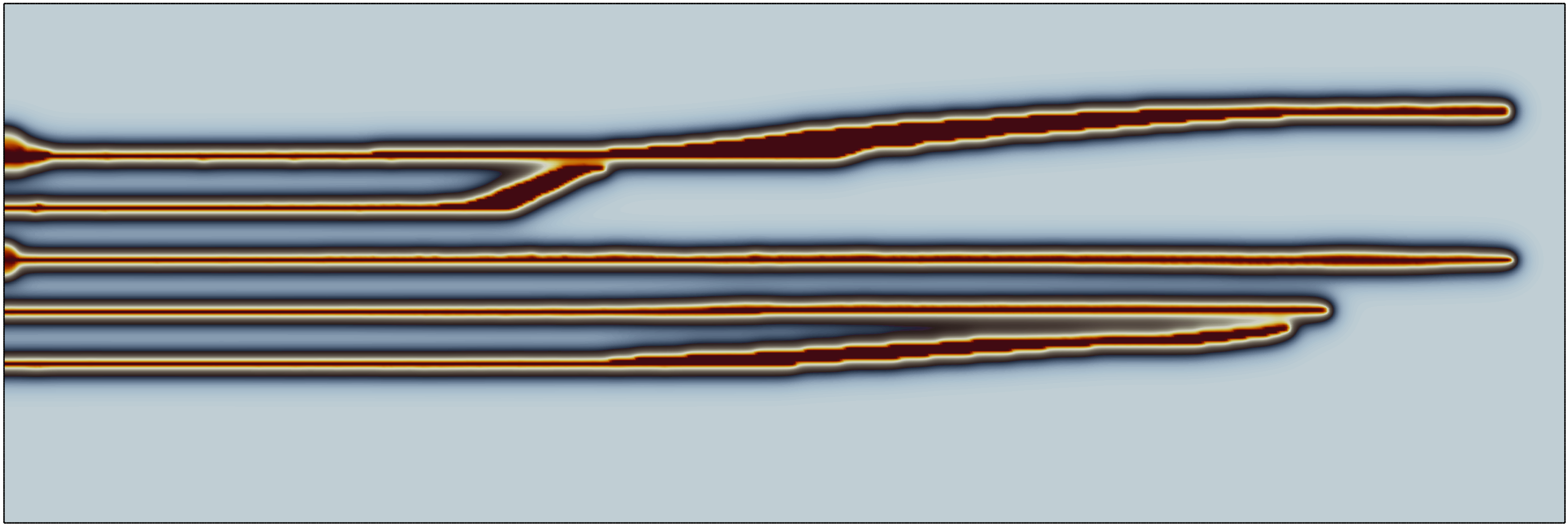}
         \caption{Time= 1000 s}
         \label{fig:Shadow-1000-Phi}
     \end{subfigure} \hspace{3mm}
     \begin{subfigure}[b]{0.06\textwidth}
         \includegraphics[width=\textwidth]{Legend-PHI.pdf}
         \vspace{0mm}
     \end{subfigure} \\ \hspace{19mm}
     \begin{subfigure}[b]{0.55\textwidth}
         \centering
         \includegraphics[width=\textwidth]{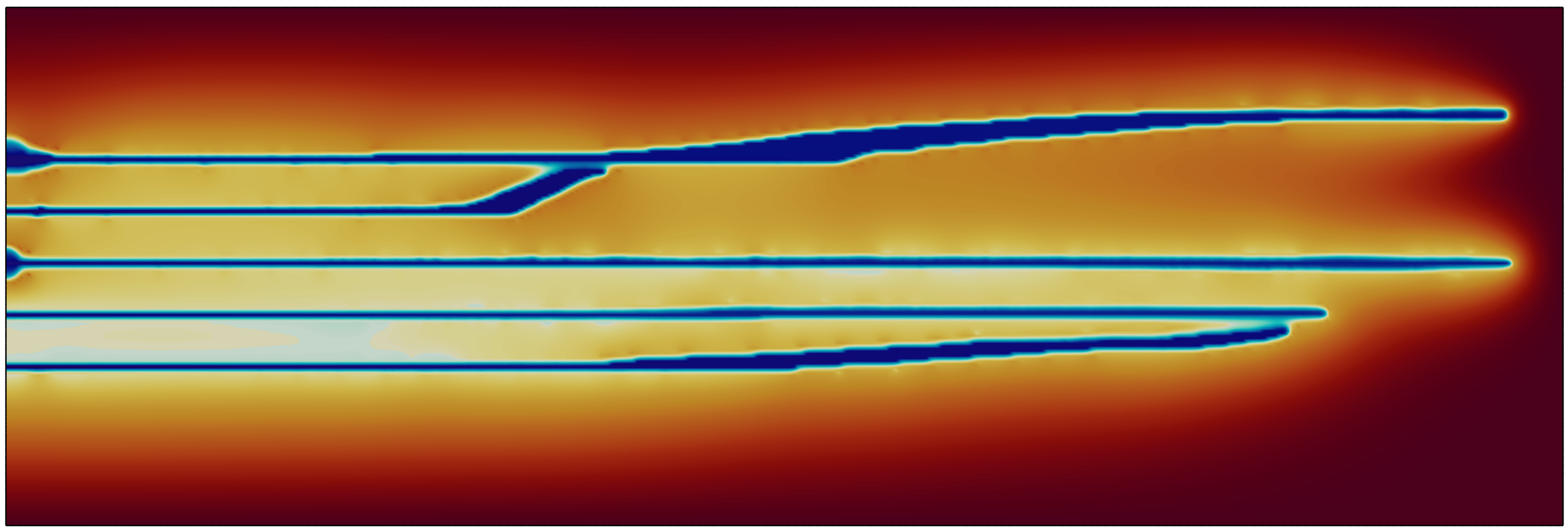}
         \caption{Time= 1000 s}
         \label{fig:Shadow-1000-P}
     \end{subfigure} \hspace{3mm}
     \begin{subfigure}[b]{0.09\textwidth}
         \includegraphics[width=\textwidth]{Shadow-legend-1000-P.pdf}
         \vspace{0mm}
     \end{subfigure}
    \caption{ Injection into an axisymmetric boundary: (a) Geometry and boundary conditions, (b) contour of phase field at time $t=500$ s, (c) contour of pore pressure at $t=500$ s, (d) contour of phase field at $t=1000$ s, and (e) contour of pore pressure at $t=1000$ s.}
    \label{fig:Shadow1}
\end{figure}

To demonstrate the robustness of the current implementation, we analyse a complex problem involving multiple injections along an axisymmetric boundary, as depicted in Fig. \ref{fig:Shadow-config}. The material properties are identical to those used in the second case study, with the following exceptions: Young’s modulus \( E=53 \) GPa, Poisson’s ratio \( \nu=0.19 \), critical fracture energy release rate \( G_c=500 \) N/m, and characteristic length scale \( \ell=0.29 \) m. The geometry is discretised with over 55,000 bilinear quadrilateral elements, and the mixed staggered scheme is applied with a time increment of 1 s over a total injection period of 2400 s. The hybrid permeability method is used, with initial stresses applied in both horizontal and vertical directions. Specifically, a horizontal stress of \( \sigma_{xx}=34.9 \) MPa and a vertical stress of \( \sigma_{yy}=20.9 \) MPa are applied in the first step to establish the initial stress state. In the second step, five pre-existing cracks are introduced, followed by five simultaneous injections at a rate of $q=126 \, \text{kg} \cdot \text{s}^{-1} \cdot \text{m}^{-2} $ over 2400 s in the third step. The top, bottom, and right edges of the domain are considered to be pore-pressure-free boundaries (\( p=0 \)).

Crack growth is observed at each of the five injection points. Figures \ref{fig:Shadow-500-Phi} and \ref{fig:Shadow-500-P} show the initial propagation of cracks horizontally up to 500 s. Beyond this point, the crack originating from injection point 4 begins to approach the crack from injection point 5. At this stage, the crack originating from injection point 4 is coalescing with the crack from injection point 5 through crack interactions, leading to increased fluid flux within the crack at injection point 5 and further propagation through other cracks. Crack propagation from injection points 2 and 3 remains horizontal throughout the injection, while the crack from injection point 1 deviates after 610 s. Figures \ref{fig:Shadow-1000-Phi} and \ref{fig:Shadow-1000-P} illustrate the phase field and pressure contours at 1000 s, highlighting continued propagation and interaction between cracks.

\begin{figure}[H]
    \centering
    \begin{subfigure}[b]{0.55\textwidth}
         \centering
         \includegraphics[width=\textwidth]{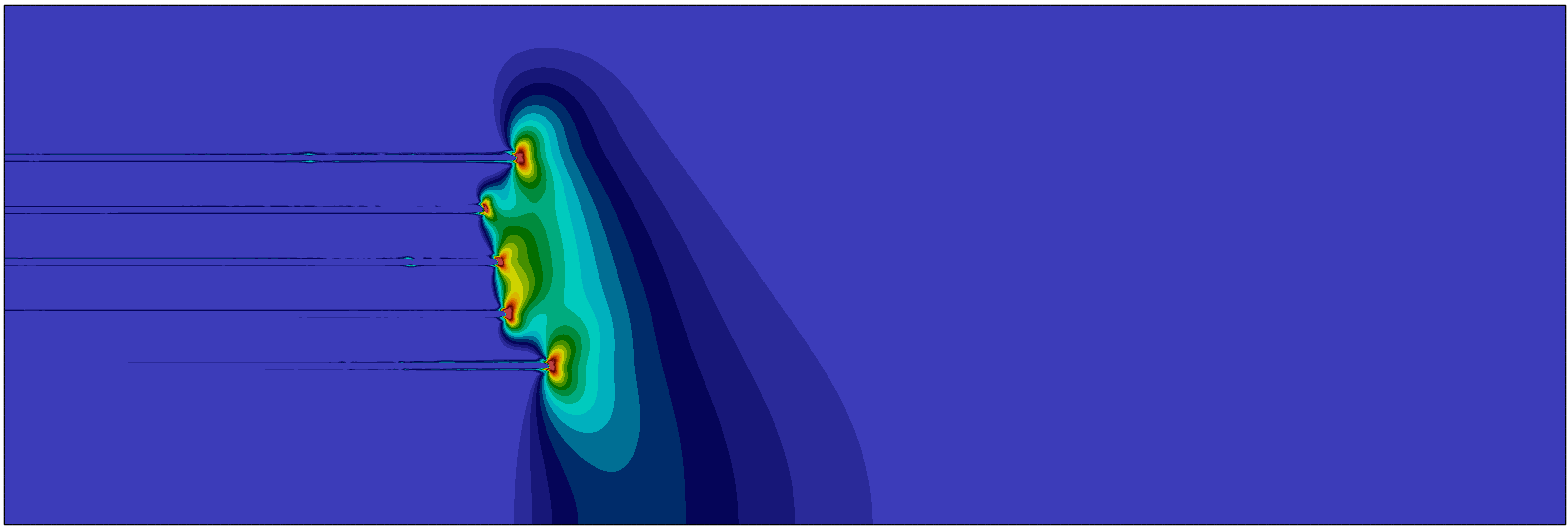}
         \caption{Time= 500 s}
     \end{subfigure} \\ \hspace{20mm}
    \begin{subfigure}[b]{0.55\textwidth} 
         \centering
         \includegraphics[width=\textwidth]{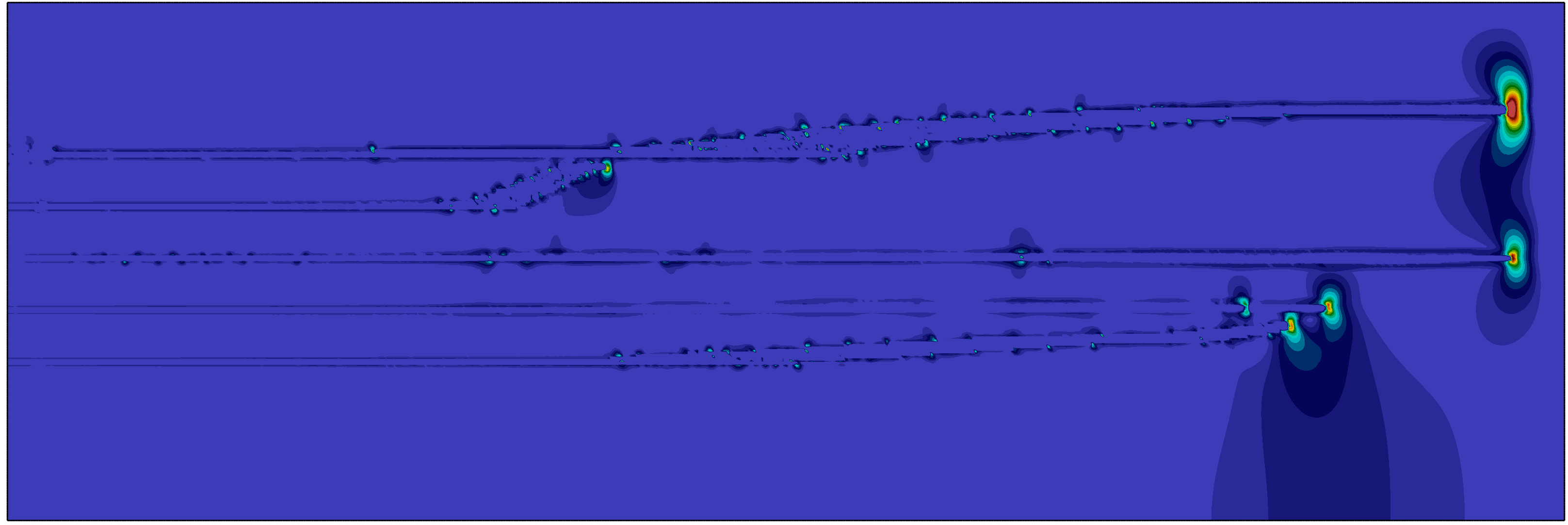}
         \caption{Time= 1000 s}
     \end{subfigure} \hspace{3mm}
     \begin{subfigure}[b]{0.09\textwidth}
         \includegraphics[width=\textwidth]{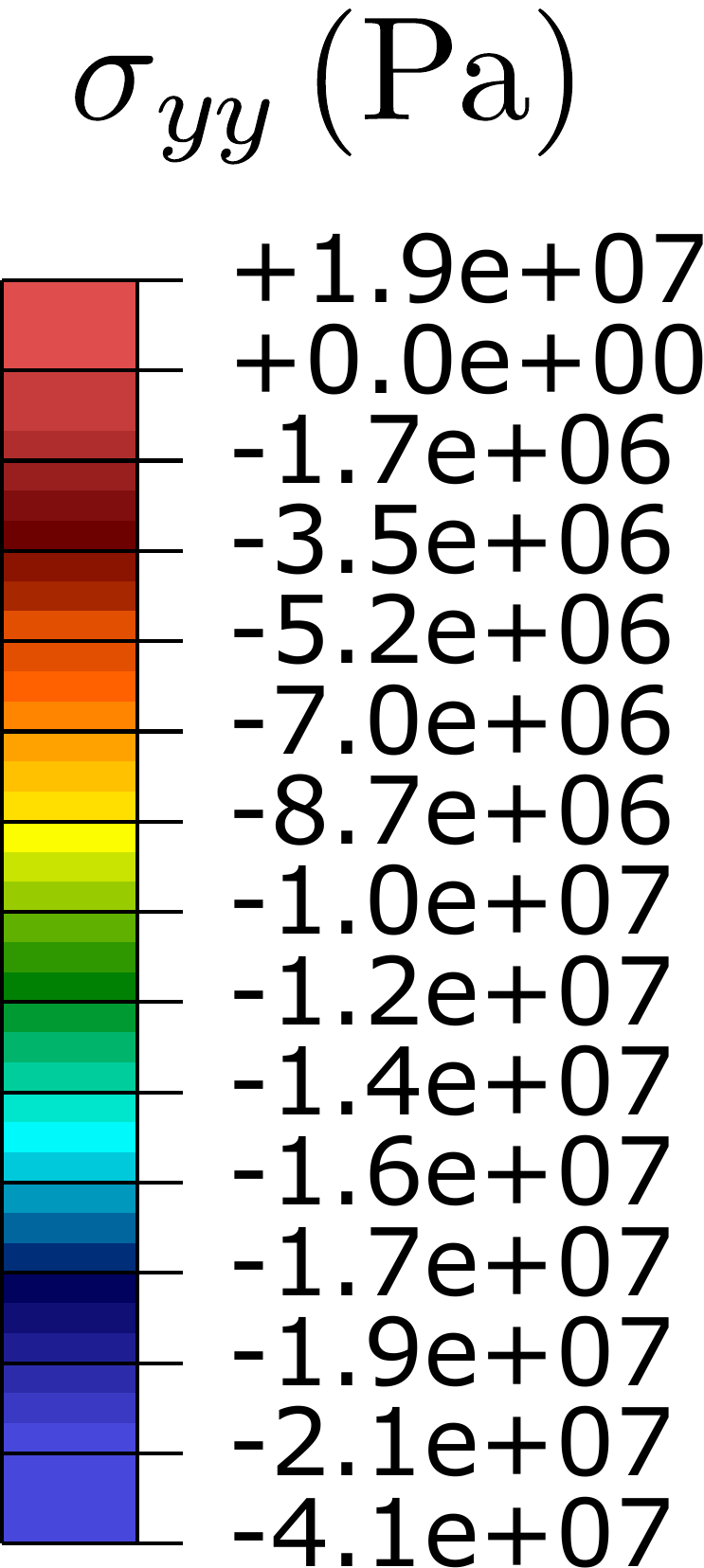}
         \vspace{0mm}
     \end{subfigure} \\ \hspace{12mm}
    \caption{Vertical stress \( \sigma_{yy} \) contour at: (a) Time= 500 s, and (b) Time= 1000 s.}
    \label{fig:Shadow2}
\end{figure}

Finally, vertical stress \( \sigma_{yy} \) contours at 500 s and 1000 s are shown in Fig. \ref{fig:Shadow2}. Throughout the domain, compressive vertical stress increases, except at the crack tip, where tensile stress drives further crack propagation.

\section{Conclusions}
\label{Sec:Conclusions}

We have presented a novel, theoretical and computational framework to simulate hydraulic fractures based on the phase field method. The model encompasses a number of relevant features, from constitutive choices to solution schemes, and introduces two key innovations that enhance the accuracy and adaptability of hydraulic fracture simulations. First, a novel hybrid coupling approach is introduced to link the phase field evolution equation with the pore pressure equation. This approach offers refined control over permeability transitions, making it especially effective in capturing fluid flow and fracture propagation interactions in complex geomechanical problems. Second, we incorporate a Drucker-Prager-based strain energy split to model stick-slip behaviour accurately, a critical aspect in hydraulic fracturing and fault activation scenarios.\\

Through a series of representative case studies, we demonstrated the robustness and versatility of our proposed formulation. The results highlighted the significant impact of different coupling strategies on fluid behaviour within fractures and underscored the influence of strain energy decomposition on fracture propagation paths and crack interactions. Our hybrid permeability approach effectively addressed the limitations of existing methods, providing a more flexible solution for hydraulic fracturing scenarios characterised by complex geometries and evolving fracture patterns. Additionally, the Drucker-Prager-based split effectively modelled the transition between elastic, frictional, and fully fractured states, yielding new insights into the role of shear stress in hydraulic fracture propagation.\\

The phase field framework developed in this work not only extends the current capabilities of hydraulic fracture modelling but also serves as a valuable tool for a wide range of geomechanical applications, including reservoir engineering, fault activation, and fracture interaction under multiaxial stress conditions.

\section*{Acknowledgments}
\label{Sec:Acknowledge of funding}

\noindent Y. Navidtehrani and C. Betegón acknowledge financial support from the FUO contract 22-364 and from the Ministry of Science, Innovation and Universities of Spain through grant PGC2018-099695-B-I00.  C. Betegón acknowledges financial support from the Ministry of Science, Innovation, and Universities of Spain under grant MCINN-23-PID2022-1420150B-100. J. Vallejos acknowledges the financial suport from CONICYT (Project AFB230001). E. Mart\'{\i}nez-Pa\~neda was supported by an UKRI Future Leaders Fellowship (grant MR/V024124/1). 

\appendix

\section{Abaqus implementation}
\label{Sec:Abaqus}

The generalised phase field model for hydraulic fracture presented is implemented in the commercial finite element package Abaqus in a very straightforward way, without the need for user element subroutines (i.e., at the integration point level). This is achieved by exploiting the analogy between the heat transfer balance equation and the fluid flow and phase field balance equations. The idea is similar to that exploited in Refs. \cite{navidtehrani2021unified,navidtehrani2021simple} to implement phase field fracture by means of (solely) a user material (\texttt{UMAT}) subroutine. However, on this occasion, the approach is extended to encompass an additional balance equation, as explained below.
\\

First, consider the heat transfer equation in its general form,
\begin{equation}\label{Eq:G-HEAT}
\rho\Dot{U}+{\nabla}\cdot\mathbf{f}={r},
\end{equation}

\noindent where \( U \) is the internal thermal energy, \( \mathbf{f} \) is the heat flux vector, and \( r \) is the heat source. Rearranging the phase field evolution (\ref{eq:PhaseFieldStrongForm}) and fluid flow (\ref{Eq:dMassContent3}) equations,
\begin{equation}\label{Eq:G Phase}
\left(\frac{g^{\prime}(\phi) \mathcal{H} 2 c_w}{\ell G_c}+\frac{w^{\prime}(\phi)}{2 \ell^2}\right)-\nabla \cdot \left(\nabla \phi \right)=0
\end{equation}
\begin{equation}\label{Eq:G Fluid}
\rho_{fl} \left(S\Dot{p}+\alpha\chi_r\Dot{\varepsilon}_{vol}\right)-\nabla\cdot\left(\rho_{fl}\frac{\bm{K}_{fl}}{\mu_{fl}}{\nabla p}\right)= {q}_{m}.
\end{equation}

By comparing Eqs. (\ref{Eq:G Phase})-(\ref{Eq:G Fluid}) with Eq. (\ref{Eq:G-HEAT}), we observe an analogy between these diffusion-like equations. Table \ref{tab:Analogy} summarizes the analogous variables across the heat transfer, phase field, and fluid flow equations.

\begin{table}[H]
\caption{Analogy of variables between heat transfer, phase field, and fluid flow equations.}
\label{tab:Analogy}
\begin{center}
\begin{tabular}{ c  c  c }
\hline
 Heat Transfer Equation & Phase Field Evolution Equation & Fluid Flow Equation \\ 
\hline \hline
$\rho\Dot{U}+{\nabla}\cdot\mathbf{f}={r}$ & $\left(\frac{g^{\prime}(\phi) \mathcal{H} 2 c_w}{\ell G_c}+\frac{w^{\prime}(\phi)}{2 \ell^2}\right)-\nabla \cdot \left(\nabla \phi \right)=0$ & $\rho_{fl} \left(S\Dot{p}+\alpha\chi _r\Dot{\varepsilon}_{vol}\right)-\nabla\cdot\left(\rho_{fl}\frac{\bm{K}_{fl}}{\mu_{fl}}{\nabla p}\right)= {q}_{m}$ \\
 $T$ & $\phi$ & $p$ \\
$\mathbf{f}$ & $-\nabla \phi$ & $-\rho_{fl}\frac{\bm{K}_{fl}}{\mu_{fl}}{\nabla p}$ \\
 $\rho$ & $1$ & $1$ \\
$\Dot{U}$ & $\frac{g^{\prime}(\phi) \mathcal{H} 2 c_w}{\ell G_c}+\frac{w^{\prime}(\phi)}{2 \ell^2}$ & $\rho_{fl} \left(S\Dot{p}+\alpha\chi _r\Dot{\varepsilon}_{vol}\right)$\\
 $r$  & $0$ & ${q}_{m}$\\   
\hline
\end{tabular}
\end{center}
\end{table}

The numerical implementation is carried out in Abaqus using a \texttt{UMATHT} subroutine, where several variables must be defined to establish equivalence with heat transfer and mass diffusion variables. Table \ref{tab:UMATHT} outlines the required quantities and their corresponding expressions for the heat transfer, phase field, and fluid flow equations.

\begin{table}[H]
\caption{Variables that must be defined in a \texttt{UMATHT} subroutine and their associated expressions for the heat transfer, phase field and fluid flow problems.}
\label{tab:UMATHT}
\begin{center}
\begin{tabular}{ c  c  c  c }
\hline
 UMATHT variable & Heat transfer & Phase field  & Fluid flow  \\ 
\hline
$U$ & $U_{t}+\dot{U} \Delta t$ & $U_{t}+\left(\frac{g^{\prime}(\phi) \mathcal{H} 2 c_w}{\ell G_c}+\frac{w^{\prime}(\phi)}{2 \ell^2}\right) \Delta t$ & $U_{t}+\rho_{fl} \left(S\Dot{p}+\alpha\chi _r\Dot{\varepsilon}_{vol}\right) \Delta t$ \\ 
DUDT & $\frac{\partial U}{\partial T}$ & $\left(\frac{g^{\prime\prime}(\phi) \mathcal{H} 2 c_w}{\ell G_c}+\frac{w^{\prime\prime}(\phi)}{2 \ell^2}\right) \Delta t$ & $\rho_{fl} S \Delta t$ \\
DUDG & $\frac{\partial U}{\partial\left(\nabla T\right)}$ & $0$ & $0$ \\
 FLUX &$\mathbf{f}$ & $-\nabla \phi$ & $-\rho_{fl}\frac{\bm{K}_{fl}}{\mu_{fl}}{\nabla p}$ \\
 DFDT & $\Dot{\mathbf{f}}$ & $0$ & $0$ \\
 DFDG & $\frac{\partial \mathbf{f}}{\partial\left(\nabla T\right)}$ & $-\bm{I}$ & $-\rho_{fl} \frac{\bm{K}_{fl}}{\mu_{fl}}$ \\
\hline
\end{tabular}
\end{center}
\end{table}

By exploiting this analogy, the temperature variable \( T \) becomes equivalent to the phase field variable \( \phi \), which varies between 0 and 1, or to the pore pressure variable \( p \). To account for these analogies, a user material (\texttt{UMAT}) subroutine is employed to degrade both the material stiffness and the stress tensor with respect to the phase field variable, while incorporating as well the effect of pore pressure on the total stress using Biot's coefficient \( \alpha \). The evolution equation for the phase field and the fluid flow equation are subsequently addressed using the \texttt{UMATHT} subroutine, which defines the internal heat energy \( U \) and the heat flux vector \( \mathbf{f} \) along with their respective variations concerning temperature \( T \) and the temperature gradient \( \nabla T \), as shown in Table \ref{tab:UMATHT}. \\

Different to Refs. \cite{navidtehrani2021unified,navidtehrani2021simple}, the temperature field is now used to describe two fields: phase field $\phi$ and pore pressure $p$. This can be accomplished by defining two identical Abaqus \texttt{Part}s with the same geometry and mesh. Since only one temperature can be defined per integration point, this approach enables effective data transfer between the two parts. Local numbering for elements and nodes remains consistent across both parts if the same meshing algorithm is used. \\

The proposed procedure is as follows: for a given element, Abaqus provides the \texttt{UMAT} integration point-level subroutine with values of strain and phase field (temperature) in the first part, interpolated from the nodal solutions. The pore pressure, represented as the temperature field in the second part, is stored in a FORTRAN module and transferred to the \texttt{UMAT} subroutine using local element numbering. Within each integration point and loop, the \texttt{UMAT} is first called. Inside the \texttt{UMAT}, the material Jacobian \( \bm{C} \) and effective stress \( \bm{\sigma}^{eff} \) are computed from the strain tensor. The current phase field value \( \phi \) (temperature in the first part) is then used to account for the degradation of these quantities, while the pore pressure value \( p \) (temperature in the second part) is used to compute the total stress \( \bm{\sigma} \). The dissipation part of the strain energy density \( \psi_d \) is stored in solution-dependent state variables (\texttt{SDV}s), enabling enforcement of the irreversibility condition. The rate of volumetric strain \( \Dot{\varepsilon}_{vol} \) is stored within a FORTRAN module and transferred to the \texttt{UMATHT} subroutine to solve the fluid flow equation. In the \texttt{UMATHT}  subroutine, definitions of internal heat energy \( U \), heat flux vector \( \mathbf{f} \), and their variations (\( \partial U / \partial T \), \( \partial U / \partial \nabla T \), \( \partial \mathbf{f} / \partial T \), \( \partial \mathbf{f} / \partial \nabla T \)) are performed for the phase field equation in the first part and for the fluid flow equation in the second part. The \texttt{UMATHT}  subroutine distinguishes between parts based on the material name, where MATERIAL-1 denotes the first part (deformation and phase field problems) and MATERIAL-2 denotes the second part (fluid flow equation). \\

The updated \texttt{SDV}s are transferred to the \texttt{UMATHT} subroutine to carry the current value of the history field \( \mathcal{H} \) without requiring external FORTRAN modules. Additionally, the values of volumetric strain rate and the phase field variable are transferred from the \texttt{UMAT} subroutine to the \texttt{UMATHT} subroutine, which manages the fluid flow equation, via a FORTRAN module. This process is repeated for each integration point, allowing Abaqus to assemble the element stiffness matrices and residuals externally and subsequently form the global system of equations, as per the procedure outlined in Algorithm \ref{alg:Solution Scheme}. The coupled problem can be approached using either a \emph{monolithic} or \emph{staggered} scheme. In the monolithic scheme, all variables are updated simultaneously, resulting in unconditional stability. In contrast, the staggered scheme updates variables sequentially, with some equations utilizing variables from the previous increment or iteration instead of the current one, as described in Section \ref{Sec:Solution}. \\

This implementation offers two types of schemes: the Mixed monolithic scheme and the mixed staggered scheme. In the mixed monolithic scheme, equilibrium and phase field evolution equations are solved using a monolithic approach, while the pore pressure is derived from the previous iteration. In the mixed staggered scheme, the history field \( \mathcal{H} \) is not updated within an increment for the phase field evolution equation; instead, the previous increment’s history field \( \mathcal{H} \) is applied. However, as in the mixed monolithic scheme, pore pressure is updated based on the solution of the previous iteration.


\begin{thebibliography}{10}
\expandafter\ifx\csname url\endcsname\relax
  \def\url#1{\texttt{#1}}\fi
\expandafter\ifx\csname urlprefix\endcsname\relax\def\urlprefix{URL }\fi
\expandafter\ifx\csname href\endcsname\relax
  \def\href#1#2{#2} \def\path#1{#1}\fi

\bibitem{barati2014review}
R.~Barati, J.-T. Liang, A review of fracturing fluid systems used for hydraulic
  fracturing of oil and gas wells, Journal of Applied Polymer Science 131~(16)
  (2014).

\bibitem{Rojas2017}
E.~Rojas, P.~Landeros, Hydraulic fracturing applied to tunnel development at el
  teniente mine, in: Proceedings of the Ninth Rockburst and Seismicity in Mines
  Conference, 2017, pp. 251--257.

\bibitem{LEI2023102692}
Z.~Lei, Y.~Zhang, S.~Zhang, Y.~Shi, Numerical study of hydraulic fracturing
  treatments and geothermal energy extraction from a naturally fractured
  granitic formation, Geothermics 111 (2023) 102692.

\bibitem{SAMPATH2018251}
K.~Sampath, M.~Perera, P.~Ranjith, Theoretical overview of hydraulic fracturing
  break-down pressure, Journal of Natural Gas Science and Engineering 58 (2018)
  251--265.

\bibitem{LECAMPION201866}
B.~Lecampion, A.~Bunger, X.~Zhang, Numerical methods for hydraulic fracture
  propagation: A review of recent trends, Journal of Natural Gas Science and
  Engineering 49 (2018) 66--83.

\bibitem{2205_85_Vallejos}
R.~Navarrete, C.~Soto, F.~Henriquez, H.~Godoy, Hydraulic fracturing in the
  construction of andes norte project, in: Caving 2022: Proceedings of the
  Fifth International Conference on Block and Sublevel Caving, Australian
  Centre for Geomechanics, pp. 1227--1240.

\bibitem{2205_81_Amorer}
G.~Amorer, S.~Duffield, J.~De~Ross, G.~Viegas, Surface hydraulic fracturing
  trial at cadia east, in: Caving 2022: Proceedings of the Fifth International
  Conference on Block and Sublevel Caving, Australian Centre for Geomechanics,
  pp. 1173--1188.

\bibitem{YI2020113396}
L.-P. Yi, H.~Waisman, Z.-Z. Yang, X.-G. Li, A consistent phase field model for
  hydraulic fracture propagation in poroelastic media, Computer Methods in
  Applied Mechanics and Engineering 372 (2020) 113396.

\bibitem{Bourdin2012a}
B.~Bourdin, C.~Chukwudozie, K.~Yoshioka, {A variational approach to the
  numerical simulation of hydraulic fracturing}, Proceedings - SPE Annual
  Technical Conference and Exhibition 2 (2012) 1442--1452.

\bibitem{Wheeler2014}
M.~F. Wheeler, T.~Wick, W.~Wollner, {An augmented-Lagrangian method for the
  phase-field approach for pressurized fractures}, Computer Methods in Applied
  Mechanics and Engineering 271 (2014) 69--85.

\bibitem{mikelic2015phase}
A.~Mikeli{\'c}, M.~F. Wheeler, T.~Wick, Phase-field modeling of a fluid-driven
  fracture in a poroelastic medium, Computational Geosciences 19 (2015)
  1171--1195.

\bibitem{Mikelic2015}
A.~Mikelic, M.~Wheeler, T.~Wick, {A phase-field method for propagating
  fluid-filled fractures coupled to a surrounding porous medium}, Multiscale
  Modeling and Simullation 13 (2015) 367--398.

\bibitem{mikelic2015quasi}
A.~Mikeli{\'c}, M.~F. Wheeler, T.~Wick, A quasi-static phase-field approach to
  pressurized fractures, Nonlinearity 28~(5) (2015) 1371.

\bibitem{hageman2023phase}
T.~Hageman, E.~Mart{\'\i}nez-Pa{\~n}eda, A phase field-based framework for
  electro-chemo-mechanical fracture: Crack-contained electrolytes, chemical
  reactions and stabilisation, Computer Methods in Applied Mechanics and
  Engineering 415 (2023) 116235.

\bibitem{miehe2015minimization}
C.~Miehe, S.~Mauthe, S.~Teichtmeister, Minimization principles for the coupled
  problem of darcy--biot-type fluid transport in porous media linked to phase
  field modeling of fracture, Journal of the Mechanics and Physics of Solids 82
  (2015) 186--217.

\bibitem{Miehe2016}
C.~Miehe, S.~Mauthe, {Phase field modeling of fracture in multi-physics
  problems. Part III. Crack driving forces in hydro-poro-elasticity and
  hydraulic fracturing of fluid-saturated porous media}, Computer Methods in
  Applied Mechanics and Engineering 304 (2016) 619--655.

\bibitem{Wilson2016}
Z.~A. Wilson, C.~M. Landis, {Phase-field modeling of hydraulic fracture},
  Journal of the Mechanics and Physics of Solids 96 (2016) 264--290.

\bibitem{EHLERS2018429}
W.~Ehlers, C.~Luo, A phase-field approach embedded in the theory of porous
  media for the description of dynamic hydraulic fracturing, part ii: The
  crack-opening indicator, Computer Methods in Applied Mechanics and
  Engineering 341 (2018) 429--442.

\bibitem{Heider2017}
Y.~Heider, B.~Markert, {A phase-field modeling approach of hydraulic fracture
  in saturated porous media}, Mechanics Research Communications 80 (2017)
  38--46.

\bibitem{Lee2016}
S.~Lee, M.~F. Wheeler, T.~Wick, Pressure and fluid-driven fracture propagation
  in porous media using an adaptive finite element phase field model, Computer
  Methods in Applied Mechanics and Engineering 305 (2016) 111--132.

\bibitem{Zhou2018c}
S.~Zhou, X.~Zhuang, T.~Rabczuk, A phase-field modeling approach of fracture
  propagation in poroelastic media, Engineering Geology 240 (2018) 189--203.

\bibitem{LI2021107887}
P.~Li, D.~Li, Q.~Wang, K.~Zhou, Phase-field modeling of hydro-thermally induced
  fracture in thermo-poroelastic media, Engineering Fracture Mechanics 254
  (2021) 107887.

\bibitem{LEE2025126487}
S.~Lee, M.~F. Wheeler, T.~Wick, A phase-field diffraction model for
  thermo-hydro-mechanical propagating fractures, International Journal of Heat
  and Mass Transfer 239 (2025) 126487.

\bibitem{WANG2025117750}
X.~Wang, P.~Li, D.~Lu, Phase-field hydraulic fracturing operator network based
  on en-deeponet with integrated physics-informed mechanisms, Computer Methods
  in Applied Mechanics and Engineering 437 (2025) 117750.

\bibitem{lee2017iterative}
S.~Lee, M.~F. Wheeler, T.~Wick, Iterative coupling of flow, geomechanics and
  adaptive phase-field fracture including level-set crack width approaches,
  Journal of Computational and Applied Mathematics 314 (2017) 40--60.

\bibitem{yoshioka2020crack}
K.~Yoshioka, D.~Naumov, O.~Kolditz, On crack opening computation in variational
  phase-field models for fracture, Computer Methods in Applied Mechanics and
  Engineering 369 (2020) 113210.

\bibitem{https://doi.org/10.1002/2016JB013572}
D.~Santillán, R.~Juanes, L.~Cueto-Felgueroso, Phase field model of
  fluid-driven fracture in elastic media: Immersed-fracture formulation and
  validation with analytical solutions, Journal of Geophysical Research: Solid
  Earth 122~(4) (2017) 2565--2589.

\bibitem{Zhou2019c}
S.~Zhou, X.~Zhuang, T.~Rabczuk, {Phase-field modeling of fluid-driven dynamic
  cracking in porous media}, Computer Methods in Applied Mechanics and
  Engineering 350~(8) (2019) 169--198.

\bibitem{SHAHOVEISI2024117113}
S.~Shahoveisi, M.~Vahab, B.~Shahbodagh, S.~Eisenträger, N.~Khalili,
  Phase-field modelling of dynamic hydraulic fracturing in porous media using a
  strain-based crack width formulation, Computer Methods in Applied Mechanics
  and Engineering 429 (2024) 117113.

\bibitem{yang2024phase}
L.~Yang, Y.~Ma, G.~Yang, Z.~Liu, K.~Kang, M.~Zhang, Z.~Wang, Phase field
  modeling of hydraulic fracturing with length-scale insensitive degradation
  functions, Energies 17~(20) (2024) 5210.

\bibitem{Lo2023}
Y.-S. Lo, T.~J. Hughes, C.~M. Landis, Phase-field fracture modeling for large
  structures, Journal of the Mechanics and Physics of Solids 171 (2023) 105118.

\bibitem{Aldakheel2021}
F.~Aldakheel, N.~Noii, T.~Wick, P.~Wriggers, {A global–local approach for
  hydraulic phase-field fracture in poroelastic media}, Computers and
  Mathematics with Applications 91 (2021) 99--121.

\bibitem{maurini2014crack}
C.~Maurini, B.~Bourdin, G.~Gauthier, V.~Lazarus, Crack patterns obtained by
  unidirectional drying of a colloidal suspension in a capillary tube:
  experiments and numerical simulations using a two-dimensional variational
  approach, in: Fracture Phenomena in Nature and Technology: Proceedings of the
  IUTAM Symposium on Fracture Phenomena in Nature and Technology held in
  Brescia, Italy, 1-5 July 2012, Springer, 2014, pp. 75--91.

\bibitem{LUO2023115962}
C.~Luo, L.~Sanavia, L.~{De Lorenzis}, Phase-field modeling of drying-induced
  cracks: Choice of coupling and study of homogeneous and localized damage,
  Computer Methods in Applied Mechanics and Engineering 410 (2023) 115962.

\bibitem{Heider2021}
Y.~Heider, {A review on phase-field modeling of hydraulic fracturing},
  Engineering Fracture Mechanics 253~(June) (2021) 1--24.

\bibitem{chen2022review}
B.~Chen, B.~R. Barboza, Y.~Sun, J.~Bai, H.~R. Thomas, M.~Dutko, M.~Cottrell,
  C.~Li, A review of hydraulic fracturing simulation, Archives of Computational
  Methods in Engineering (2022) 1--58.

\bibitem{2063_15_Rimmelin}
R.~Rimmelin, G.~Chitombo, E.~Rojas, Hydraulic fracturing in cave mining:
  Opportunities for improvement, in: MassMin 2020: Proceedings of the Eighth
  International Conference \& Exhibition on Mass Mining, University of Chile,
  pp. 275--288.

\bibitem{Pardo201611}
C.~Pardo, E.~Rojas, Selection of exploitation method based on the experience of
  hydraulic fracture techniques at the el teniente mine, in: Proceedings of the
  Seventh International Conference and Exhibition on Mass Mining (Massmin
  2016), The Australasian Institute of Mining and Metallurgy, pp. 97--103.

\bibitem{Navidtehrani2022}
Y.~Navidtehrani, C.~Beteg{\'{o}}n, E.~Mart{\'{i}}nez-Pa{\~{n}}eda, {A general
  framework for decomposing the phase field fracture driving force,
  particularised to a Drucker–Prager failure surface}, Theoretical and
  Applied Fracture Mechanics 121 (2022) 103555.

\bibitem{molnar2020toughness}
G.~Moln{\'a}r, A.~Doitrand, R.~Estevez, A.~Gravouil, Toughness or strength?
  regularization in phase-field fracture explained by the coupled criterion,
  Theoretical and applied fracture mechanics 109 (2020) 102736.

\bibitem{de2022nucleation}
L.~De~Lorenzis, C.~Maurini, Nucleation under multi-axial loading in variational
  phase-field models of brittle fracture, International Journal of Fracture
  237~(1) (2022) 61--81.

\bibitem{lopez2025classical}
O.~Lopez-Pamies, J.~E. Dolbow, G.~A. Francfort, C.~J. Larsen, Classical
  variational phase-field models cannot predict fracture nucleation, Computer
  Methods in Applied Mechanics and Engineering 433 (2025) 117520.

\bibitem{Griffith1920}
A.~A. Griffith, {The Phenomena of Rupture and Flow in Solids}, Philosophical
  Transactions A, 221 (1920) 163--198.

\bibitem{kristensen2021assessment}
P.~K. Kristensen, C.~F. Niordson, E.~Mart{\'\i}nez-Pa{\~n}eda, An assessment of
  phase field fracture: crack initiation and growth, Philosophical Transactions
  of the Royal Society A 379~(2203) (2021) 20210021.

\bibitem{navidtehrani2021unified}
Y.~Navidtehrani, C.~Beteg{\'o}n, E.~Martinez-Paneda, A unified abaqus
  implementation of the phase field fracture method using only a user material
  subroutine, Materials 14~(8) (2021) 1913.

\bibitem{haghighat2023efficient}
E.~Haghighat, D.~Santill{\'a}n, An efficient phase-field model of shear
  fractures using deviatoric stress split, Computational Mechanics 72~(6)
  (2023) 1263--1278.

\bibitem{HESAMMOKRI2023112080}
P.~Hesammokri, H.~Yu, P.~Isaksson, An extended hydrostatic–deviatoric strain
  energy density decomposition for phase-field fracture theories, International
  Journal of Solids and Structures 262-263 (2023) 112080.

\bibitem{vicentini2024energy}
F.~Vicentini, C.~Zolesi, P.~Carrara, C.~Maurini, L.~De~Lorenzis, On the energy
  decomposition in variational phase-field models for brittle fracture under
  multi-axial stress states, International Journal of Fracture 247~(3) (2024)
  291--317.

\bibitem{Amor2009}
H.~Amor, J.~J. Marigo, C.~Maurini, {Regularized formulation of the variational
  brittle fracture with unilateral contact: Numerical experiments}, Journal of
  the Mechanics and Physics of Solids 57~(8) (2009) 1209--1229.

\bibitem{Miehe2010a}
C.~Miehe, M.~Hofacker, F.~Welschinger, {A phase field model for
  rate-independent crack propagation: Robust algorithmic implementation based
  on operator splits}, Computer Methods in Applied Mechanics and Engineering
  199~(45-48) (2010) 2765--2778.

\bibitem{Freddi2010}
F.~Freddi, G.~Royer-Carfagni, {Regularized variational theories of fracture: A
  unified approach}, Journal of the Mechanics and Physics of Solids 58~(8)
  (2010) 1154--1174.

\bibitem{DelPiero1989}
G.~{Del Piero}, {Constitutive equation and compatibility of the external loads
  for linear elastic masonry-like materials}, Meccanica 24~(3) (1989) 150--162.

\bibitem{darcy1856fontaines}
H.~Darcy, Les fontaines publiques de la ville de dijon; paris: Dalmont (1856).

\bibitem{Jaeger2009}
J.~Jaeger, N.~Cook, R.~Zimmerman, {Fundamentals of rock mechanics}, Blackwell
  Publishing, Oxford, UK, 2009.

\bibitem{XIE2012919}
N.~Xie, Q.-Z. Zhu, J.-F. Shao, L.-H. Xu, Micromechanical analysis of damage in
  saturated quasi brittle materials, International Journal of Solids and
  Structures 49~(6) (2012) 919--928.

\bibitem{jia2021experimental}
C.~Jia, S.~Zhang, W.~Xu, Experimental investigation and numerical modeling of
  coupled elastoplastic damage and permeability of saturated hard rock, Rock
  Mechanics and Rock Engineering 54 (2021) 1151--1169.

\bibitem{Ulloa2022}
J.~Ulloa, N.~Noii, R.~Alessi, F.~Aldakheel, G.~Degrande, S.~Fran{\c{c}}ois,
  {Variational modeling of hydromechanical fracture in saturated porous media:
  A micromechanics-based phase-field approach}, Computer Methods in Applied
  Mechanics and Engineering 396 (2022) 115084.

\bibitem{Zhou2018}
S.~Zhou, X.~Zhuang, H.~Zhu, T.~Rabczuk, {Phase field modelling of crack
  propagation, branching and coalescence in rocks}, Theoretical and Applied
  Fracture Mechanics 96 (2018) 174--192.

\bibitem{CHUKWUDOZIE2019957}
C.~Chukwudozie, B.~Bourdin, K.~Yoshioka, A variational phase-field model for
  hydraulic fracturing in porous media, Computer Methods in Applied Mechanics
  and Engineering 347 (2019) 957--982.

\bibitem{Bryant2018}
E.~C. Bryant, W.~C. Sun, {A mixed-mode phase field fracture model in
  anisotropic rocks with consistent kinematics}, Computer Methods in Applied
  Mechanics and Engineering 342 (2018) 561--584.

\bibitem{ZIAEIRAD2016304}
V.~Ziaei-Rad, L.~Shen, J.~Jiang, Y.~Shen, Identifying the crack path for the
  phase field approach to fracture with non-maximum suppression, Computer
  Methods in Applied Mechanics and Engineering 312 (2016) 304--321, phase Field
  Approaches to Fracture.

\bibitem{gerasimov2019penalization}
T.~Gerasimov, L.~De~Lorenzis, On penalization in variational phase-field models
  of brittle fracture, Computer Methods in Applied Mechanics and Engineering
  354 (2019) 990--1026.

\bibitem{kristensen2020phase}
P.~K. Kristensen, E.~Mart{\'\i}nez-Pa{\~n}eda, Phase field fracture modelling
  using quasi-newton methods and a new adaptive step scheme, Theoretical and
  Applied Fracture Mechanics 107 (2020) 102446.

\bibitem{GONZALEZ2022104975}
F.~Gonzalez, J.~Vallejos, E.~Rojas, P.~Landeros, Evaluation of the seismic rock
  mass response to mining and the impact of preconditioning using an
  epidemic-type aftershock model, International Journal of Rock Mechanics and
  Mining Sciences 150 (2022) 104975.

\bibitem{navidtehrani2021simple}
Y.~Navidtehrani, C.~Beteg{\'o}n, E.~Mart{\'\i}nez-Pa{\~n}eda, A simple and
  robust abaqus implementation of the phase field fracture method, Applications
  in Engineering Science 6 (2021) 100050.

\end{thebibliography}
\end{document}